\documentclass[useAMS,usenatbib]{mn2e}
\usepackage{graphicx}
\usepackage[hyperindex,pdftex]{hyperref}
\usepackage{lineno} 
 \usepackage{epstopdf}
 \usepackage{epsfig}
\usepackage{amsmath}
\usepackage{fixltx2e}
\usepackage{graphicx}
\usepackage{amssymb, amsmath}
\usepackage{times}
\usepackage{color}
\usepackage{lscape}
\usepackage[normalem]{ulem}

  \newcommand{\modif}[1]{\relax #1} 
    \newcommand{\modiff}[1]{\relax #1} 
\begin{document}

\title[Effect of smoothing of density field on Reconstruction.]{ The clustering of galaxies in the SDSS-III Baryon Oscillation Spectroscopic Survey: Effect of smoothing of density field  on reconstruction and anisotropic BAO analysis.}
\author[Vargas-Maga\~na et al.] {
    Mariana Vargas-Maga\~na\thanks{Email: mmaganav@fisica.unam.mx}$^{1}$, Shirley Ho$^{2,3}$, Sebastien. Fromenteau$^{2,3}$,  Antonio. J.~Cuesta$^{4}$\\
    $^{1}$ Instituto de Fisica, Universidad Nacional Aut\'onoma de M\'exico, Apdo. Postal 20-364, M\'exico. \\
    $^{2}$ Departments of Physics, Carnegie Mellon University, 5000 Forbes Ave., Pittsburgh, PA 15217. \\
    $^{3}$ McWilliams Center for Cosmology, Carnegie Mellon University, 5000 Forbes Ave., Pittsburgh, PA 15217 .\\
    $^{4}$ Institut de Ci{\`e}ncies del Cosmos (ICCUB), Universitat de Barcelona (IEEC-UB), Mart{\'\i} i Franqu{\`e}s 1, E-08028, Barcelona, Spain .\\
     }   
\maketitle
  \begin{abstract}
   {
   The reconstruction algorithm introduced by \cite{Eis07}, which is widely used in clustering analysis, is based on the inference of the first order Lagrangian displacement field from the Gaussian smoothed galaxy density field 
   in redshift space. The \modif2{smoothing scale} applied to the density field affects the inferred displacement field that is used to move {the galaxies}, and partially \modif2{erases} the nonlinear evolution {of the density field}.
     In this article, we explore this crucial step \modif2{in} the reconstruction algorithm. We study the performance of the reconstruction technique using two metrics: first, we study the performance using the anisotropic clustering, extending previous studies focused on isotropic clustering; second, we study its effect on the displacement field. 
We find that smoothing has a strong effect in the quadrupole of the correlation function and affects the accuracy and precision \modif2{with} which we can measure $D_A (z)$ and $H(z)$.  We find that the optimal smoothing scale to use in the reconstruction algorithm applied to BOSS-CMASS is between 5-10 $h^{-1}$Mpc. Varying from the ``usual" 15$h^{-1}$Mpc to $5 h^{-1}$Mpc \modif2{shows} $\sim$ 0.3\% variations in $D_A(z)$ and $\sim$ 0.4\% $H(z)$ and uncertainties are also reduced by 40\% and 30\% respectively. We also find that the accuracy of velocity field reconstruction depends strongly on the smoothing scale used for the density field. We measure the bias and uncertainties associated with different choices of smoothing length. 
  }

\end{abstract}
 \section{Introduction}
\modif{ The Baryon Acoustic Oscillations (BAO) is without any doubt a robust and promising probe for decrypting dark energy (DE). Furthermore, BAO plays a central role in current and future DE experiments devoted to understand cosmic expansion \citep{DESI,Euclid,WFIRST}. The BAO signature corresponds to the imprint left by sound waves generated during the early Universe in the baryon-photon fluid and propagated until the time of decoupling, when the photon decouples from baryons and the sound waves get frozen. While the linear physics of these baryon acoustic oscillations is well understood, 
the nonlinear evolution of the matter density field leads to a coupling of the Fourier-modes, which generates a damping of the Baryonic Acoustic Oscillations (BAO) in the power spectrum of galaxies as well as a shift in the BAO peak position.}
The damping in the power spectrum translates to a blurring of the baryonic acoustic peak in the correlation function of galaxies in configuration space. 
\modif{ This reduction of the contrast in the baryonic acoustic feature increases the uncertainty in the BAO distance derived from measurements.}

The major contribution to this damping comes from the bulk flows that shift the positions of the galaxies from the linear theory prediction \citep{Eis07}. 
Early work on reconstruction \citep{Eis06,PadWhiCoh} suggested that this effect could be partially reversed using the density field to infer the gravitational potential that sources the movement of the galaxies. 
\modif{ Running backwards the gravitational flow restores the BAO feature and reduce the errors in the distance measurements.}

 \cite{Pad12} extended the methodology  from \cite{Eis07} by applying reconstruction to galaxy {catalogues}. Since then, reconstruction has been successfully applied to galaxy clustering analysis in different galaxy {catalogues} and surveys. In most of the cases\modif{\footnote{ Few samples  in SDSS galaxy {catalogues}, such as DR9 CMASS sample \citep{And12} and the DR10 LOWZ sample \citep{Toj14}, reported  almost no improvement, but such results are consistent with the expected results using mock galaxy { catalogues} given the already small initial error compared to the average of the pre-reconstruction { catalogues}.}}, reconstruction is shown to increase the precision of the measurements in the samples analysed \citep{Pad12,And12,And13, And14,Toj14,Kaz14, Ross14} with up to 45 percent improvement in the error bars in certain samples. These results are responsible for reconstruction becoming an essential part of the clustering analysis.

\modif{ While reconstruction has shown to be a successful mechanism to obtain more precise BAO measurements, the performance of reconstruction algorithm is still under intense investigation. }
\modif{ Given the current precision in BAO distance measurements, the study of the associated uncertainties becomes essential for current and future surveys.} Some efforts to understand reconstruction analytically have appeared in the literature: \cite{PadWhiCoh} provided an analytic formalism within the context of Lagrangian perturbation theory; \cite{Noh09} extended this formalism for biased tracers. \cite{white15} developed the formalism to describe post-reconstruction correlation functions within the Zeldovich Approximation. \cite{xu12} studied the isotropic effects of reconstruction while \cite{And12} studied the anisotropic  effects. 
Reconstruction performance also has been studied using N-body simulations \citep{Seo06,Seo10}. \cite{Pad12} provides a first exploration of the robustness of reconstruction methodology against different implementation choices in the isotropic BAO analysis.  \cite{bur14} and \cite{bur15} { provided} a more detailed empirical study of the systematics (density dependence, geometry effects, RSD corrections) { in the} reconstruction algorithm. Both of these studies concentrated on the reconstruction effects on isotropic clustering. \cite{Var14} also studied the results post-reconstruction but focused on anisotropic fitting systematics. 

The reconstruction algorithm requires an estimate of the density field{ in order to} estimate the displacement field using the Zeldovich approximation \citep{Zel}. The density field is usually smoothed with a Gaussian kernel \citep{Pad12,And12,And13,And14},
whose width sets the scale that will source the displacement field.  
 A large smoothing scale could erase cosmological information, reducing the effect of reconstruction. Furthermore, information in the regions sourcing the nonlinear growth will be suppressed. On the other hand, a too small smoothing scale increases the noise in the linear density field. 
 As it is implemented now, the smoothing scale is a free parameter. In Table \ref{tab:smoothingPast}, we summarise different studies related to the smoothing scale; we show the references, scales explored, the method of analysis (i.e configuration of Fourier space), and the \sout{kind of} mocks used for the study. \cite{whi10} studied the shot noise effect on reconstruction and found that for low density tracers, 
 \modif{ a large smoothing scale performs better in terms of isotropic clustering as it generates a smaller shift in the  BAO measurement.}
 Most BAO clustering analyses \citep{Eis07,Pad12,And12,And13,And14} use a Gaussian smoothing kernel of $R$ = 10-20 $h^{-1}$ Mpc. Some small deviations from this smoothing length have been shown not to alter the results (see Appendix B of Anderson et al. 2012 and Padmanabhan et al. 2012a). \cite{bur14} studied the impact of the smoothing length on the isotropic BAO analysis. They found that the bias in the measurement of the isotropic dilation parameter $\alpha$ is reduced using a smoothing scale of $15 h^{-1}$ Mpc  for CMASS and a smoothing scale of $10h^{-1}$ Mpc for LOWZ. 
 In this paper, we extend those analyses to present a study of the effect of smoothing scale on reconstruction performance in the anisotropic BAO analysis.  
\begin{table*}
\begin{center}
\caption{Smoothing scales tested with simulations } 
\label{tab:smoothingPast}
\begin{tabular}{@{}lcccccccccc}
\hline
Reference &$R$'s ($h^{-1}$ Mpc)& Best $R$&observable&Mock {\bf catalogues}\\
\hline
\cite{Pad12}&10,15, 20, 25 & 15 &$\xi_0(r)$& N-body \\
\cite{bur14,bur15} &5,8 10 ,15 20 ,40 & 15&$P(k)$& 2LPT Mocks \\
\cite{bur15} &5 10 ,15 & -&$\Psi(r)$& 2LPT Mocks \\
\cite{Noh09}&5,10, 15, 20, 25, 30&-&$\Psi(r)$&N-body\\
 \hline
\\[-1.5ex]
\end{tabular}
\end{center}
\end{table*}

\begin{figure}
   \centering    
   \includegraphics[width=3.6in]{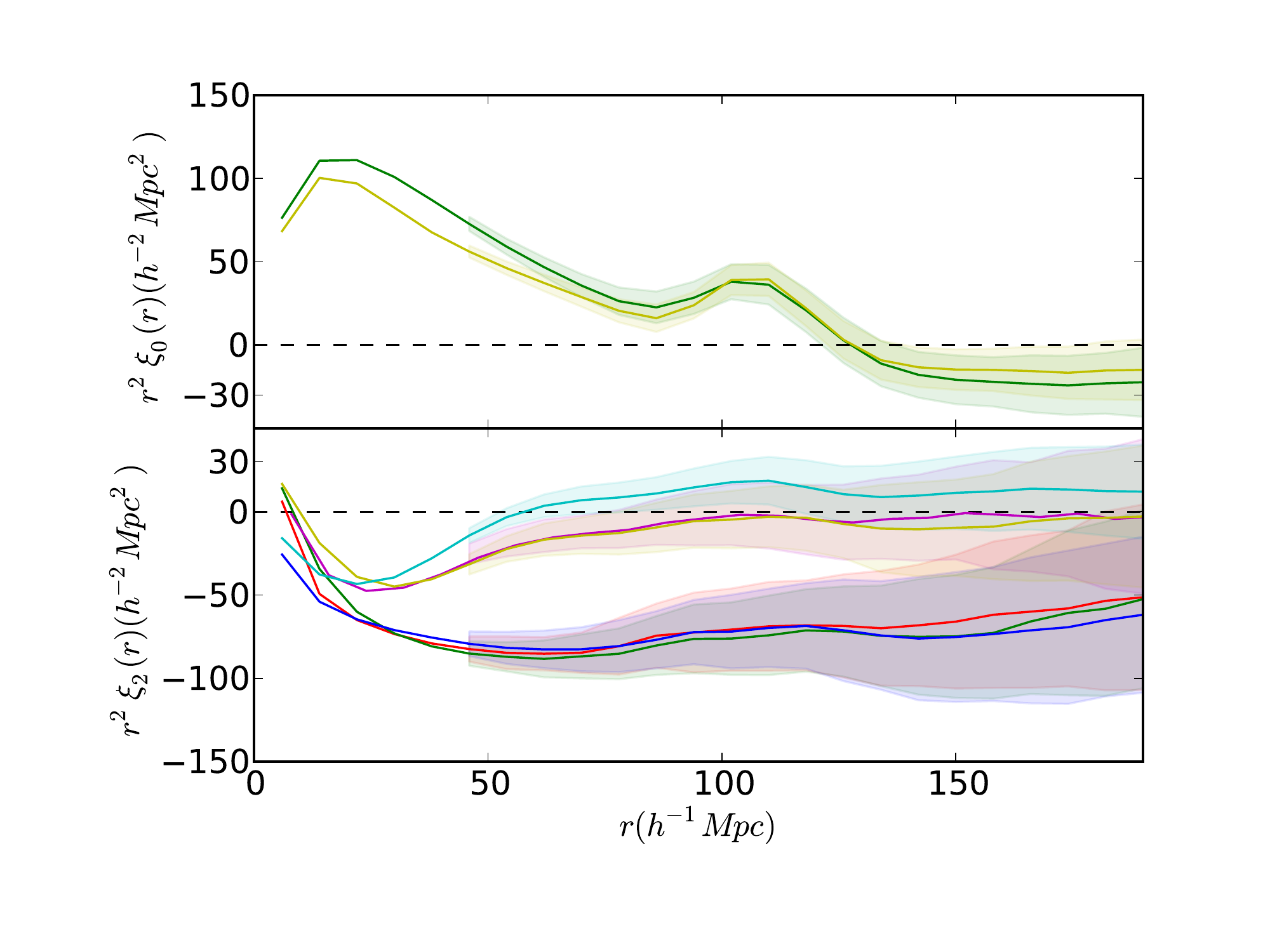}
  \caption{Performance of reconstruction tested in different kinds of sky-type mock { catalogues}. We show the mean of monopole [top panel], quadrupole [bottom panel] from 100 mocks pre-reconstruction and post-reconstruction. The different colours are different kinds of mock { catalogues}: darker shades are pre-reconstruction and lighter shades are post-reconstruction { catalogues}. In blue are the PTHALOS  \citep{man13}. In green are the Quick Particle Mesh \citep{QPM} . In red are the PATCHY (Kitaura et al 2015; companion paper). For the monopole we get pretty similar results in the BAO fitting range. However, in all cases the post-reconstruction quadrupole  is not exactly zero; there is an extra correlation (negative or positive).  
     The purpose of this work is to disentangle the relation of this residual with the smoothing length and check the effect of the smoothing length on the BAO anisotropic post-reconstruction results.}
   \label{fig:MeanDiffSimsImp}
\end{figure}

The motivation for this study is illustrated in Figure \ref{fig:MeanDiffSimsImp}. We show the mean and root mean square (RMS) of the correlation function from 100 mock {catalogues} before and after reconstruction (hereafter, pre- and post-reconstruction). The different colours are different kinds of mock {catalogues}.
Darker shades are pre-reconstruction, and lighter shades are post-reconstruction {catalogues}. In blue are the PTHALOS mock {catalogues} used for the analysis of the Baryonic Oscillations Spectroscopic Survey (BOSS) \citep{Daw13} of Sloan Digital Sky Survey III (SDSS-III) \citep{Eis11} galaxy samples Data Releases 9, 10 and 11 (DR9, DR10, DR11). PTHALOS mocks are based on second order Lagrangian Perturbation Theory \citep{man13}. In green are the Quick Particle Mesh mock {catalogues} used to analyse BOSS DR12 based on low resolution N-body simulations combined with HOD models for populating halos with galaxies. Finally, in red are the MD-PATCHY (Kitaura et al, 2015; 
companion paper) 
mocks based on Augmented Lagrangian Perturbation Theory and stochastic bias prescriptions generated for BOSS DR12 analysis.
In all types of simulations, reconstruction succeeded \modiff{in} 1) \modiff{sharpening} the BAO feature in the monopole, and 2) \modiff{reducing} the quadrupole to be consistent with zero at large scales when we remove the redshift space distortions.  
 However, different trends {are} observed in the post-reconstruction mean correlation function, {depending} on the details of the reconstruction implementation and/or the simulations.   
The PTHALOS mocks show a slightly positive quadrupole post-reconstruction compared with QPM and PATCHY, which show a slightly negative quadrupole. 
We note that the implementation used in PTHALOS is from \cite{Pad12}, while the one used in the QPM and PATCHY mocks is ours; the cosmology of the three different simulations is also different. 
 A perfect reconstruction would remove large-scale anisotropy from the correlation function when we remove the redshift space distortions from the {catalogues}. However, in the case plotted, there is a residual anisotropy in the quadrupole at large scales (negative or positive), showing that the reconstruction is not perfect.  The aim of this article is to disentangle the relation of this residual with the smoothing length and check the effect of the smoothing length in the BAO anisotropic post-reconstruction results. 
 
In this work, the main metric we use to evaluate the performance of reconstruction is the anisotropic BAO fits. Additionally, we explore the smoothing effects on the correlation functions and on the displacement field 
 \modif{ as a way  of understanding} the anisotropic fit results. We find that the smoothing length affects the quadrupole amplitude. Furthermore, we find that the differences in the amplitude also depend on the implementation. However, we show that the differences in the amplitude of the quadrupole do not determine the anisotropic results. We also explore \modiff{whether} the origin of the improvement in anisotropic fits was related to a better estimate of the displacement field.

This study addresses a crucial point in current BAO analysis, especially in the context of \modiff{the} final data release of BOSS galaxy data. The results we found are BOSS specific; however, we include a section with some additional tests to explore how these conclusions scale with the bias and number density that the results could be generalised to other surveys.
 
The layout of this paper is as follows. We introduce the reconstruction implementation in Section \ref{sec:rec}. We present the simulations used in our study in Section \ref{sec:sim} and the analysis methods for anisotropic BAO measurements in Section \ref{sec:method}. We then present the results in Sections \ref{sec:fits} and \ref{sec:disp}. Section \ref{sec:fits} presents the effect of the smoothing in anisotropic BAO fitting results and Section \ref{sec:disp} presents the effect of smoothing on the displacement estimation accuracy. We conclude in Section \ref{sec:con}.

\section{Basic Reconstruction Algorithm}\label{sec:rec}

The algorithm of density field reconstruction has been described in  \cite{Eis07}, \cite{Pad12} and \cite{bur14}. 
\modif{ We describe the most general algorithm applied to biased tracers considering redshift space distortions, the angular and radial mask of the data.}We focus on the part where the smoothing scale enters the reconstruction algorithm.
The algorithm proposed by \cite{Eis07} can be directly summarised as follows: 
\begin{itemize}
\item Estimate the over-density field from the galaxy positions using an interpolation method. We are using \modiff{the} Nearest Grid Point (NGP) interpolation method. 
\item Smooth the over-density field using a Gaussian filter with smoothing scale R in order to eliminate high $k$ non-linearities (small scales).
\begin{equation}
W_G(k)=\exp(-R^2 k^2/2)
\end{equation}
\item Solve the Eq. (\ref{Ap:ZAg}) using the Zeldovich approximation. 
\begin{equation}\label{Ap:ZAg}
\nabla \cdot \vec{\Psi}+f \nabla \cdot (\vec{\Psi} \cdot \vec{r}) \hat{r}=-\frac{\delta_g}{b},
\end{equation}
where $\vec{\Psi}$ is the displacement field, $\vec{r}$ the galaxy position, $\delta_g$ the galaxy contrast of density and $b$ the bias,

\modif{ We can solve Eq. (\ref{Ap:ZAg}) in configuration space following a finite differences approach \citep{Pad12} or in Fourier space \citep{bur14}.}
\modif{ In order to solve it in Fourier space,  it is assumed that $(\vec{\Psi} \cdot \hat{r})\hat{r}$

is irrotational, which enables us to approximate it as the gradient of a potential field, i.e  solve Eq. (\ref{Ap:ZAg}).}
\modif{ The implications of this approximation were explored in \cite{bur15}  showing that the displacement field is underestimated when the irrotational component is neglected, suggesting an empirical formula to correct this effect.} 

\modif{ In our implementation we also use the Fourier transform method  to solve for the displacements. However, we follow a simpler approach by neglecting the effect of the RSD when measuring the density field, leading to this equation :}\begin{equation} \nabla \cdot \vec{\Psi}=-\frac{\delta_g}{b},\end{equation} \modif{ instead of solving equation \ref{Ap:ZAg}. }, and we estimate the displacement field by:
\begin{equation}
\vec{\Psi}=\text{IFFT} \left [ \frac{-i\vec{k}\delta_g(\vec{k})}{k^2b}\right ].
\end{equation}

We verify that this choice does not affect the conclusions of the tests performed (see Appendix \ref{ap:imp}). 
The effect of applying different RSD corrections on the  displacement field 
is performed in \cite{bur15}. 
A study of the effects of applying different RSD corrections  in terms of anisotropic fits is performed in Vargas-Maga\~na (in preparation).  
\item \modif{ Once the displacement is computed, we move the particles positions by the corresponding displacement, $-\vec{\Psi}$, to approximate their initial Lagrangian positions.} This step provides the short-scale modes of the reconstructed density \citep{PadWhiCoh, Noh09}. 
\item Move an additional $\vec{\Psi}_{RSD}$ if we want to eliminate the redshift space distortions at large scales in the catalogue: 
\begin{equation}
\vec{\Psi}_{RSD}=-f (\vec{\Psi} \cdot  \hat{r}) \; \hat{r}. 
\end{equation}
\item Generate a uniform random sample and move the particles using the displacement field previously estimated from data. This step provides us the large-scale modes of the reconstructed density.
\end{itemize}
The  2-point correlation post-reconstruction is then  defined as:
\begin{equation}
\xi_{LS}(r,\mu)=\frac{DD(r,\mu)-2DS(r, \mu)+SS(r, \mu)}{RR(r, \mu)},
\end{equation}
where $D$ accounts for the data, $S$ for the ``shifted'' random sample and $R$ for the ``non-shifted" random sample.  The $DD$ describes the pair counts per r-$\mu$ bin with data-data, $SS$ the same for the shifted random pair-counts and $RR$ for the non-shifted randoms. $DS$ will be the pair-counts per r-$\mu$ bin taking  one point from data and one from the shifted random set.

\modif{ The current reconstruction algorithm treats separately the small and large scale modes. While small scale modes stay in the {\it shifted} galaxy field denoted by "D", the large scale modes are imprinted in the {\it shifted } random catalogue, denoted by "S".  The reconstructed density field is then defined as $\delta_{rec}=\delta_{d}-\delta_{s}$ that represents the sum of the two contributions of the large- and small-scale modes  \citep{PadWhiCoh}.}

\section{Simulations}\label{sec:sim}
In this work we use two kinds of mock {catalogues}. One set of approximate mock \modif{ catalogues that enables us to have a sufficiently large number of realizations to test the BAO anisotropic fitting results. A second set, composed of a small number of high fidelity mocks based on N-body simulations, guarantees that the velocities are more accurate, enabling  us to test the accuracy of the reconstructed velocity field for cosmological applications. }
\subsection{Quick Particle Mesh Mocks.}\label{sec:QPM} 
Quick Particle Mesh (QPM hereafter) mocks were generated for BOSS clustering analysis. These mock { catalogues} use low mass
and force resolution particle-mesh simulations employing $1280^3$ particles in a $(2560 h^{-1}$Mpc)$^3$ box run with large time steps. At select
times, the particles and their local density smoothed on $2h^{-1}$Mpc scales were dumped; these particles were then sampled (with a
density-dependent probability) to form a set of mock halos that are then populated using a halo occupation distribution \citep{QPM}.

We use ``Sky mocks,'' which match the observed number density of BOSS galaxies and follow the radial and angular selection functions. These mocks are required to study the anisotropic galaxy clustering and to extract conclusions applicable to current analysis of BOSS-DR12 galaxy samples.
We used the version of QPM mocks that matches CMASS North Galactic Cap samples for Data Release 12 of BOSS.

\subsection{RunPB Simulations}
\modiff{In Section \ref{sec:disp}, we use} RunPB Simulations (RunPB hereafter) \citep{white_simus} for the study of the velocity field.
The {catalogues} are based on high-resolution realisations of the $\Lambda -$CDM model with $\Omega_m = 0.292$ and $h = 0.69$, employing $2048^3$ particles in a periodic box of side length $1380 h^{-1} $Mpc for a total volume of $2.6 h^{-3} $Gpc$^3$. The values used at redshift $z=0.55$ for the growth factor and the power spectrum amplitude are $f=0.76$ and $\sigma_8=0.62$. The simulations were run with the TreePM code; 
the mock {catalogues} are described further in \cite{whi10}. \modif{ Briefly, halos were found using the friends-of-friends algorithm. We use a cut in the halo mass ($M_{halo} > 1.10^{13} \; M\odot$) which mimics the bias property of the CMASS survey.}We can do that under the assumption that most of the CMASS galaxies are Brightest Central Galaxies (BCG) ($90\%$) and so have the same velocities as the halos.

\section{Methodology}\label{sec:method}
\subsection{Anisotropic Analysis}
In this paper, we follow the multipole fitting procedure described in \cite{xu12} and \cite{And13}, which extracts measurements
of the isotropic dilation of the coordinates parametrized by $\alpha$ and the anisotropic warping of the coordinates parametrized by $\epsilon$. 
$\alpha$ and $\epsilon$ parametrize the geometrical distortion derived from assuming a ``wrong" cosmology when calculating the galaxy correlation function. 

The parameters $\alpha$ and $\epsilon$ are defined as
\begin{equation}
\alpha = \alpha_{\perp}^{2/3} \alpha_{||}^{1/3} \,, 
\end{equation}
and
\begin{equation}
1+ \epsilon = \left( \frac{\alpha_{||}}{\alpha_{\perp}} \right)^{1/3} \,
\end{equation}
where $\alpha_\perp$ and $\alpha_{||}$ are defined by
\begin{align}
r_{\perp} & = \alpha_{\perp} r_{\perp, {\rm obs}} 
\label{eqn:aperpdef} \\
r_{\parallel} & = \alpha_{\parallel} r_{\parallel, {\rm obs}} .
\label{eqn:apardef}
\end{align}
Here, the `obs" subscript denotes the observed coordinates. The coordinates without "obs" subscripts
\modif{ correspond to the assumed cosmology. $r_{\perp}$ and $r_{||}$ are respectively the
transverse and parallel to the line line-of-sight galaxy separations.}

The transverse shift, $\alpha_\perp$, allows us to measure $D_A(z)/r_s$,
where $D_A(z)$ is the angular diameter distance to redshift $z$ and $r_s$
is the sound horizon scale. The line-of-sight shift, $\alpha_{||}$, allows us
to measure $cz/(H(z)r_s)$, where $H(z)$ is the Hubble parameter. This is
done using
\begin{equation}
\alpha_{\perp} = \frac{D_{A} (z) r_{s,{\rm fid}}}{D^{\rm fid}_{A} r_{s}},
\end{equation}
and
\begin{equation}
\alpha_{||} = \frac{H^{\rm fid}(z) r_{s,{\rm fid}}}{H(z) r_{s}}.
\end{equation}
Note that when analyzing the mocks in the cosmology in which they are define,  $\alpha =  1$ and $\epsilon = 0$. 

\subsection{Nonlinear Models for the Correlation Function}\label{sec:NLM}

The template for the 2D nonlinear power spectrum, \modiff{following} from  \cite{Fis94} reads as follows:
\begin{equation}
P(k, \mu) =(1+\beta \mu^2 )^2 F(k, \mu, \Sigma_s)P_{NL} (k, \mu).
\end{equation}
\modif{ where the term $(1+\beta \mu^2)$  corresponds to the Kaiser model for large-scale redshifts distortions, which produce an anisotropic damping and $P_{NL}(k)$ is the nonlinear power spectrum.}
$F(k, \mu, \Sigma_s)$ is the streaming model for FoG  given by: 
\begin{equation}\label{eq:streamming}
F(k, \mu, \Sigma_s)=\frac{1}{(1+k^2\mu^2 \Sigma_s^2)},
\end{equation}
where $\Sigma_s$ is the streaming scale. In this work, we 
consider the de-wiggled power spectrum $P_{dw}$. To get the templates for the multipoles, we decompose the 2D power spectrum into its Legendre moments:
\begin{equation}
P_{l,t}(k)=\frac{2l+1}{2}\int_{-1}^{1}P_t(k, \mu)L_l(\mu)d\mu,
\end{equation}
which can then be transformed to configuration space using
\begin{equation}
\xi_{l,t}(r) = i^l \int \frac{k^3 d log(k)}{2 \pi^2}P_{l,t} j_l(kr),
\end{equation}
where $j_l(kr)$ is the l-th spherical Bessel function and $L_l(\mu)$ is the l-th Legendre polynomial.

\subsection{De-wiggled Template} \label{sec:dewiggled}

The de-wiggled template is a  power spectrum prescription widely used in clustering analysis \citep{xu12,And12,And13}. This phenomenological prescription takes a linear power spectrum template to which we add the nonlinear growth of structure.
The de-wiggled power spectrum is defined as:
\begin{equation}
\begin{array}{ll}
P_{dw}(k, \mu)=&[P_{lin}(k) -P_{nw}(k)] \\
&\times exp\left[ -\frac{k^2 \mu^2 \Sigma_{||}^2+k^2(1-\mu^2)\Sigma_{\perp}^2}{2}\right ] +P_{nw} ,
\end{array}
\end{equation}
\modif{ where $P_{lin}(k)$ is the linear theory power spectrum and $P_{nw}(k)$ is a power spectrum without the acoustic oscillations (\cite{Eis98}). $\Sigma_{||}$ and $\Sigma_\perp$ are the radial and transverse components of the standard Gaussian damping of BAO, $\Sigma_{NL}$ :}
\begin{equation}
\Sigma_{NL}^2=(\Sigma_{||}^2+\Sigma_\perp^2)/2.
\end{equation}
 $\Sigma_{NL}$ models the degradation of signal because of nonlinear growth of structure.

\subsubsection{Multipole Fitting}
In order to measure $\alpha$ and $\epsilon$, we define a fiducial
cosmology and then fit template multipoles (monopole and quadrupole) of the correlation function in
this fiducial cosmology to the observed data.\modif { We use the model described in \cite{xu12} to generate the templates for multipoles of the correlation function.}
This model is summarized in detail in \cite{xu12}, \cite{Var14}, \cite{And12} and \cite{And13}.
The models we fit to our observed multipoles, $\xi_0(r)$ and
$\xi_2(r)$, are:
\begin{align}
\xi_{0}(r) &= B_0^2 \xi_{0, \rm obs}(r) + A_0(r) \nonumber, \\ 
\xi_{2}(r) &= \xi_{2, \rm obs}(r) + A_2(r),
\label{eqn:monoquadt}
\end{align}
where
\begin{equation}
A_{\ell}(r) = \frac{a_{\ell,1}}{r^2} + \frac{a_{\ell,2}}{r} + a_{\ell,3}; \,
\ell=0,2.
\label{eqn:fida}
\end{equation}
These $A_\ell(r)$ terms are used to marginalize out broadband (shape)
information through the $a_{\ell, 1}\ldots a_{\ell, 3}$ nuisance parameters. 
\modif{ We use a fit he monopole and the quadrupole in a range of $50<r<200h^{-1}$Mpc, with $40$ bins of 8 $h^{-1}$ Mpc each. 
Knowing our model uses  $10$ we have finally $30$ degrees of freedom for the fit.

In order to find the best-fitting values for $\alpha$ and $\epsilon$, we 
minimize the $\chi^2$ function}
\begin{equation}
\chi^2 = (\vec{m} - \vec{d})^T C^{-1} (\vec{m}-\vec{d}),
\end{equation}
\modif{ where $\vec{m}$ is the model vector and $\vec{d}$ is the vector of data.} We scale the
inverse sample covariance matrix, $C^{-1}_s$, using
\begin{equation}
C^{-1} = C^{-1}_s \frac{N_{\rm mocks} - N_{\rm bins} - 2}{N_{\rm mocks} - 1}.
\end{equation}
\modif{ to correct the bias in the estimate of the true inverse covariance matrix $C^{-1}$ \citep{Har07}. }

Error estimates for $\alpha$ and $\epsilon$ are obtained by walking a
grid in these two parameters to map out the
likelihood surface. Assuming that the likelihood surface is Gaussian, this allows
us to estimate $\sigma_\alpha$ and $\sigma_\epsilon$ as the standard
deviations of the marginalized 1D likelihoods of $\alpha$ and $\epsilon$
respectively. 

\section{Smoothing effect on anisotropic BAO analysis.}\label{sec:fits}

First, we study the effect of the smoothing scale in the BAO analysis described in section \ref{sec:method}. 
The results presented in this section were obtained using ``sky mocks'' as described in section \ref{sec:QPM}.
\modif{ We apply the reconstruction algorithm described in section \ref{sec:rec} using four different smoothing scales: 
 R=$5$, $10$, $15$ and $40 h^{-1}$Mpc.
Then, we compute the correlation function of the reconstructed { catalogues}. Finally we apply the BAO fitting methodology and evaluate reconstruction performance using the fits on $\alpha$ and $\epsilon$ parameters and their respective uncertainties as a metric.}
\footnote{The effective smoothing scale is a convolution of the Gaussian smoothing and the grid smoothing. Hereafter we refer only to the Gaussian smoothing. We considered that the grid smoothing is not \modiff{significantly changing} the effective smoothing. In the 1D case, using a Gaussian smoothing equal to a grid size, the effective smoothing is $\sim15\%$ larger for NGP, or $\sim30\%$ larger for CIC. The 5 Mpc smoothing, assuming a 5 Mpc grid, is really $\sim6$ Mpc for NGP and 6.5 Mpc for CIC, which is not very different.} We summarize the details of the interpolation method used in the density estimation  in Table \ref{tab:inter}.

\begin{table}
\begin{center}
\caption{ Interpolation parameters }
\label{tab:inter}
\begin{tabular}{@{}cccccccc}
\hline
Simulation & Interpolation& Box Size & Grid &Pix Size\\
&Method&($h^{-1}$Mpc)& &($h^{-1}$Mpc)\\
\hline
\\[-1.5ex]
QPM Sky & NGP& 3400&512& 6.6\\
RunPB Mocks & NGP& 1380 &512  & 2.7\\

\hline
\\[-1.5ex]
\end{tabular}
\end{center}
\end{table}
\subsection{Multipoles Results}

\begin{figure*}
   \centering
      \includegraphics[width=3.8in]{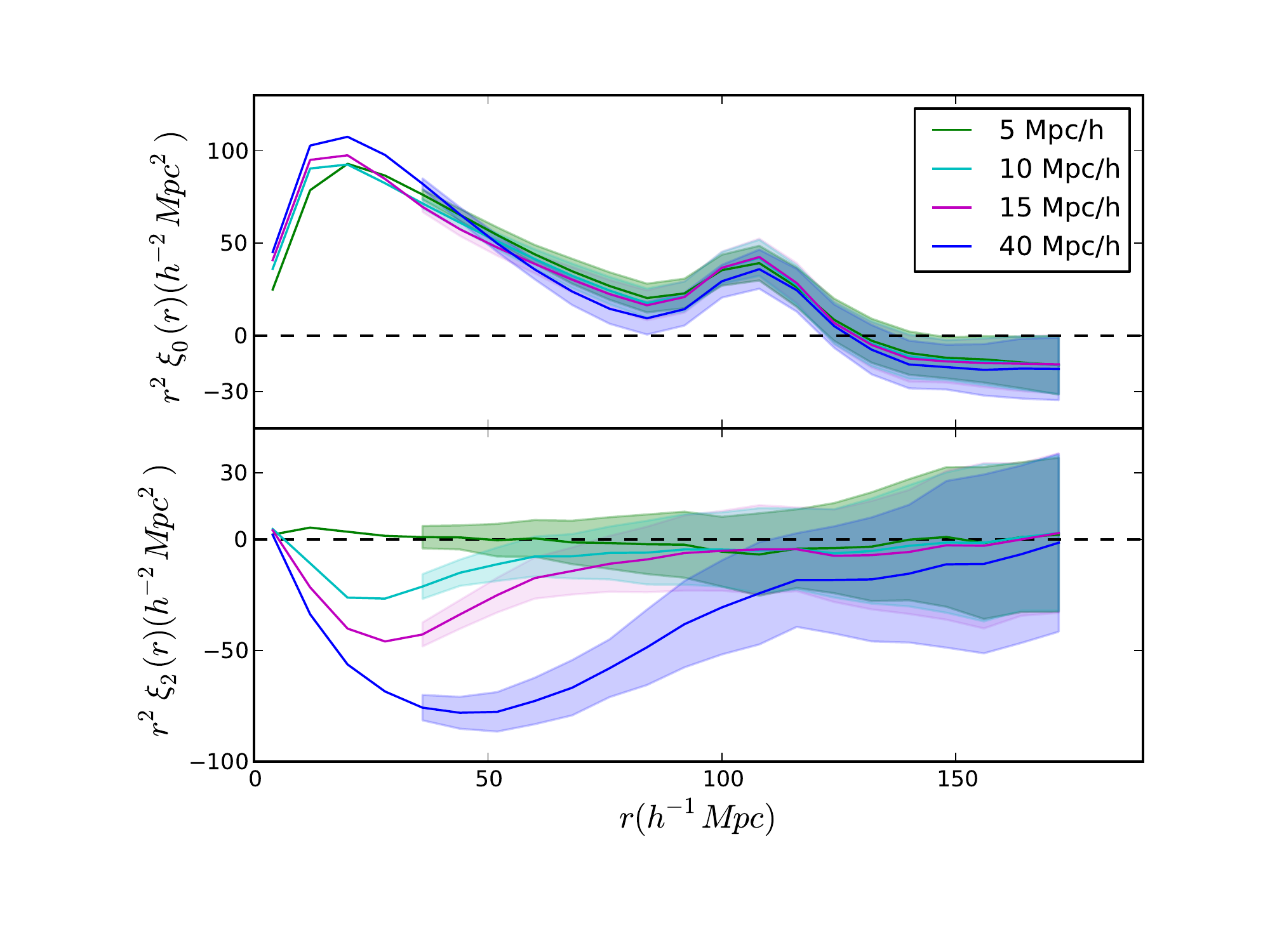}\includegraphics[width=3.8in]{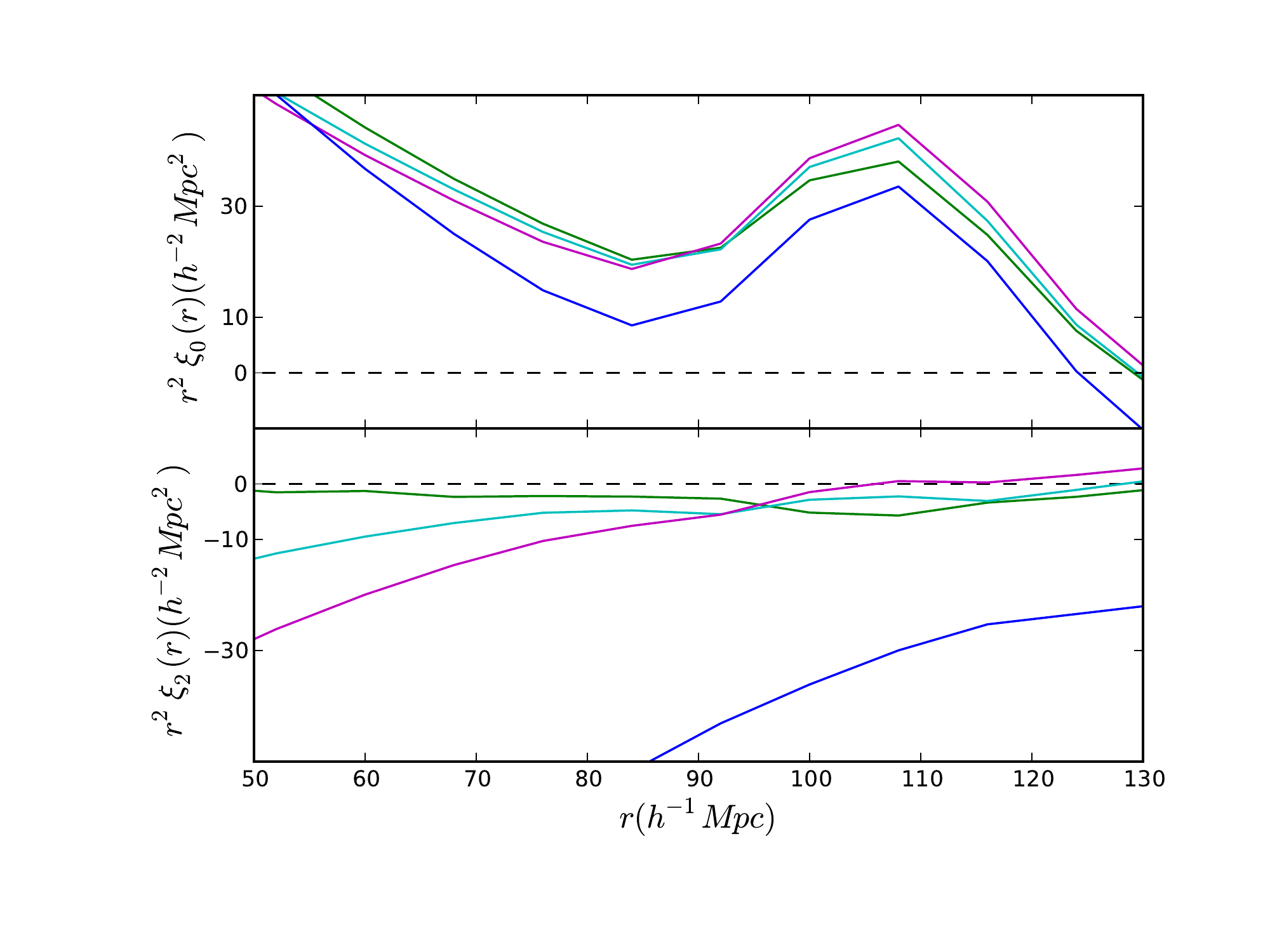}
   \caption{  Mean of $200$ QPM mocks NGC reconstructed with different smoothing scale  $5  h^{-1}$Mpc [green], $10 h^{-1}$Mpc [cyan], $15 h^{-1}$Mpc [magenta] and $40 h^{-1}$Mpc [blue]. The smoothing scale is correlated with the negative correlation observed in the quadrupole; a smaller smoothing scale decreases the correlation observed; a $5 h^{-1}$Mpc smoothing scale erases the quadrupole almost completely. Right panel: Zoom to the monopole [top panel] and quadrupole [bottom panel] in the BAO range.}
   \label{fig:meanSmooth}
\end{figure*}

\modif{ In Figure \ref{fig:meanSmooth} we show the mean multipoles from $200$ QPM reconstructed mocks with the four smoothing scales.} The upper panel is for the mean monopole and the lower panel for the mean quadrupole. \modif{The shaded regions correspond to the square root of the diagonal elements of the covariance matrix. }

\modif{ For the monopole we observe two effects: 1) At large scales we find,  as expected, that the different smoothing scales affect the level of sharpening of the mean monopole.  The cases 5, 10, and 15 Mpc/h show an enhanced peak compared with the pre-reconstruction monopole.
 However the R=5 is slightly lower compared to the 10-15 smoothing scales. The R=40 is clearly decreasing the contrast in the BAO peak. 
2) At small scales we find that the different smoothing scales affect the shape of the monopole. These scales  are not important for the fitting of the BAO feature. The differences at those scales are mostly related to the residual redshift space distortions.}

\modif{  Concerning the quadrupole, a perfect reconstruction would show a mean quadrupole consistent with zero (without any large-scale anisotropies). Figure \ref{fig:meanSmooth} illustrates that  the typical smoothing scale of $15 $ $h^{-1}$ Mpc gives a quadrupole not completely consistent with zero. 
The disagreement appears worse when using a large smoothing scale where the correlation increases in the interval $50-100 h^{-1}$Mpc. Using a smaller smoothing scale reduces the negative correlation observed in the quadrupole to be in agreement with zero within the fitting range. }

\modif{ In summary, just by comparing the results in terms of correlation functions, it is not clear that the reconstruction is performing better with any smoothing scale. Two effects affect differently the BAO anisotropic fits: 
\begin{enumerate}
\item The precision of $\alpha$ is driven by the sharpening of the BAO feature. Thus we expect, that 10-15, gives best precision on $\alpha$ followed very closely by 5, and that 40 smoothing scale increases the error on $\alpha$.
\item The precision on $\epsilon$ depends on the capability that our nuisance parameters ($B_0, A_{0,2}$) to absorb the residual quadrupole.  We can expect that a quadrupole consistent with zero is likely to give a good fit, thus we expect that 5 Mpc/h will provide best constraints on $\epsilon$, followed by 10 and 15. 
\end{enumerate}
Our next task is to test these hypotheses and check how the smoothing scales affect the anisotropic fits.}

\subsection{Anisotropic Fits on the Mean Correlation Function}

\modif{ In order to determine which smoothing scale is performing better, we fit the nonlinear damping parameters $\Sigma_{\parallel, \perp}$ and the streaming parameter $\Sigma_s$ using the mean information over the mocks. In Table \ref{tab:fixpar} we summarize the results for different values of $R$. We find a minimum value for $\Sigma_{\parallel, \perp}$  for a smoothing scale of 5 $h^{-1}$Mpc, indicating that the mean correlation function is less nonlinear using this filter. When we increase the smoothing scale, the value $\Sigma_{\parallel, \perp}$ increases for 10 and 15 Mpc/h, indicating that the function is becoming more non linear. }
\modif{ We note that the values we get for the $\Sigma_{\parallel, \perp}$ for the 40  case are smaller compared to 15. At 40 Mpc/h we expect to have a poor performance of reconstruction, since we lose information as we are smoothing scales that are in linear regime. 
 We suspect that the unexpected values obtained for the nonlinear damping are generated because we are using the fitting template for the post-reconstruction case, i.e. we are assuming that the nonlinear damping is isotropic. But this is most likely not a good approximation, since with this smoothing we are not removing most of the nonlinear evolution of the density. 
 Thus, it would be more accurate to do the fitting with both damping parameters free instead . }
  \modif{ In order to test that the assumption $\Sigma_{\parallel}=\Sigma_{\perp}$ is breaking down for R=40, we performed for all cases a fit without this assumption (see bottom panel of Table \ref{tab:fixpar}). Following this methodology we found that the get the more symmetric values for R=5, however the smaller value for $\Sigma_{NL}$ is given by 10 Mpc/h followed by R=5 Mpc/h. And the larger value is coming for R=40 Mpc/h. We notice also that the best fits gives very asymmetric values. Thus probably indicating that post-reconstruction is not necessarily well described by this assumption.  The lower values of $\chi^2$ just indicates that the data is too good to be true, thus usually can indicate two different possibilities: 1) our model is valid but that a statistically improbable excursion of $\chi^2$, 2) we overestimate the errors. We do not consider our errors overestimated as they are coming from the sample variance of the simulations renormalized by the $\sqrt{N_{sim}}$.}

We also notice that the $\beta$ value fitted is the smallest for $5h^{-1}$Mpc, indicating that the correction for the Kaiser effect is less important for a smaller smoothing scale. 
In the case of the streaming parameter $\Sigma_s$, which is related to the Fingers-of-God (FOG) effect,  we do not expect to find any improvement (to find any change) in the best fitting value for this parameter, since reconstruction does not take care of virial velocities.

\begin{table}
\begin{center}
\caption{ Values for the fixed parameters {\bf [$\Sigma_\parallel= \Sigma_\perp$ ($h^{-1}$Mpc),$\Sigma_s$($h^{-1}$Mpc) , $B_0$ and $\beta$] } fitted to the mean of 200 QPM mocks varying the smoothing scale of  reconstruction {\bf R ($h^{-1}$Mpc) }}
\label{tab:fixpar}
\begin{tabular}{@{}ccccccc}
\hline
R & $\Sigma_\parallel= \Sigma_\perp$ & $\Sigma_s$& $\chi^2/d.o.f$ & $B_0$ & $\beta$\\
\hline
\\[-1.5ex]
5 & 4.9 & 0.0 &18.4/30 & 0.98&-0.06\\
10 &5.7& 0.0 &9.4/30   & 1.01&-0.12\\
15&6.5 & 0.0 &14.3/30&  1.28&-0.11\\
40& 5.9&-6.0 & 11.3/30 &1.31&0.29\\
\hline
R & $\Sigma_\parallel$&$\Sigma_\perp$ & $\Sigma_{NL}$&  $\chi^2/d.o.f$ & $B_0$ & $\beta$\\
\hline
5&4.8& 5.6 &5.2&7.0/30 &0.9 &0.01\\
10&5.9& 1.5 &4.3&8.2/30 &1.0 &-0.13\\
15&6.9& 3.6 &5.5&10.3/30 &1.2& -0.10\\
40&5.6 &8.8&7.3&14.7/30 &1.3 &0.35\\
\hline
\hline
\\[-1.5ex]
\end{tabular}
\end{center}
\end{table}

\subsection{Anisotropic Fits Results}\label{sec:anisofits}

\begin{figure}
   \centering     
   \includegraphics[width=3.7in]{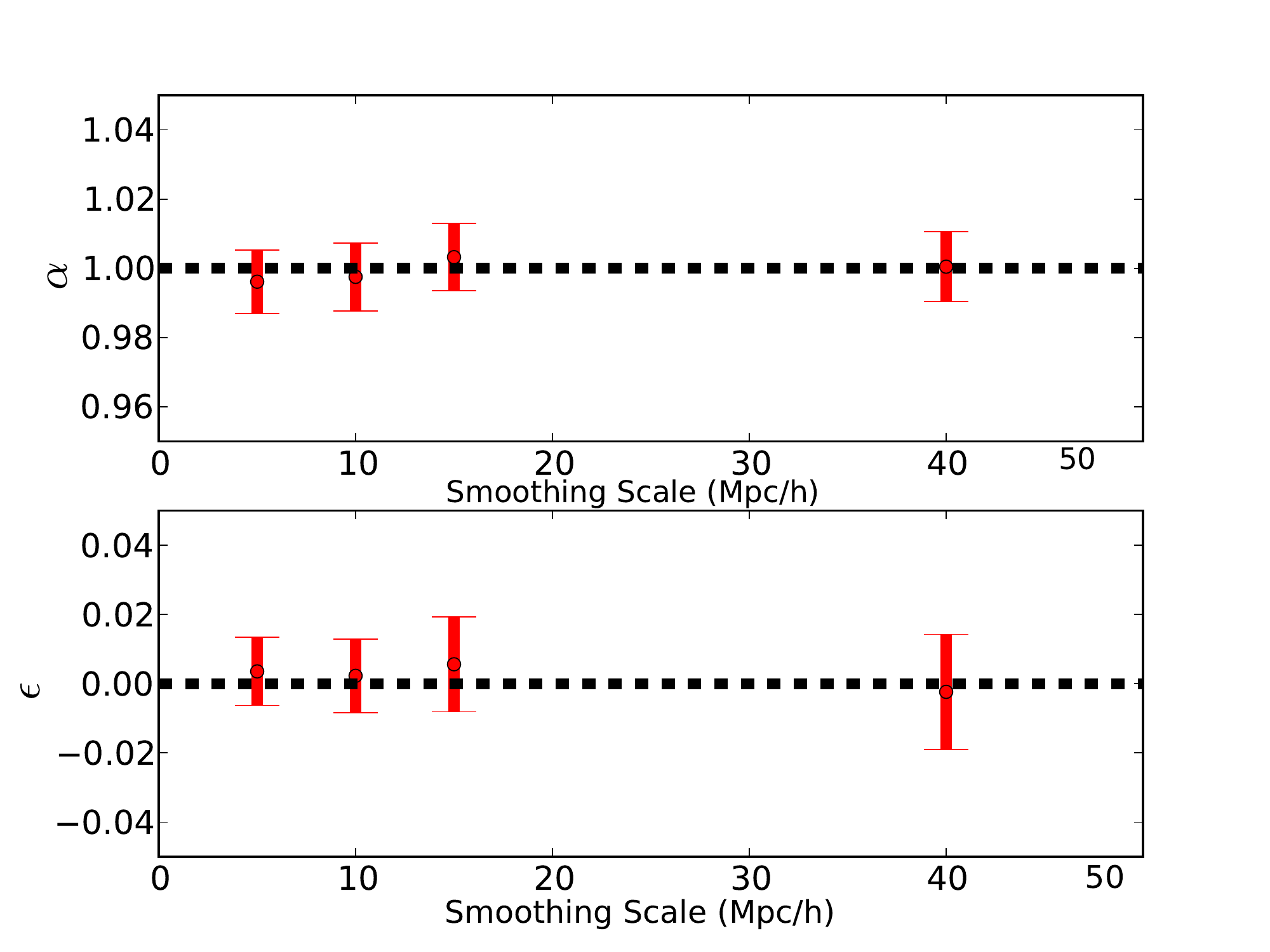}
    \caption{ Mean of best fits for $\alpha$ and $\epsilon$ for $200$ QPM mocks NGC analysed with different smoothing scales. The error bars are given by the standard deviation from $200$ realisations. For $\alpha$, the smaller dispersion is for 10-15 $h^{-1}$Mpc; however, the 15 $Mpc/h$ is significantly biased (5.3 $\sigma$). The best is 5 $h^{-1}$Mpc, \modiff{which} has less significant bias with small dispersion (0.8$\sigma$). For $\epsilon$, the less significant bias ($b_\epsilon/\sigma_\epsilon$ lower) is for \modiff{the} 5 $h^{-1}$Mpc smoothing scale given their small bias and dispersion. }
   \label{fig:MeanParSmooth}
\end{figure}

 \modif{ In Figure \ref{fig:MeanParSmooth} we show the results from fitting  200 QPM mocks reconstructed using the four smoothing scales.} We show the mean value from the best fits for BAO-related parameters $\alpha$ [top panel], $\epsilon$ [bottom panel].  We also show in Figure \ref{fig:MeanParSmooth2}  results for $\beta$ [bottom panel], and the $\chi^2$  [top panel] results for the fits.  The dotted line indicates the expected values from the mocks. The quantitative fitting results are summarized in Table \ref{tab:fits} for the best fits in anisotropic BAO analysis.

\begin{figure}
   \centering     
   \includegraphics[width=3.3in]{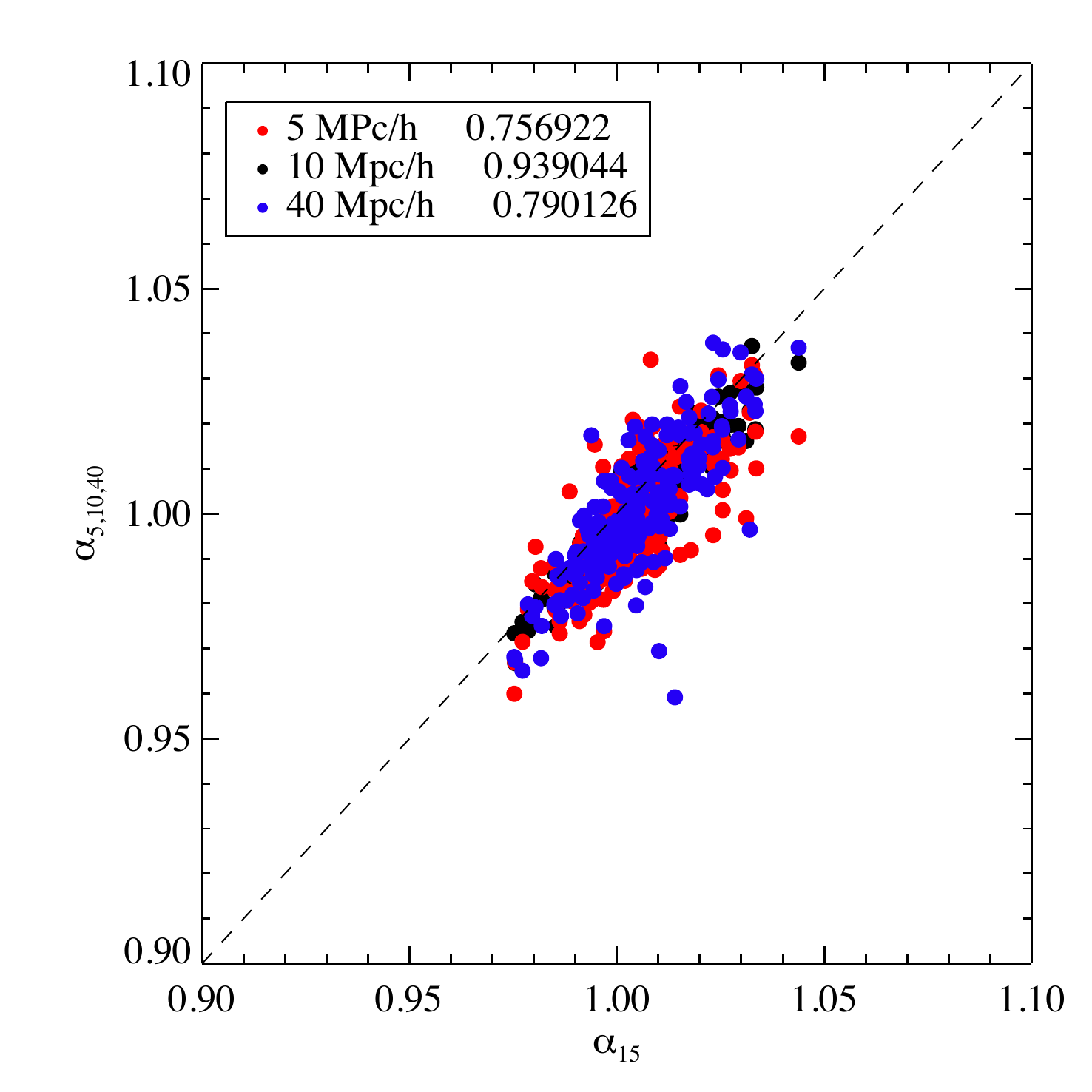}
    \caption{ Dispersion plots of $\alpha_{15}$ vs $\alpha_{5,10,40}$. The legend includes the values of the correlation coefficient in the three cases. The large values of the  correlation coefficient just shows that we are effectively using the same set of mocks  for the test and that changing the smoothing scale of reconstruction affects the correlation between the results slightly, but remains large. }
   \label{fig:disp}
\end{figure}

The accuracy in the $\alpha$  value is very similar between $5$ and $15 h^{-1}$Mpc, but visibly degrades when going to higher smoothing scales. 
The smaller dispersion for $\alpha$ is for 10-15 $h^{-1}$Mpc; however, the 15 $h^{-1}$Mpc is significantly biased (5.3 $\sigma$),  \modif{ thus, 10$h^{-1}$Mpc is the best option if we considered only $\alpha$.} 
 \modif{ As we find a relative large bias in 15 $h^{-1}$Mpc$h^{-1}$Mpc compared to $5,10,40 h^{-1}$Mpc, we perform a cross-check between the results using the 4 smoothing scales as a sanity check.  We included in Figure \ref{fig:disp} the dispersions plots of the $\alpha_{15}$ fits from the mocks, and the R=5,10, and 40 Mpc/h. The legend includes the values of the correlation coefficient in the three cases. The results show that effectively we are using the same set of mocks as they show very hight correlation coefficient and that changing the smoothing scale of reconstruction affects slightly the correlation between the results.}

For $\epsilon$, the least significant bias ($b_\epsilon/\sigma_\epsilon$ lower) is for \modiff{the} 5 $h^{-1}$Mpc smoothing scale.  \modif{ Considering $\alpha$ and $\epsilon$ simultaneously, the best option is 5 $h^{-1}$Mpc which gives the less significant bias with smaller dispersion (0.8$\sigma$ in $\alpha$ and $0.9\sigma$ in $\epsilon$ ).} 

We find maximal variations in the mean value for best fit parameters, $\sim$0.5\% for $\alpha$ and 0.3\% for $\epsilon$, producing $\Delta D_A , \Delta H\sim 0.5$\%. 
In the case of  $\beta$, the precision seems to be very similar for the $5$ and $15 h^{-1}$Mpc smoothing scales and slightly lower for the 40 $h^{-1}$Mpc. The accuracy is also similar for $5$ and $15h^{-1}$ smoothing scales and clearly degrades when using a higher smoothing scale. The $\chi^2/d.o.f$ values are very similar between the different smoothing scales explored\footnote{The values  of the $\chi^2\sim$0.8  are similar to those obtained with the reconstruction implementation used in previous analysis. We add more discussion about the results using different implementations in Appendix \ref{ap:imp}.}.

 \modif{ Figure \ref{fig:errors} presents the distributions of the uncertainties for $\alpha$ and $\epsilon$  parameters for different smoothing scales and the quantitative fitting results for the uncertainties are summarised in Table \ref{tab:err}. We can see that the smoothing scale strongly affects the error distributions. and that the error bars decrease using  5 $h^{-1}$Mpc smoothing scale. The mean error  is reduced by 13\% for $\alpha$ and 24\% for $\epsilon$ when passing from 15 to 5 $h^{-1}$Mpc.}

 \modif{ In this section, we have shown how the smoothing scale affects the anisotropic clustering results and we found that  the smaller bias and error bars are obtained using a smoothing scale of  $5 h^{-1}$Mpc.  In order to associate a systematic error, we used the scale $15 h^{-1}$Mpc as the fiducial smoothing.} We observe that varying the smoothing length from $15$ to $5h^{-1}$Mpc generates a $\Delta \alpha=0.005$ and $\Delta \epsilon=0.001$ (Table \ref{tab:fits}). Concerning the errors, the variation observed is $\Delta \sigma_\alpha=0.002$ and $\Delta \sigma_\epsilon=0.012$. Expressing these variations in the best fits and their uncertainties to the final measurements of the angular diameter distance and Hubble parameter, we get variations of $\Delta \sigma_{D_A}\sim$ 40\% and $\Delta \sigma_H\sim$33 \%.

\begin{figure}
   \centering     
      \includegraphics[width=3.7in]{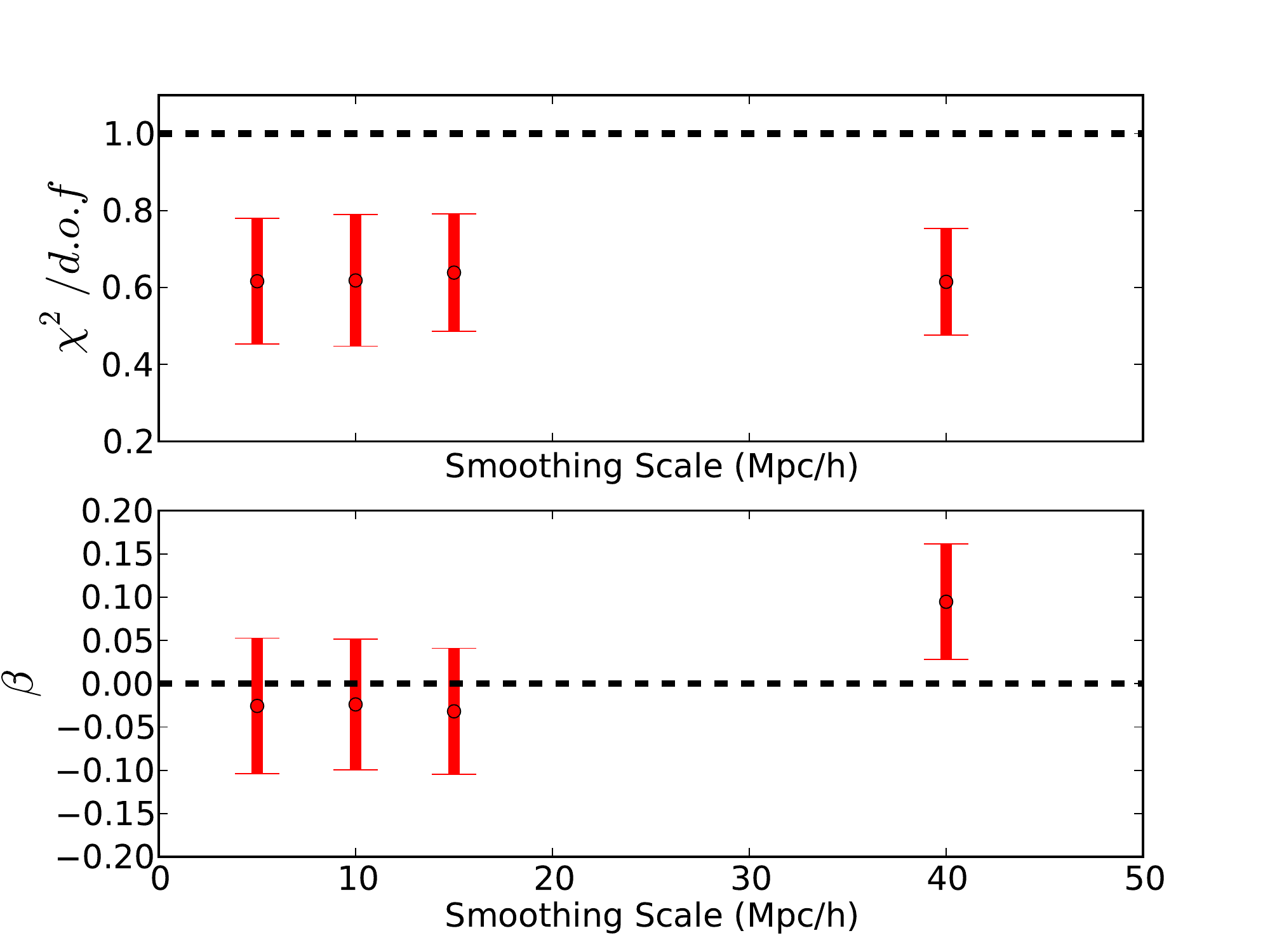}
     \caption{  Mean of best fits for $\beta$ and $\chi^2/d.o.f.$ for $200$ QPM mocks analysed with different smoothing scales. The error bars are given by the standard deviation from $200$ realisations. The 5,10 $h^{-1}$ Mpc smoothing scale gives similar RMS and bias for linear redshift distortion parameter $\beta$ but slightly more bias for larger smoothing. Even though $\beta$ is a nuisance parameter in BAO analysis, it is interesting to observe the value fitted, as it indicates the level at which the redshift corrections are using the right value of the velocity field. The $\chi^2/d.o.f$ values are very similar between the different smoothing scales explored. }
   \label{fig:MeanParSmooth2}
\end{figure}

\begin{table}
\begin{center}
\caption{ Best fits for $\alpha$ and $\epsilon$. Mean and RMS from 200 reconstructed QPM mocks. Second block refers to the results using the covariance matrix from 1000 mocks with a fixed smoothing scale of 15 Mpc/h.} \label{tab:fits}
\begin{tabular}{@{}lcccccccccc}
\hline
$R$&
$\widetilde{\alpha}$&RMS&$\frac{b_\alpha}{\sigma_\alpha}$&
$\widetilde{\epsilon}$&RMS&$\frac{b_\epsilon}{\sigma_\epsilon}$&
$\widetilde{\beta}$&RMS\\\\

\hline
5 &      0.9992 &     0.0136 &        0.8 &     0.0007 &     0.0109 &        0.9 &    -0.035 &     0.090 \\
10 &      1.0004 &     0.0128 &        0.5 &     0.0024 &     0.0142 &        2.4 &    -0.025 &     0.088\\
15 &      1.0048 &     0.0128 &        5.3 &     0.0021 &     0.0169 &        1.7 &    -0.019 &     0.085 \\
40 &      1.0002 &     0.0147 &        0.2 &    -0.0048 &     0.0224 &        3.0&     0.1020 &     0.077 \\
\hline
5 &        1.0003 &     0.0138 &        0.2 &     0.0002 &     0.0115 &        0.5 &    -0.028 &     0.079 \\
10 &      1.0012 &     0.0137 &        1.2 &     0.0021&     0.0143 &        2.1 &    -0.042 &     0.089\\
15 &      1.0027 &     0.0142 &        2.7 &     0.0029 &     0.0185 &       2.2 &    -0.030 &     0.093 \\
40 &      0.9980 &     0.0158 &        0.5 &     -0.0011 &     0.0256 &        0.6&     0.1493&     0.102\\

\hline
\\[-1.5ex]
\end{tabular}
\end{center}
\end{table}

\begin{table}
\begin{center}
\caption{  Uncertainties in $\alpha$ and $\epsilon$ parameters. Median, 16 and 84 percentiles of the uncertainties distributions from 200 reconstructed QPM mocks.} 
\label{tab:err}
\begin{tabular}{@{}lcccccccccc}
\hline
R&
$\widetilde{\sigma_\alpha}$& $\Delta \sigma_\alpha$\%&
$\widetilde{\sigma_\epsilon}$&$\Delta \sigma_\epsilon$ \% \\
\hline
$5$&$0.0136^{+0.0021}_{-0.0019}$&-9.9&$0.0118^{+0.0028}_{-0.0017}$&-45.9\\
\\[-1.5ex]
$10$&$0.0142^{+0.0030}_{-0.0019}$&-5.9&$0.0167^{+0.0040}_{-0.0024}$&-23.4\\
\\[-1.5ex]
15&$0.0151^{+0.0030}_{-0.0023}$&-&$0.0218^{+0.0062}_{-0.0039}$&-\\
\\[-1.5ex]
$40$&$0.0150^{+0.0042}_{-0.0023}$&$-0.7$&$0.0233^{+0.0080}_{-0.0041}$&6.9\\
\\[-1.5ex]

 \hline
\\[-1.5ex]
\end{tabular}
\end{center}
\end{table}

\begin{figure}
   \centering     
   \includegraphics[width=1.6in]{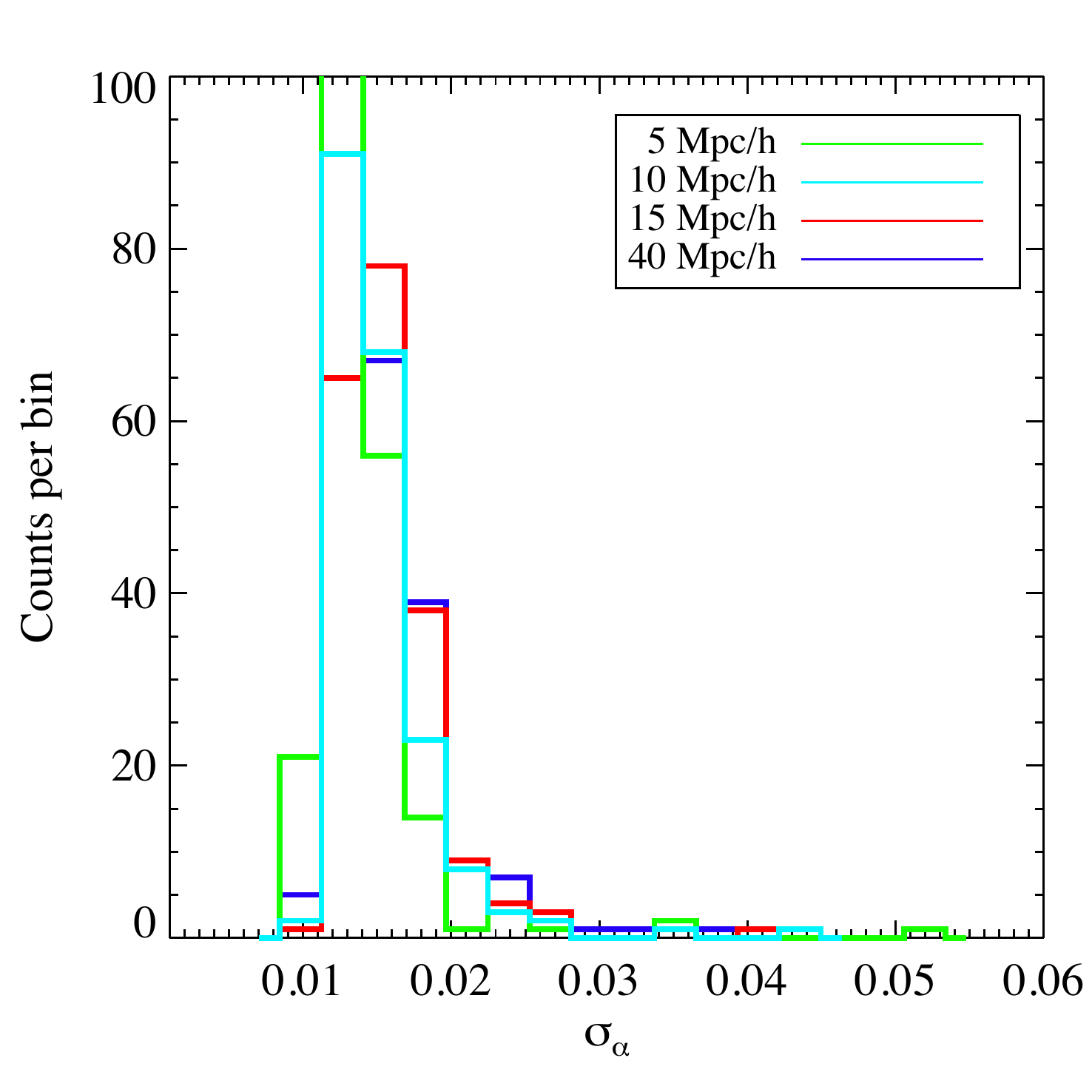}
   \includegraphics[width=1.6in]{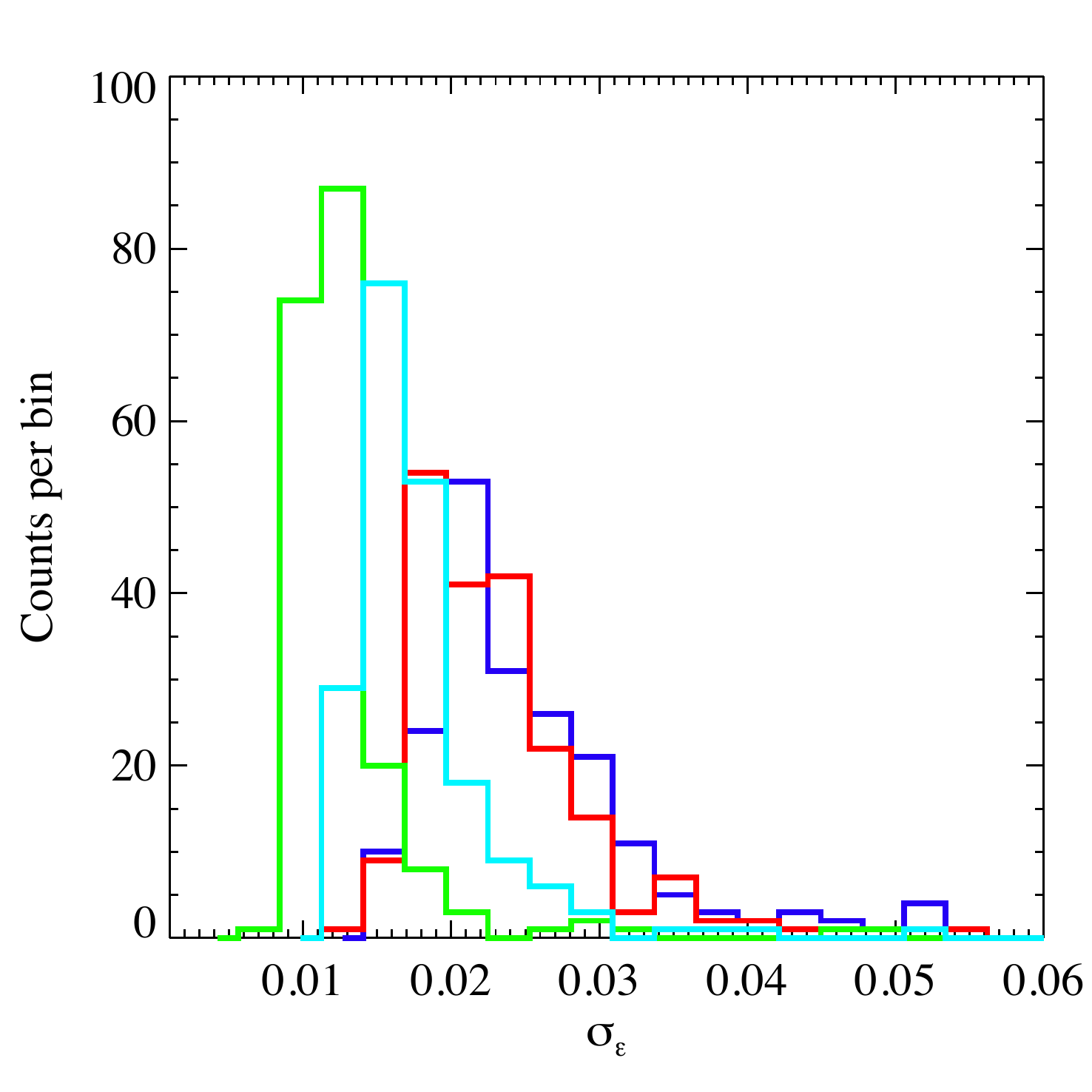}
  \caption{   Histograms of uncertainties in $\alpha$ and $\epsilon$ measured for the individual reconstructed mocks for different smoothing scales, 5 (green), 10 (cyan), 15 (red), and 40 (blue). The distributions depend strongly on the smoothing scale used in the reconstruction. }
   \label{fig:errors}
\end{figure}

\subsection{Dependence on Reconstruction Implementation}

 In Appendix \ref{ap:imp}, we show that  the results obtained in this work for the best smoothing scale, in terms of anisotropic fits performance, are independent of the particular implementation of reconstruction algorithm. We compare our implementation of the reconstruction algorithm to the method implemented by \cite{Pad12}.
 We compare results in terms of the multipoles. Figure \ref{fig:mean_smooth_bias21} shows the mean multipoles for the different implementations using three different smoothing lengths, 5, 10, and 15 $h^{-1}$ Mpc.  \modif{ We observe that the monopole behaviour is consistent between different implementations in the fitting range.} The differences observed are only important at scales smaller than 20 $h^{-1}$ Mpc.  The amplitude of the quadrupole between different implementations is significantly different.  \modif{ RP implementation shows a positive quadrupole which means that the reconstruction implementation is over-correcting the anisotropy. 
 On the other hand, our implementation of reconstruction shows a negative quadrupole, which means that our implementation is under-correcting the anisotropy; }however, both quadrupoles are very similar in shape. Differences in the quadrupole are generated by two effects, redshift distortions correction and effects of the angular and radial selection functions, which are implemented in slightly different ways. Further exploration and quantification of these two contributions \modiff{which} generate differences in the quadrupole amplitude are treated in Vargas-Maga\~na et al (in prep).

 \modif{ We fitted the multipoles post-reconstruction from both implementations and compared the results. The results are completely consistent between both implementations, the differences between the mean of  the two implementations is  ~0.1\% for $\alpha$ and $\epsilon$. }The RMS of the best fits are similar at 0.1\% in both quantities. Concerning the errors, the dispersion of the distributions are also very similar; however, {there is a} systematic larger error in RP implementation compared to RV in $\alpha$ and $\epsilon$. However, the same trends are observed in both implementations; the error in $\epsilon$ is monotonically decreasing as we apply a smaller smoothing scale.  In the case of $\alpha$ the errors we get with \modiff{the} 5 and 10 $h^{-1}$ Mpc smoothing scale are very similar, but smaller than using the 15 $h^{-1}$ Mpc that is regularly applied in BAO analysis.
 
\subsection{Dependence on Covariance Noise}

We also test the effect of the covariance noise on the fitting results. The fitting results presented in previous sections use the covariance matrix generated from the  200 reconstructed mocks with \modiff{a} different smoothing scale. In this section, we substitute this covariance matrix by the covariance  generated with 1000 mocks with \modiff{the} $15$ Mpc/h smoothing scale and we fit the fourth sets of multipoles post-reconstruction of 200  with different smoothing scales. This choice is motivated \modiff{by} previous results \cite{Var16} indicating the covariance noise could affect the stability of the fitting results. \modiff{By} performing this change we are testing the impact in the fitting results just from the noise observed in the covariance. Even this procedure is neglecting the effect of smoothing scale in the covariance post-reconstruction, we consider it \modiff{to be} testing the confidence we can have in the fitting results given the reduced number of mocks we used for \modiff{these} tests.

\begin{figure}
   \centering     
   \includegraphics[width=3.7in]{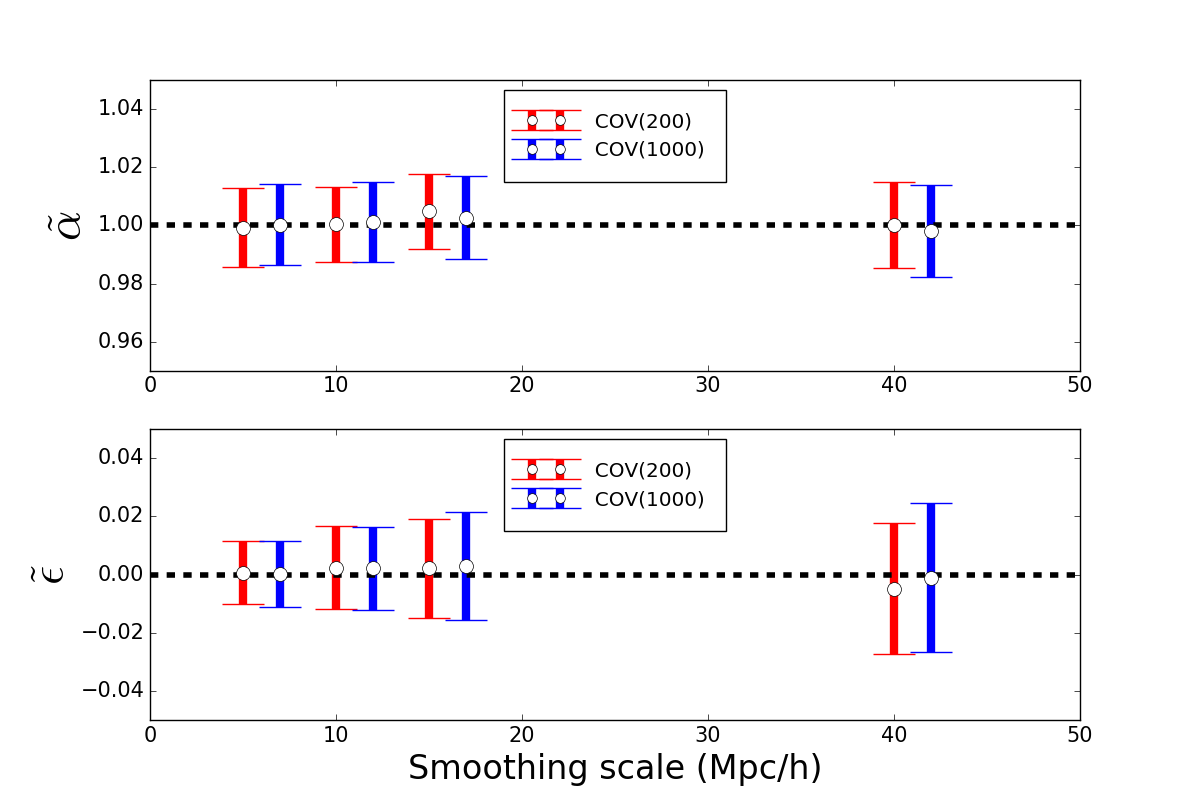}
    \caption{ { Mean of best fits for $\alpha$ and $\epsilon$ for $200$ QPM mocks NGC analysed with different smoothing scales  fitted with the 200-covariance matrix using the reconstructed mocks with different smoothing scales [in red] compared with the fits performed with the 1000-covariance matrix generated from post-reconstrucion mocks using a fixed smoothing scale of 15$h^{-1}$Mpc [in blue]. The error bars are given by the standard deviation from the $200$ realizations. For $\alpha$, the smaller dispersion is for 5-10 $h^{-1}$Mpc. The best is 5 $h^{-1}$Mpc, which has less significant bias with small dispersion (0.2$\sigma$). For $\epsilon$, the less significant bias ($b_\epsilon/\sigma_\epsilon$ lower) is for the 5 $h^{-1}$Mpc smoothing scale given their small bias and dispersion.} }
   \label{fig:covnoise}
\end{figure}

Results are shown in \modiff{the} second block of Table \ref{tab:fits} and also in Figure \ref{fig:covnoise}. The results follow similar trends than using the ``noisy'' covariance matrix (even though the fourth significant figure changes slightly). The less biased results  for $\alpha$ and $\epsilon$ are obtained \modiff{from the} smaller smoothing scale 5 Mpc/h followed by the 10 Mpc/h and 15 Mpc/h smoothing scales and \modiff{then} degrades for 40 Mpc/h. The dispersions  shows variations for some cases by $\sim$0.001 related  the error bars themselves are noisy estimates, i.e. with 200 mocks there should be more scatter in the error bars than there would be with 1,000 mocks. %

An interesting feature of the fits performed with the 1000-covariance is that 
the significant bias observed in \modiff{the} 15 Mpc/h case for the previous sections reduces from 5.3$\sigma$ to just 2.7$\sigma$ \modiff{which} indicates the noise in the covariance was generating this large bias. 
We highlight that even though the variations derived from the noise in the covariance, the main conclusions about the performance of the fits for the different smoothing scales remain unchanged. The smaller smoothing scale of 5 Mpc/h is giving the best results.  Summarizing the results, the maximal variations in the mean value for best fit parameters considering this new covariance reduces to 0.002 for $\alpha$ and $\epsilon$, producing $\Delta D_A , \Delta H\sim 0.003$ and $0.004$\%.

\begin{figure}
   \centering     
      \includegraphics[width=3.7in]{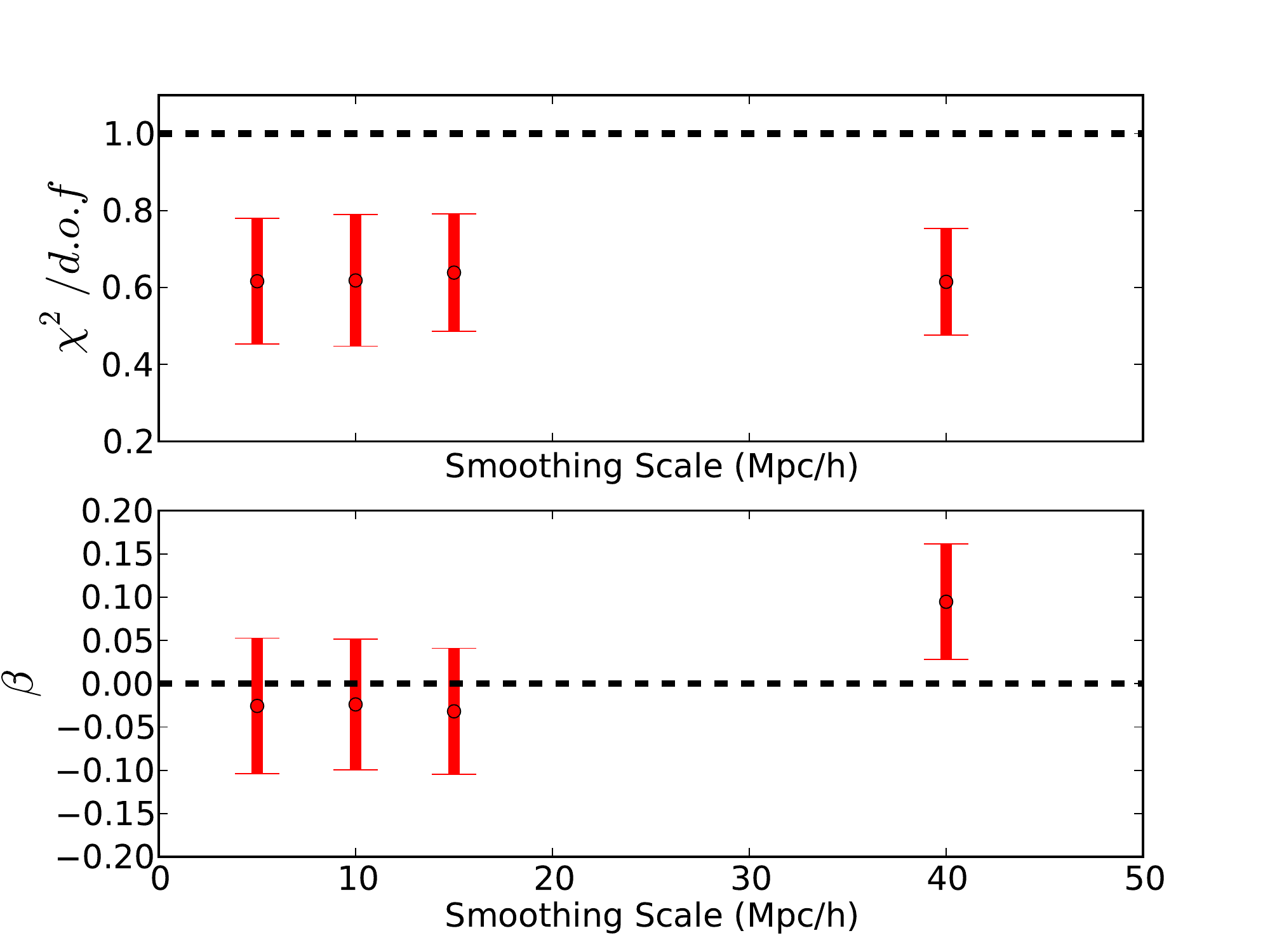}
     \caption{  Mean of best fits for $\beta$ and $\chi^2/d.o.f.$ for $200$ QPM mocks analysed with different smoothing scales. The error bars are given by the standard deviation from $200$ realisations. The 5,10 $h^{-1}$ Mpc smoothing scale gives similar RMS and bias for linear redshift distortion parameter $\beta$ but slightly more bias for larger smoothing. Even though $\beta$ is a nuisance parameter in BAO analysis, it is interesting to observe the value fitted, as it indicates the level at which the redshift corrections are using the right value of the velocity field. The $\chi^2/d.o.f$ values are very similar between the different smoothing scales explored. }
   \label{fig:MeanParSmooth2}
\end{figure}

\section{Smoothing effect on accuracy of displacement field estimation.}\label{sec:disp}
 In this section, we explore the effect of the smoothing scale on the reconstructed displacement field  to figure out the origin of the improvement  observed in the post-reconstruction anisotropic clustering. In particular, we investigate if the improvement is related to the better estimation of the displacement field when we reduce the smoothing scale applied in the reconstruction algorithm.

{Moreover,} the peculiar velocity of galaxies is a valuable quantity in cosmology, since it contains complementary information to that enclosed in the galaxies' positions.
In the literature, there are many strategies with different approximations to obtain the velocity field.
Reconstruction provides the simplest approach to get \modif{ a model dependent  reliable velocity field at large scales}. For reconstruction, the velocity field is obtained using the well-known Zeldovich approximation, in which the displacement field is given by the gradient of the gravitational potential at $\vec{q}$.
The range of applicability of this method is limited to very large scales, as it fails to describe the dynamics of a nonlinear field \citep{Kit12}.  \modif{ Even though it has a limited range of applicability, the velocity field from reconstruction provides a direct method to constrain gravitational model comparing with velocity measurements}

In this section, we analyze the accuracy of the velocity field derived from reconstruction and then the effect of the smoothing length  on the estimated velocity field. Instead of working with \modiff{a} velocity field we transformed the velocity to displacements using the continuity equation:
\begin{equation}
\Psi_i =v_i/f(z)H(z),
\end{equation}
\modiff{where} $f$ is the growth factor at redshift of the snapshot and $H(z)$ is the Hubble parameter. 

First we evaluate the quality of the recovered displacement field from reconstruction compared with simulations.  \modif{ This test is important for validating the use of displacement field for other cosmological applications such as the measurement of the kSZ effect \citep{Scha16,ksZPlanck,ksZ3}.
 Secondly, we present an estimate of the uncertainties associated with the reconstructed displacements for individual galaxies and the bias of the displacement estimations with respect to the true displacements as a function of the smoothing length.}

 \modif{ For this section we use a different set of mock {catalogues}: the cubic RunPB real space mocks. We use RunPB mocks because they are expected to have more accurate velocities; even if we do not have a large number of these mocks to study the statistical properties for BAO fits, we have enough galaxies in each of them to accurately compare the individual velocities with the reconstructed field\footnote{The ideal set of mocks for testing the displacement field accuracy would be high fidelity mocks with reliable velocities with the properties of the survey; however, we only have available high fidelity mocks without the mask of the simulations (RunPB mocks).}.}

\subsection{Reconstructed Displacement Field}

\begin{figure*}
   \centering     
    \includegraphics[width=6.8in]{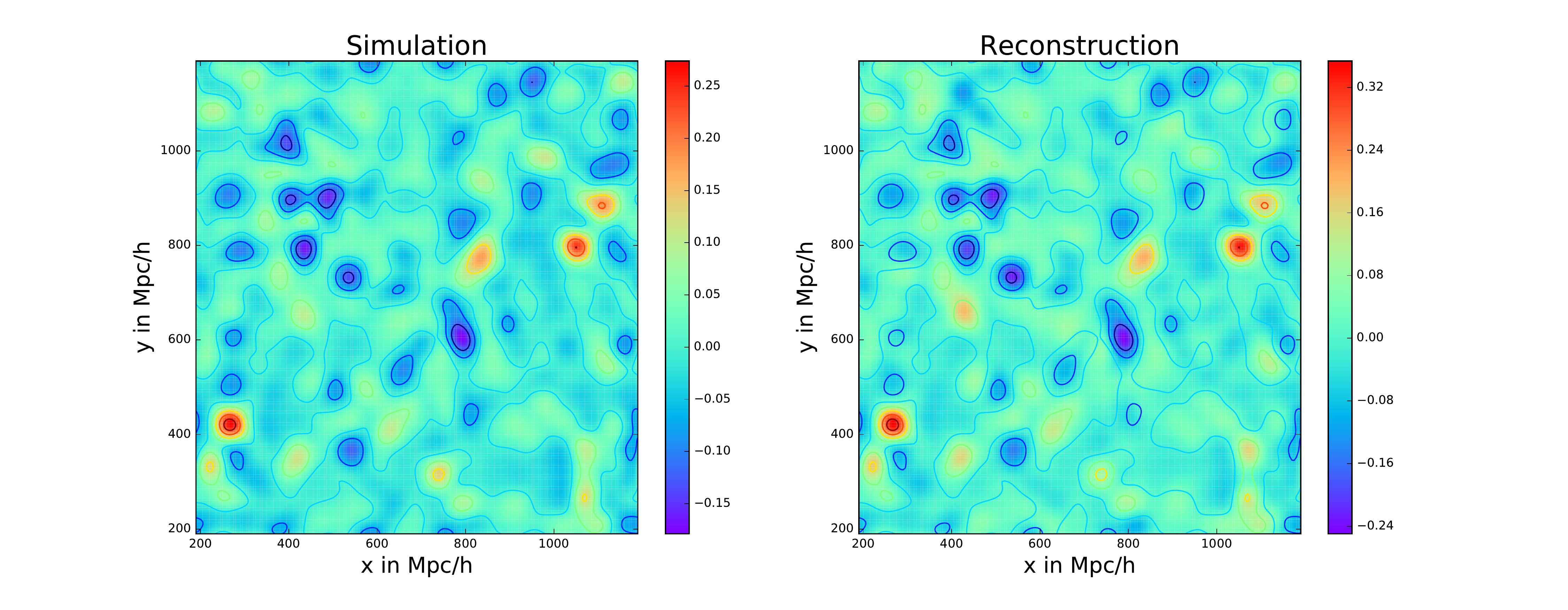}
  \caption{  Comparison of the displacement field from a RunPB simulation [left] and from reconstruction [right]. The reconstructed field reproduces the structure of the true velocity field, demonstrating the accuracy of the reconstructed displacement field. Both of the fields are smoothed with a 10 Mpc/h Gaussian filter. The color bars are in Mpc/h.
} \label{fig:velfield_smooth}
\end{figure*}

\modif{The purpose of this subsection is  to show that reconstruction is providing a reliable estimate of the true displacement field but it is limited to reproduce the large scales.  We analyze the quality of the reconstructed displacement field as compared to the true displacement field.}
\modif{ We compare the individual displacement of each galaxies with the reconstructed one in the corresponding position in a grid of $276^3$ pixels of size (5 $h^{-1}$ Mpc)$^3$.} In Figure \ref{fig:velfield_smooth}, we show the x-component of the displacement field, $\Psi_x$, from reconstruction (right panels) and the true corresponding displacement from simulations (left panels) using a Gaussian filter of 10 $h^{-1}$Mpc. We \modiff{only} show one component of the vector field, as  the other two components are \modiff{quite} similar. The $\Psi_x$ is a 3D scalar field; for the illustration, we show the average value in a slice of 50 $h^{-1}$Mpc over one direction. 
 The plot shows the reconstructed displacement field reproduces the simulation displacement field at scales of 10$h^{-1}$Mpc, demonstrating the accuracy of the reconstructed displacement field. However, we observe that, at small scales, the reconstructed displacement field shows \modiff{fewer} structures compared with the mock {catalogue }displacement field.  

\subsection{Error and Bias of Reconstructed Displacements}

Up to now, we have been using the values of the displacement field in pixels. Now we are interested in comparing the individual values of the displacement for each galaxy to evaluate their agreement to the simulations. We expect only be correct on the linear component of the velocity field. We are also interested in associating an error \modiff{with} those velocities;
this is interesting if we use those velocities for inferring cosmological information for applications other than BAO.

\begin{figure*}
   \centering
\includegraphics[width=1.8in]{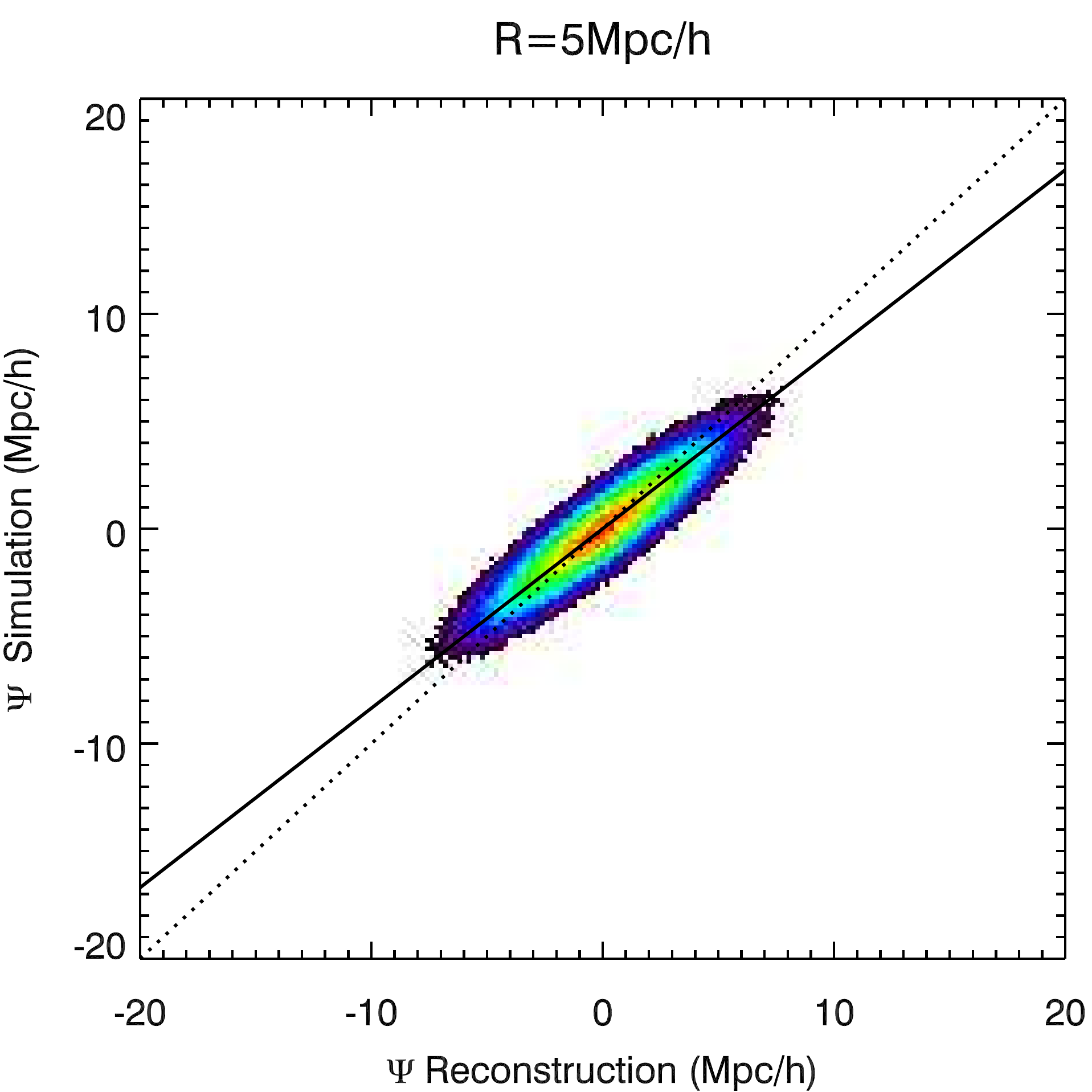}\includegraphics[width=1.8in]{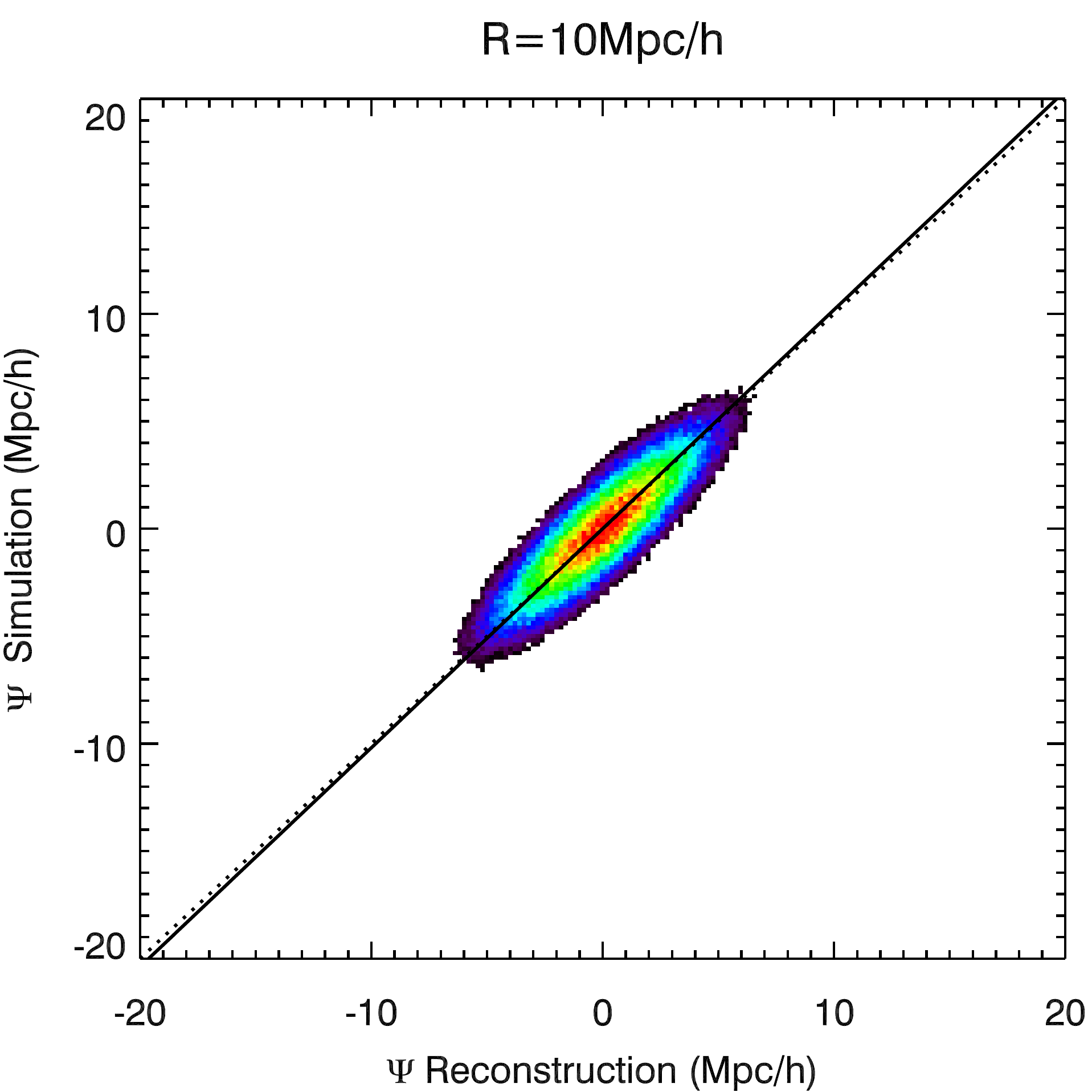}\includegraphics[width=1.8in]{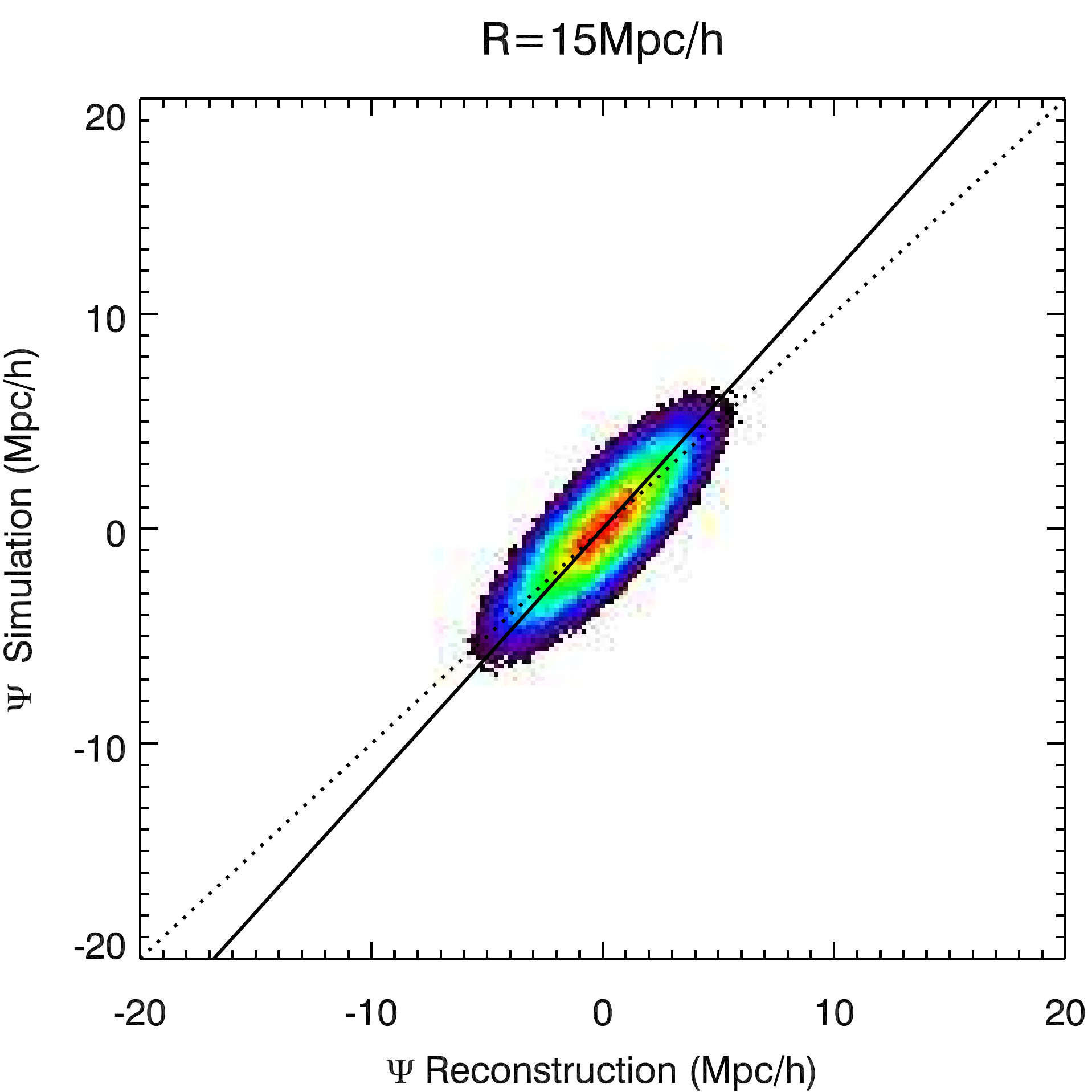}\includegraphics[width=1.8in]{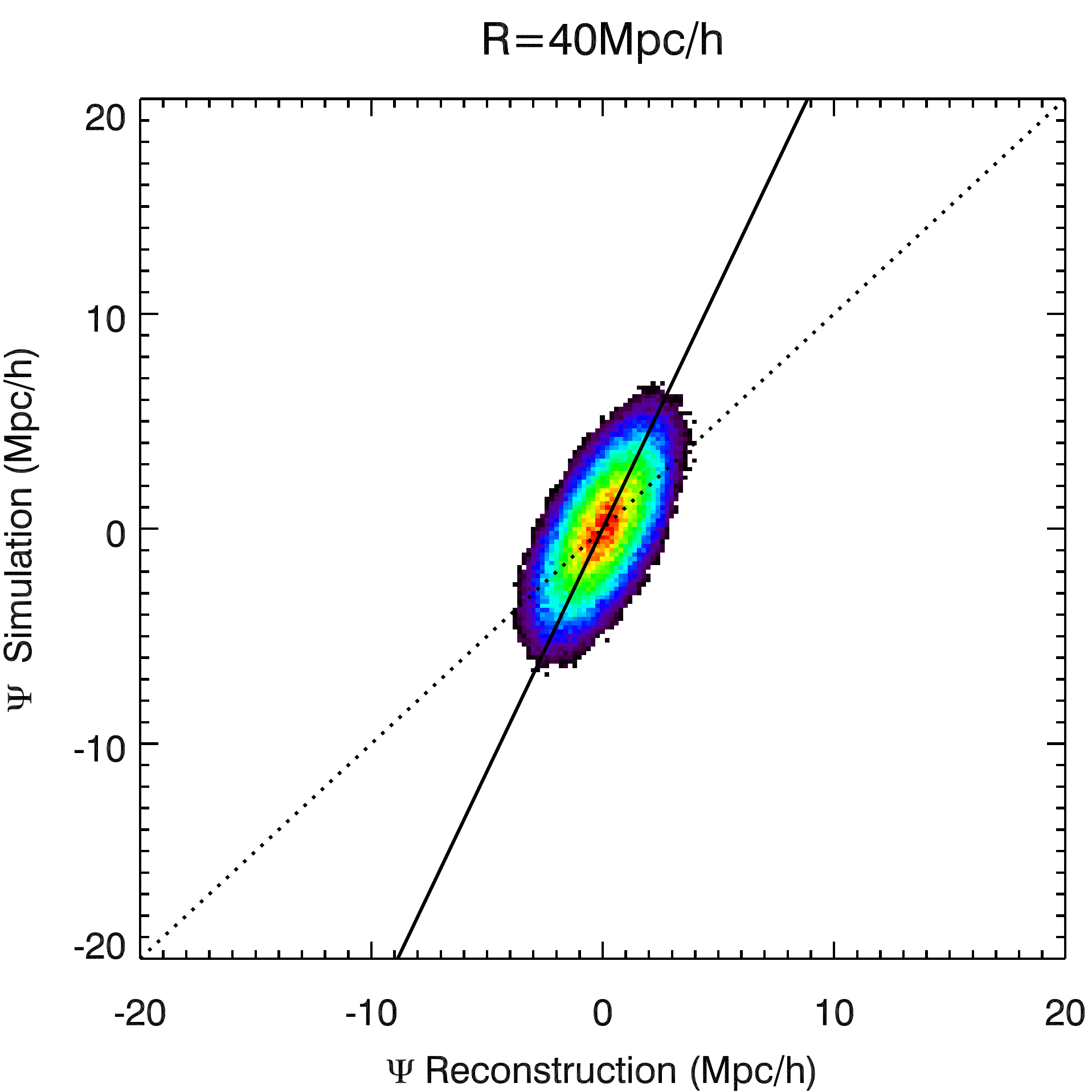}
	  \caption{RunPB mocks. Scatter plots between true and reconstructed displacements from different smoothing scales, from left to right R=5,10,15,40 $h^{-1}$ Mpc. The mocks used in this test were in real space. The dashed white line indicates the 1-1 relation.
	 The solid line comes from fitting a 2D-Gaussian to the 2D histogram; it represents the angle between the major axis and the ordinate axis. This angle does not  indicate the bias of the reconstructed displacements compared with the true displacements but provides an illustration of the effect. 
	  The best result is obtained for the 5 $h^{-1}$ Mpc according to our quantitative results shown in Table \ref{tab:noise_bias_pbrun}. These mocks have more realistic velocities compared to QPM mocks.}
	    	\label{fig:hist2dPBrun}
\end{figure*}

We study the dispersion between true and reconstructed displacements, for which we estimate the 2D histograms mock-by-mock for different smoothing scales using the RunPB cubic mocks (Figure \ref{fig:hist2dPBrun}). 
 
 The dotted black line indicates the 1-1 relation. The solid line comes from a 2D Gaussian fitting to the 2D histogram; it represents the angle between the major axis and the ordinate axis. This angle cannot be interpreted as the bias of the reconstructed displacements compared with the true displacements, but it provides an illustration of the effect. In Section \ref{biasrec} we describe the methodology we follow to characterise the bias and noise in the displacement estimation and we present the results in \ref{biasrecres}. The derivation of the equations is presented in Appendix \ref{ap:biasnoise}.

\begin{table}
\begin{center}
\caption{ \modiff{RunPB Cubic} Error and bias from equations \ref{eq:postulat}, \ref{eq:alpha}, and \ref{eq:sigma}; from 1 reconstructed RunPB Cubic mock with different smoothing lengths. The best results  are obtained for the 5$h^{-1}$ Mpc smoothing scale.}
\label{tab:noise_bias_pbrun}
\begin{tabular}{@{}lcccccc}
\hline
R & $\mathcal{C}(\tilde{\psi}_R,\psi)$ & $\alpha$ & 
$\sigma_N$ & 
$\frac{\sigma_N}{\alpha}$   \\
\hline
5    & 0.87     &1.01  &    2.23    &   2.21\\
10   &  0.83   &  0.77 &    2.08  &     2.70\\
15    & 0.77&   0.64  &      2.05  &    3.20\\
40    & 0.60 &  0.33 &     1.74  &      5.27\\
\hline
\\[-1.5ex]
\end{tabular}
\end{center}
\end{table}

\subsubsection{Theoretical Estimation of Bias and Noise}\label{biasrec}
In order to get a sensible estimate of the bias and noise level, we follow an approach in terms of probability distributions. We  write the reconstructed displacement as a biased estimation of the "true" displacement $\psi$ plus a Gaussian noise: 
\begin{equation}
\label{eq:postulat}
\tilde{\psi}_R = \alpha\times\psi + \mathcal{N}(0,\sigma_N),
\end{equation}
where $\alpha$ is the bias term and $\sigma_N$ characterises entirely the noise. Because the noise is uncorrelated to the displacement, \modiff{we then} have:
\begin{equation}
\label{eq:first_rel}
\tilde{\sigma}_R^2 = \alpha^2\sigma_{\psi}^2 + \sigma_N^2,
\end{equation}
where $\tilde{\sigma}_R^2$ is the variance of the ``reconstructed" displacements for the smoothing scale $R$ and $\sigma_{\psi}^2$ is the variance of the ``true" displacements. We must notice we do not have direct access to  $\alpha$ and $\sigma_N$. The observable quantities are the variance or standard deviation from $\psi$ and $\tilde{\psi}_R$ (i.e. $\sigma_{\psi}$ and $\tilde{\sigma}_R$) and $\tilde{\alpha}$ the angle of the correlation between $\psi$ and $\tilde{\psi}_R$\footnote{$\tilde{\sigma}_R$ using directly the standard deviation of the reconstructed values. In the case of simulations, we have access to $\sigma_{\psi}$, which is obviously not the case using data.}.

While the equation (\ref{eq:first_rel}) provides a relation between the observable quantities and the intrinsic bias and noise, we still need to break the degeneracy between the two contributions. We propose \modiff{using} the scatter plot between $\tilde{\psi}_R$ and $\psi$. Following the calculus of Appendix \ref{ap:biasnoise}, we get the following expressions for the bias of the reconstructed displacements and the associated noise ($\alpha$, $\sigma_N$) for a smoothing scale $R$ given the measurable quantities ($\tilde{\alpha}$, $\sigma_{\psi}$, $\sigma_{\tilde{\psi}_R}$): \begin{equation}
\label{eq:alpha}
\alpha = \frac{\sigma_{\tilde{\psi}_R}^2}{\tilde{\alpha}\sigma_{\psi}^2},
\end{equation}
\begin{equation}
\sigma_N^2 = \tilde{\sigma}_R^2 - \frac{\tilde{\sigma}_R^4}{\tilde{\alpha}^2\sigma_{\psi}^2}.
\label{eq:sigma}
\end{equation}

The dispersion of the de-biased displacement is then given by $\sigma_N/\alpha$. This quantity enables us to evaluate the quality of the estimator depending only on the smoothing scale $R$, $\sigma_N$, and $\alpha$.

\subsubsection{Measurement of Bias and Noise}\label{biasrecres}

We measure the $\sigma_{\tilde{\psi}_R}$ and $\sigma_{\psi}$ standard deviation from reconstructed and simulation displacements. We measure the correlation coefficient $\mathcal{C}(\tilde{\psi}_R,\psi)$. 
We calculate the biased coefficient $\tilde{\alpha}$ as:
\begin{equation}
\tilde{\alpha}=\frac{\sigma_{\tilde{\psi}_R}}{\sigma_{\psi} \mathcal{C}(\tilde{\psi}_R,\psi)}.
\end{equation} 
We deduce the two quantities, the bias $\alpha$ and the noise, $\sigma_N$, using equations (\ref{eq:alpha}) and (\ref{eq:sigma}). Once we have the values for the bias and the noise, we can determine which smoothing scale is the best. We expect a \modiff{result similar to} the correlation coefficient \modiff{with extra information from the contribution of the} intrinsic noise, $\sigma_N$, for the different smoothing scales. We expect a lower noise for the larger smoothing scale, as we are deleting the nonlinear scales more efficiently. However, we also expect to have a poorer precision in the displacement reconstruction because we are losing some structures corresponding to the linear theory. \modiff{Therefore}, we expect a lower value for the bias $\alpha$ as we  increase the smoothing scale. The best smoothing scale is therefore the one that balances the two effects and corresponds to the minimum value of the ratio $\sigma_N / \alpha$ as demonstrated in the appendix.

\begin{figure}
   \centering    
   \includegraphics[width=3.8in]{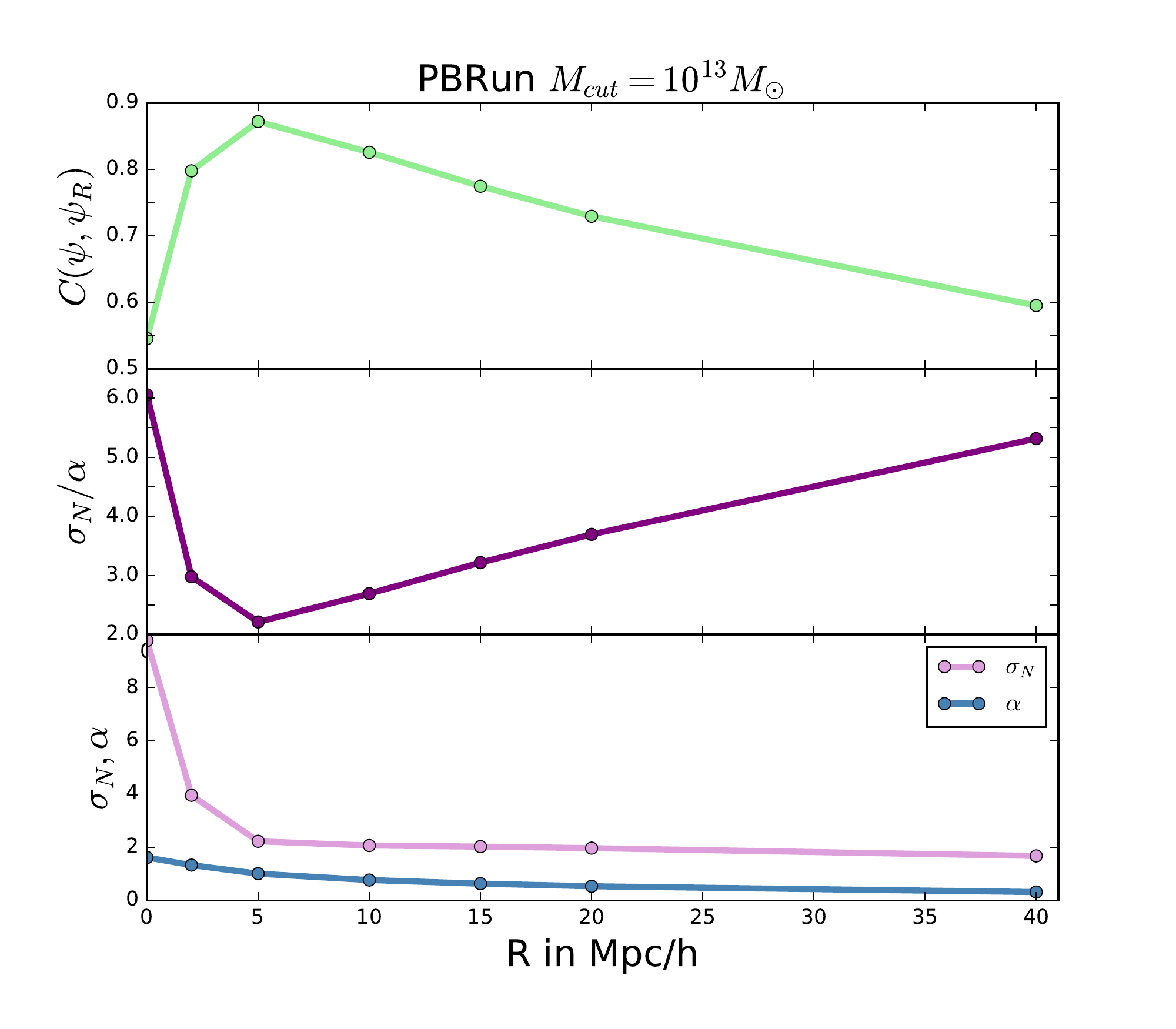}
   	  \caption{ Top panel shows the correlation coefficient $<\mathcal{C}(\tilde{\psi}_R,\psi)>$ , central panel the quality factor $\sigma_N / \alpha$, and bottom panel the bias $\alpha$ and the noise $\sigma$, as a function of the smoothing scale. The best results for cubic mocks  RunPB are for 5 $h^{-1}$ Mpc, while the best result for the Sky mocks are obtained for 10-15 $h^{-1}$ Mpc.}
	    	\label{fig:sumbiasnoise}
\end{figure}

The results are shown in Table \ref{tab:noise_bias_pbrun} for RunPB cubic mocks in real space. Figure \ref{fig:sumbiasnoise} summarises bias and noise results.
\modiff{The top} panel shows the correlation coefficient $<\mathcal{C}(\tilde{\psi}_R,\psi)>$ as a function of the smoothing scale, \modiff{the} intermediate panel shows the quality factor $\sigma_N / \alpha$ as a function of the smoothing scale. Finally, \modiff{the} bottom panel shows the noise $\sigma$ and the bias $\alpha$ as a function of the smoothing scale.
 
 \modif{  The best results are obtained for the highest correlation coefficient and the lowest quality factor. We find that the quality value is a monotonic function of the smoothing scale. The ratio $\sigma_N/\alpha$ is minimized for $R=5 h^{-1}$ Mpc and also that the maximal correlation coefficient value ($C=0.87$) corresponds to the $R=5 h^{-1}$ Mpc. 
We can see also from the bottom panel of Figure \ref{fig:sumbiasnoise}, that both the bias, $\alpha$, and the noise, $\sigma_N$, are decreasing with the smoothing scale, but at a different ratio. As shown in the appendix B, the ratio is the way to measure the accuracy of the estimator (equation B.10). Figure  \ref{fig:sumbiasnoise} also allows to explain why $R=5 h^{-1}$ Mpc is performing better, while a smaller smoothing results in a improved correlation, the noise level increases more abruptly than the correlation, thus giving a poorer displacement field.
From the top and intermediate panels of Figure  \ref{fig:sumbiasnoise}, we can see that the correlation coefficient and the ratio $\sigma_N/\alpha$ are symmetric and so provide equivalent information. However, while the correlation coefficient provides a single information, with the methodology we proposed, we have access to the bias and the noise of the estimator separately. This point is important when one wants to use the reconstruction methodology to derive the velocity field directly and associate errors to it.}

\subsubsection{Halo Bias Dependence}

In this section, we explore how the conclusions we \modiff{got} for BOSS-like samples in the last section scale for other biases and number densities so the results could be applied to other surveys. We show the effects in terms of the displacement field as the bias effect in the correlation function; anisotropic fits are treated in another appendix. For this test, we generate several halo samples from RunPB simulation, applying different halo mass cuts. The sample information  is summarised in Table \ref{tab:halocut_spec}; we show the number density, shot noise level, and halo bias. The halo bias is determined by comparing the power spectrum of halos with the dark matter power spectrum from the simulations.
\begin{table}
\begin{center}
\caption{ RunPB mocks samples generated from different halo mass cuts. } 
\label{tab:halocut_spec}
\begin{tabular}{@{}lccccc}
\hline
Mcut $(h^{-1}M_{\odot})$	 & number density&	 Shot Noise & Halo Bias& $k_{max}$\\	
\hline
1.0e12  & 0.00287320     &348.044 &1.29069 &0.568604\\
3.0e12   &0.00105167     & 950.872&1.55206&0.364857\\
5.0e12  &0.000616345    &  1622.47&1.71838&0.326531 \\
7.0e12  &0.000425321   &   2351.17&1.85922&0.261644\\
9.0e12  &0.000319666   &   3128.27&1.96803&0.234149\\
1.0e13  &0.000281704   &   3549.83&2.02379&0.234149\\
2.0e13  &0.000117975   &   8476.39&2.41207&0.167890\\
3.0e13  &6.69662e-05   &   14932.9&2.71752&0.134476\\
4.0e13  &4.35019e-05  &     22987.5&2.95412& 0.107754\\
5.0e13  &3.05197e-05  &      32765.7&3.14878&0.0964309\\
\hline
\end{tabular}
\end{center}
\end{table}

\begin{figure*}
   \centering    
\includegraphics[width=1.1in]{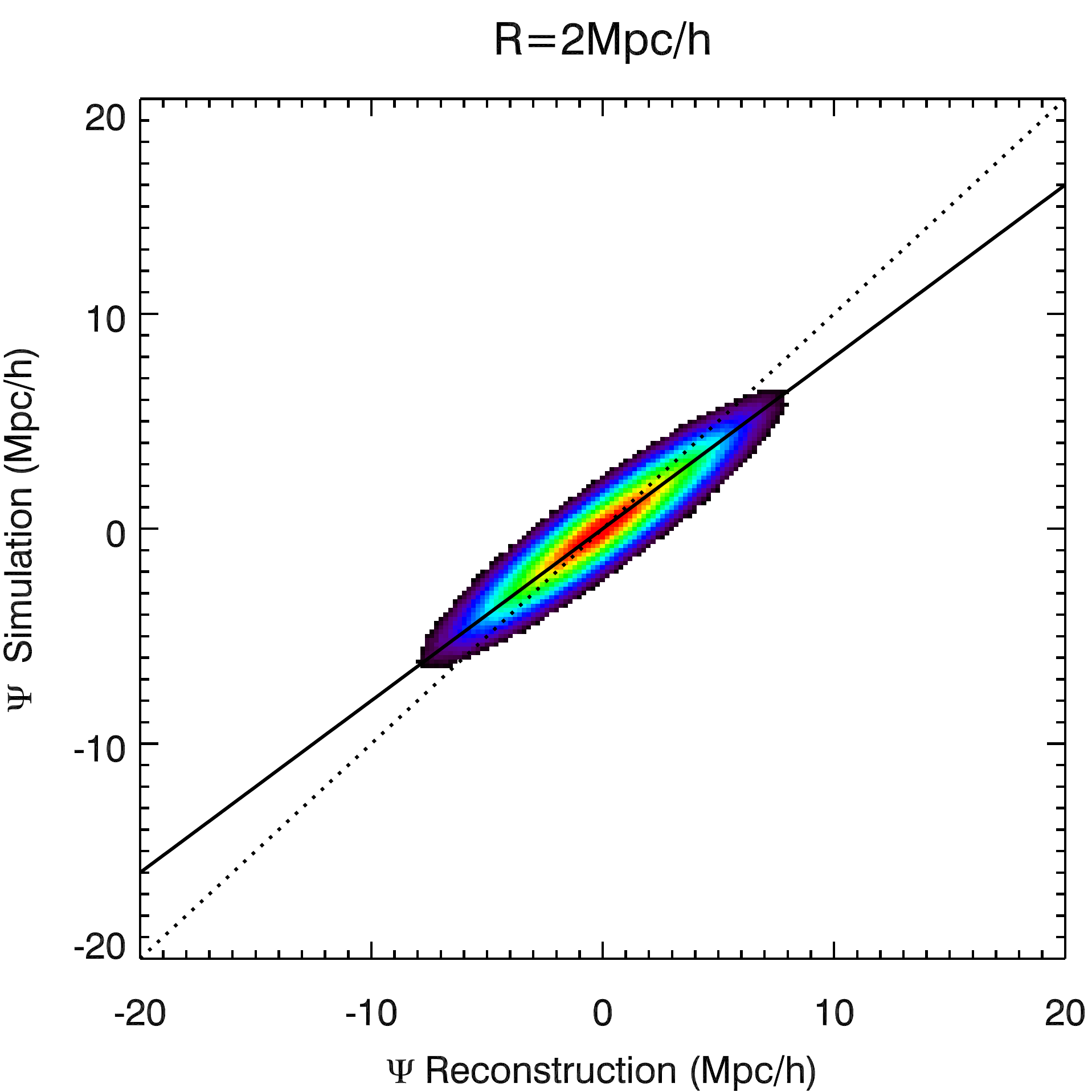}\includegraphics[width=1.1in]{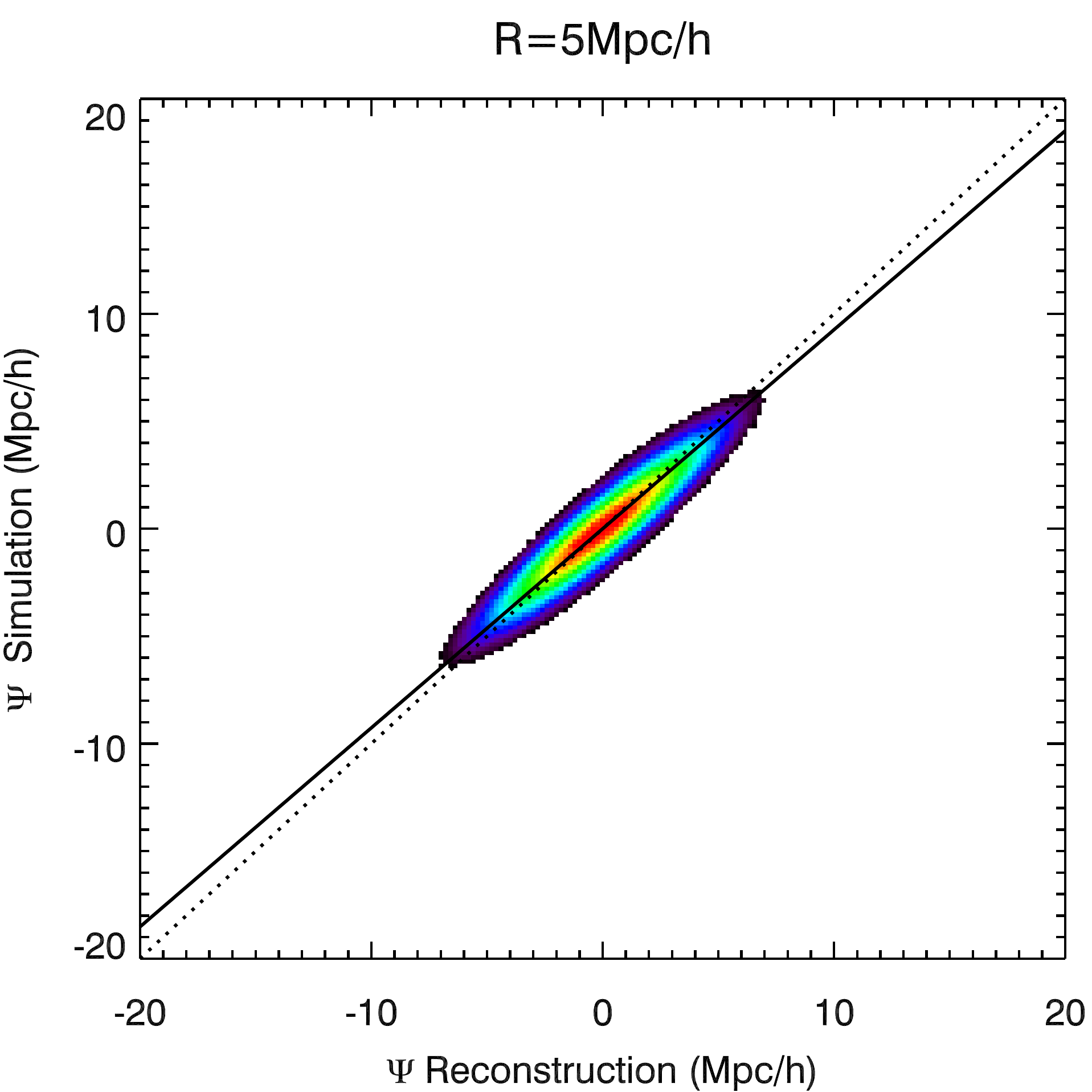}\includegraphics[width=1.1in]{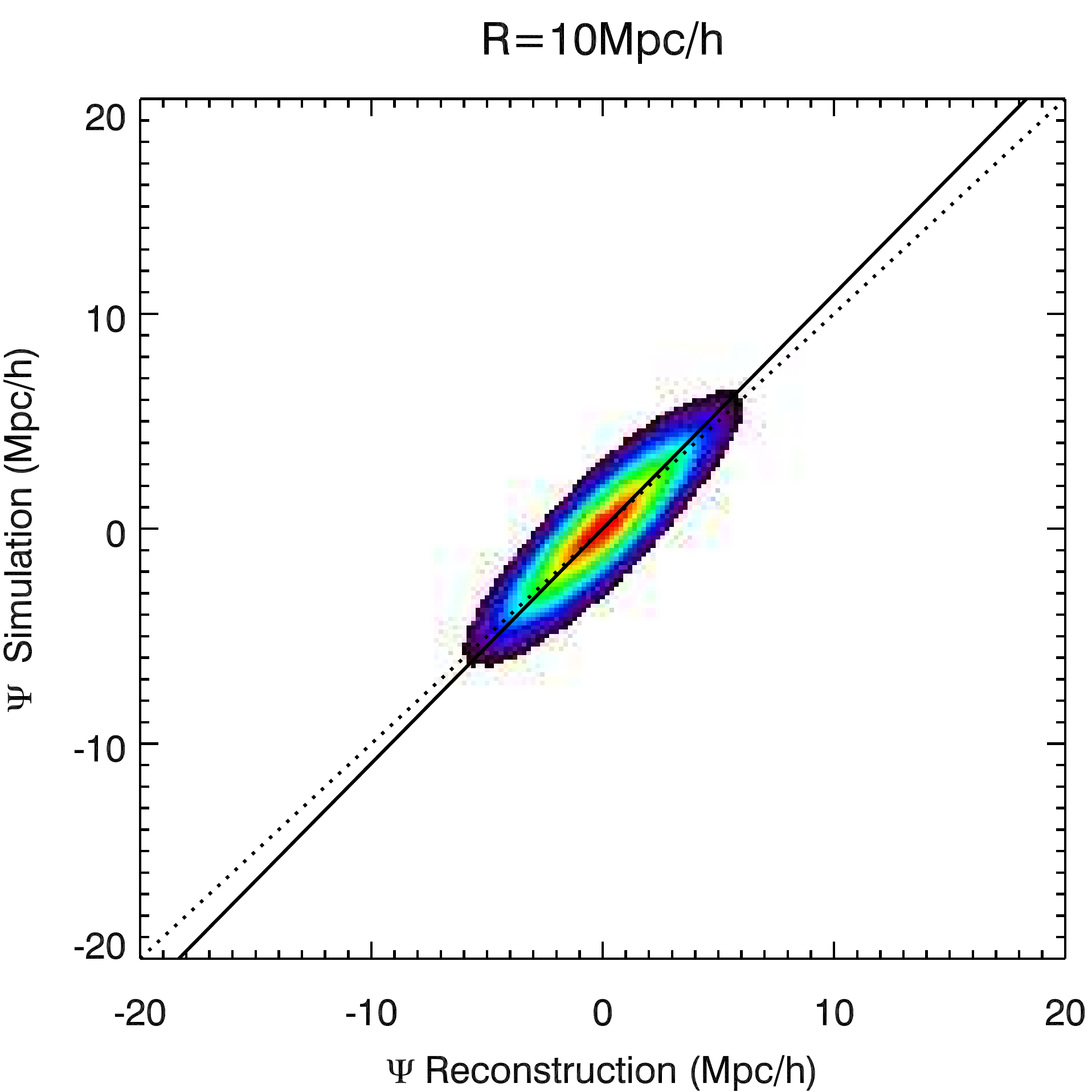}\includegraphics[width=1.1in]{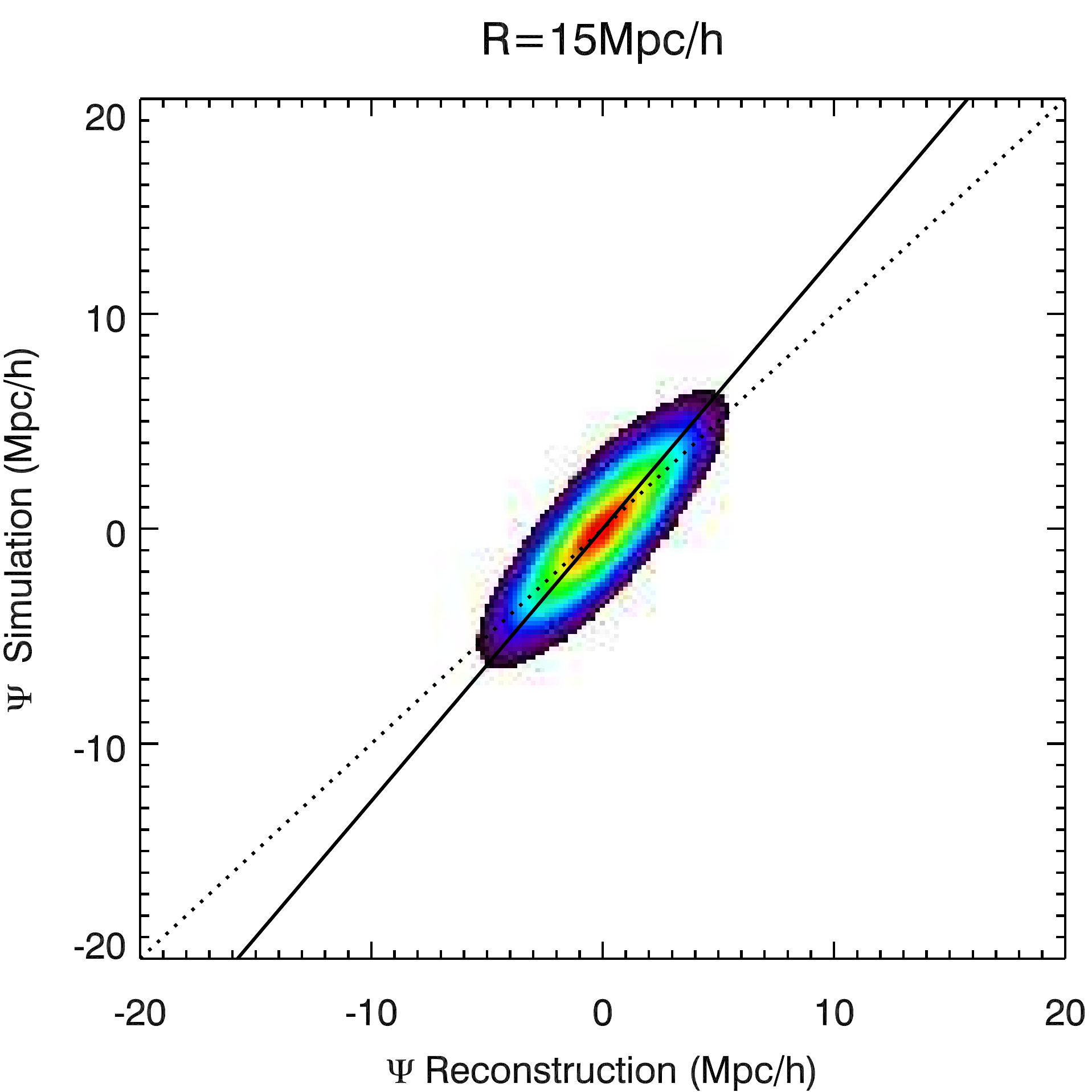}\includegraphics[width=1.1in]{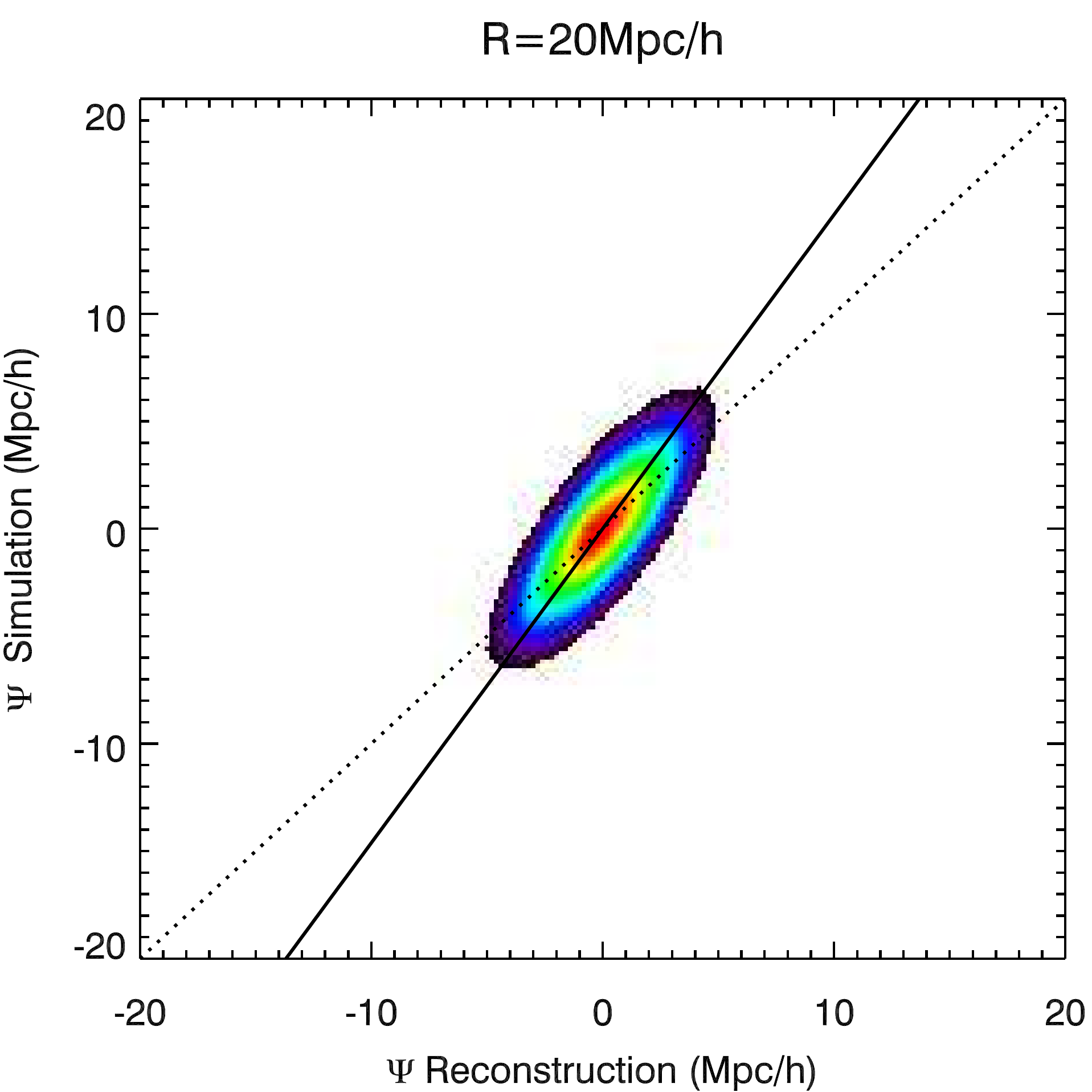}\includegraphics[width=1.1in]{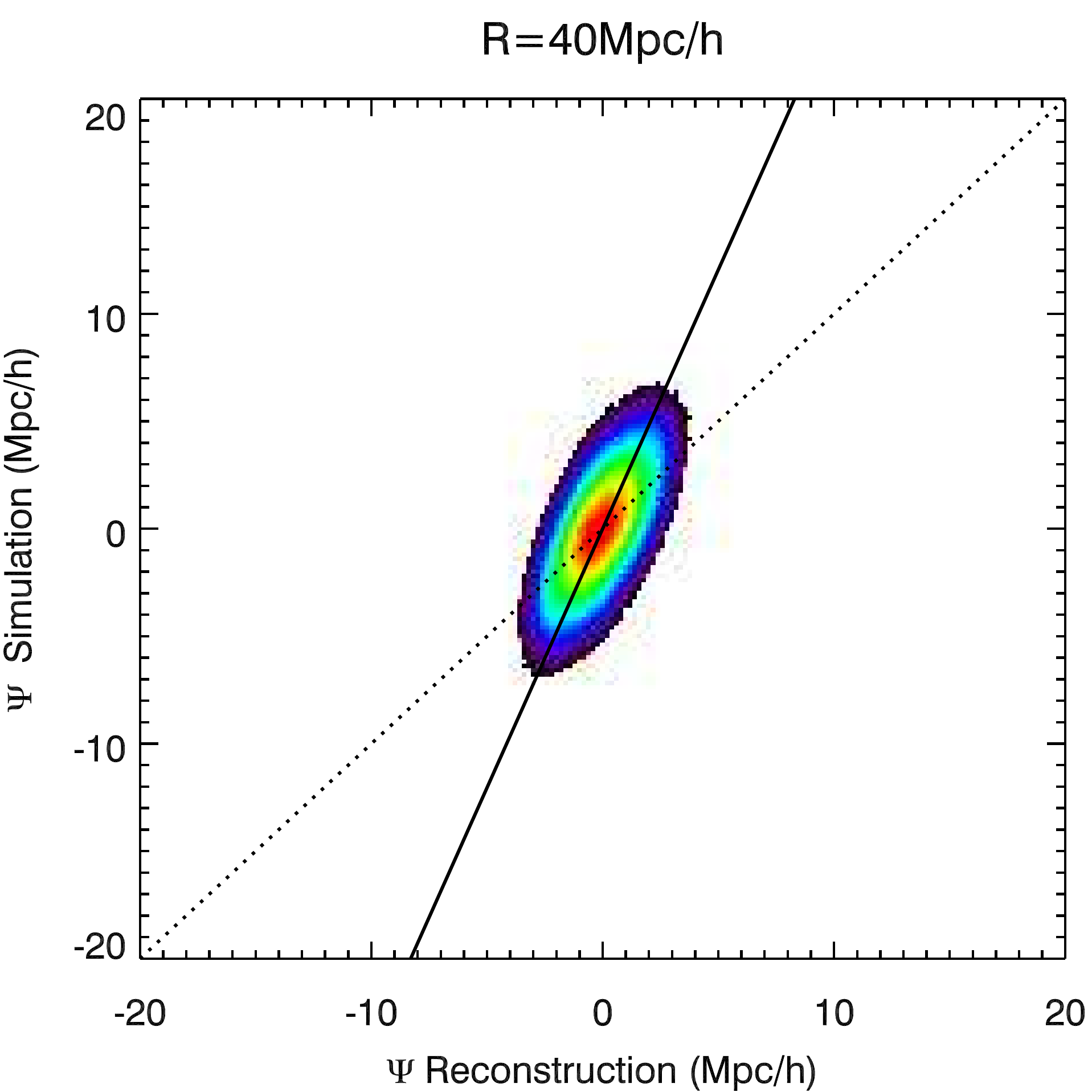}
\includegraphics[width=1.1in]{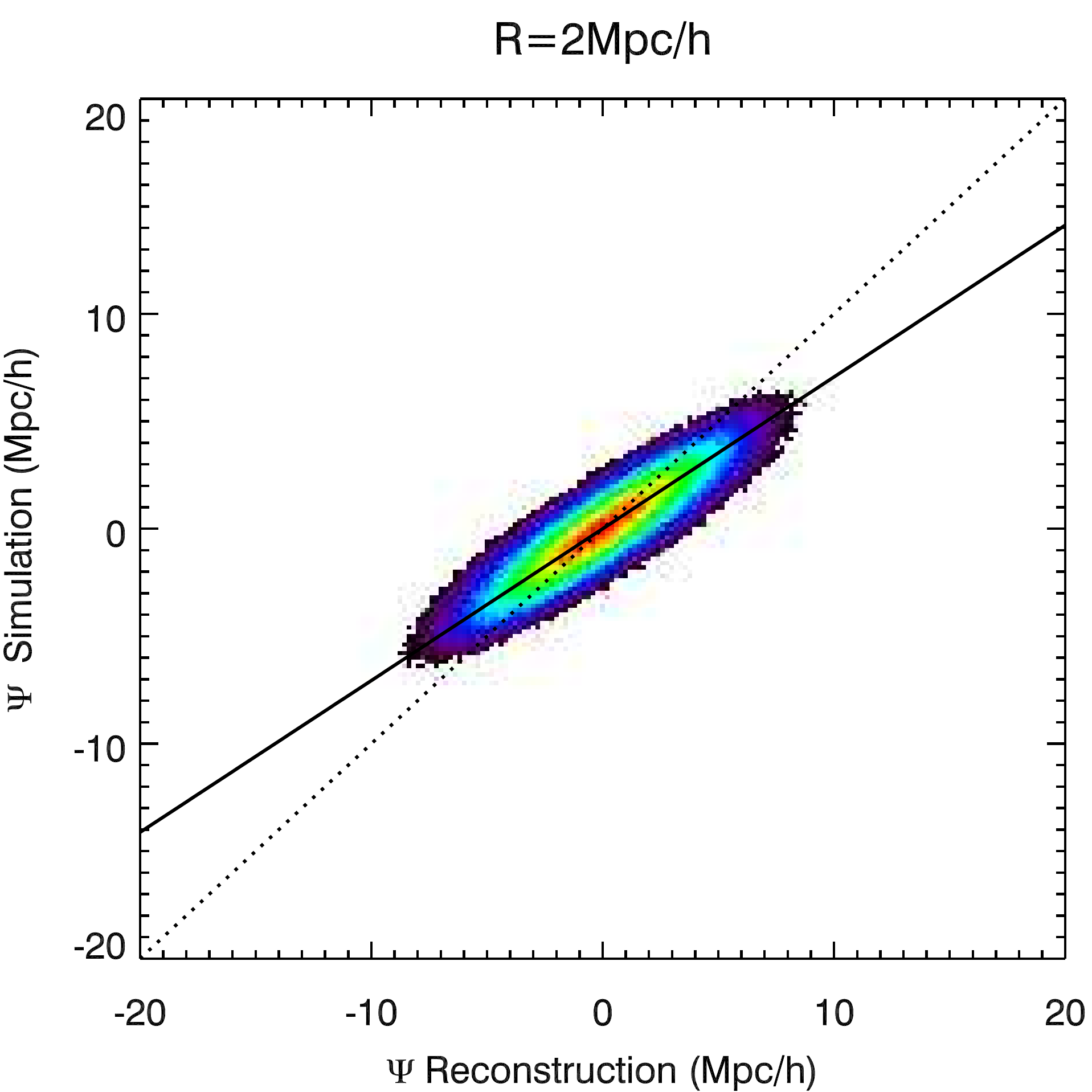}\includegraphics[width=1.1in]{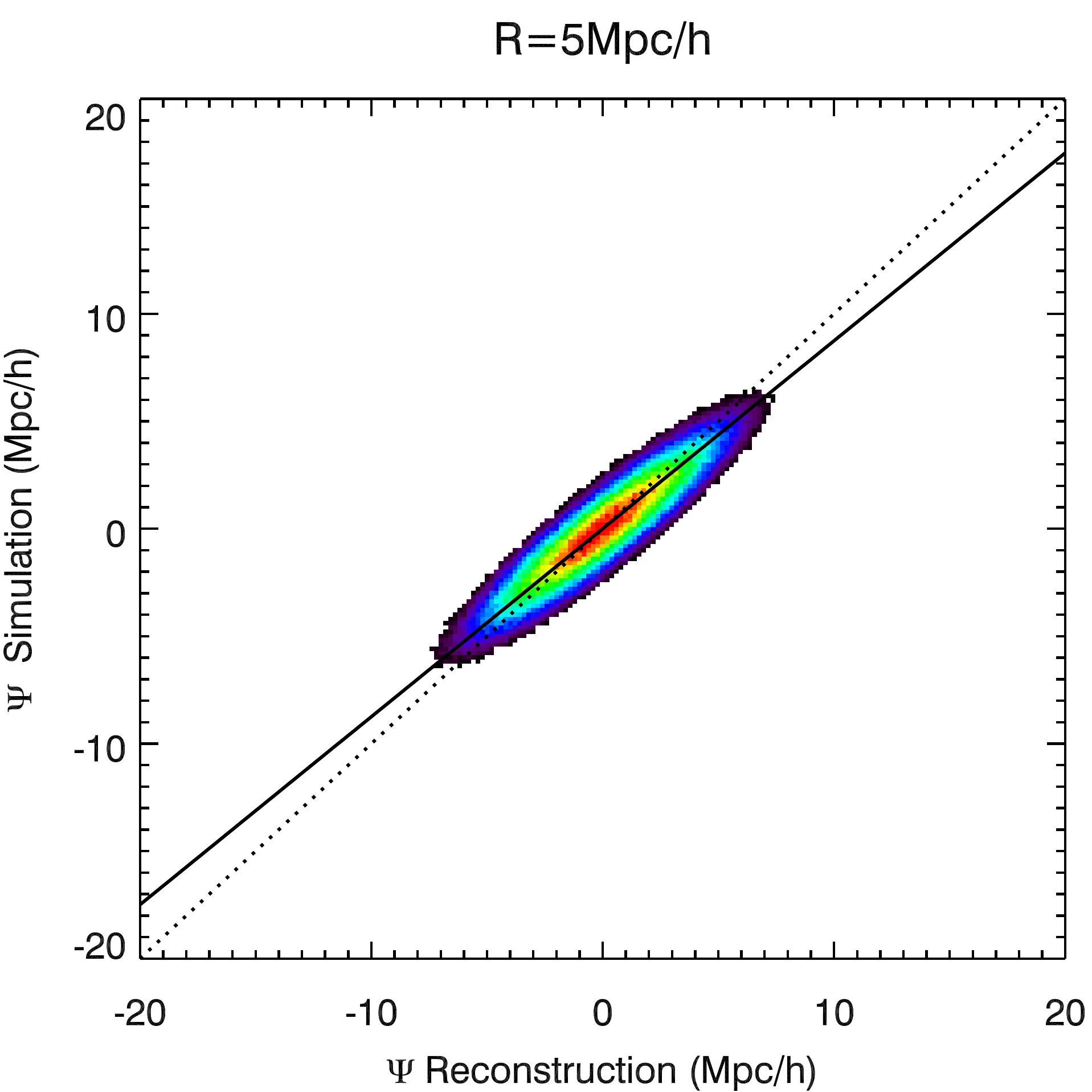}\includegraphics[width=1.1in]{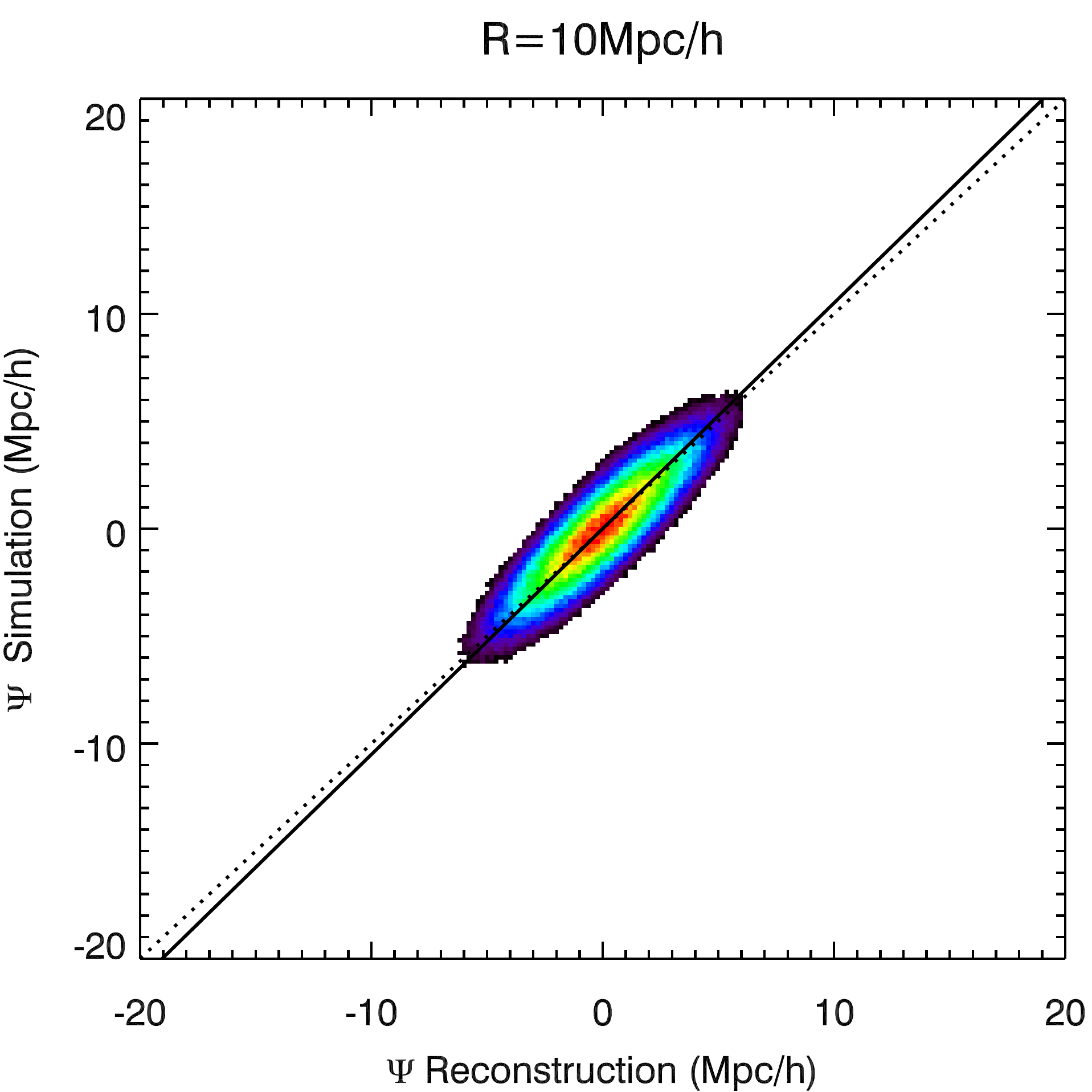}\includegraphics[width=1.1in]{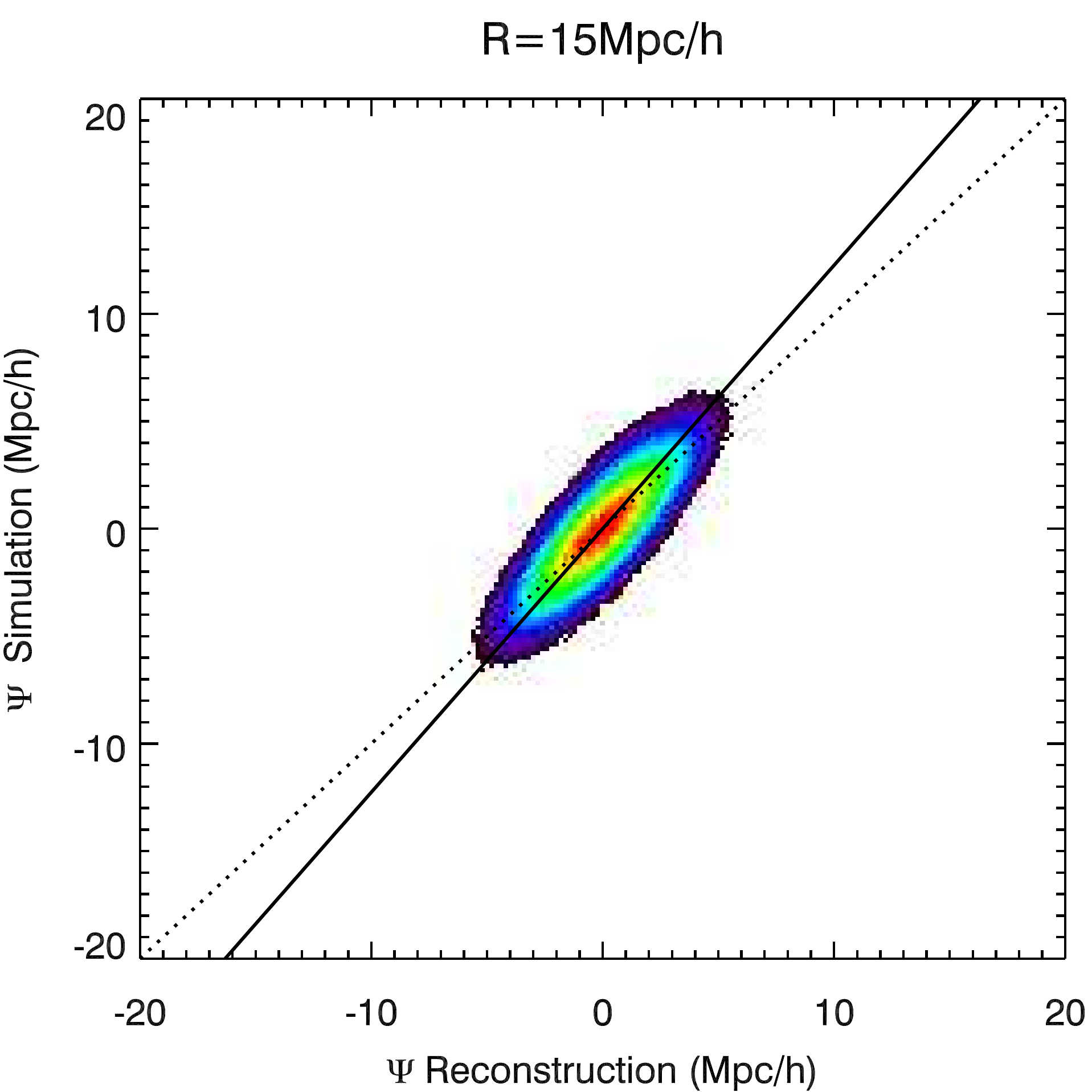}\includegraphics[width=1.1in]{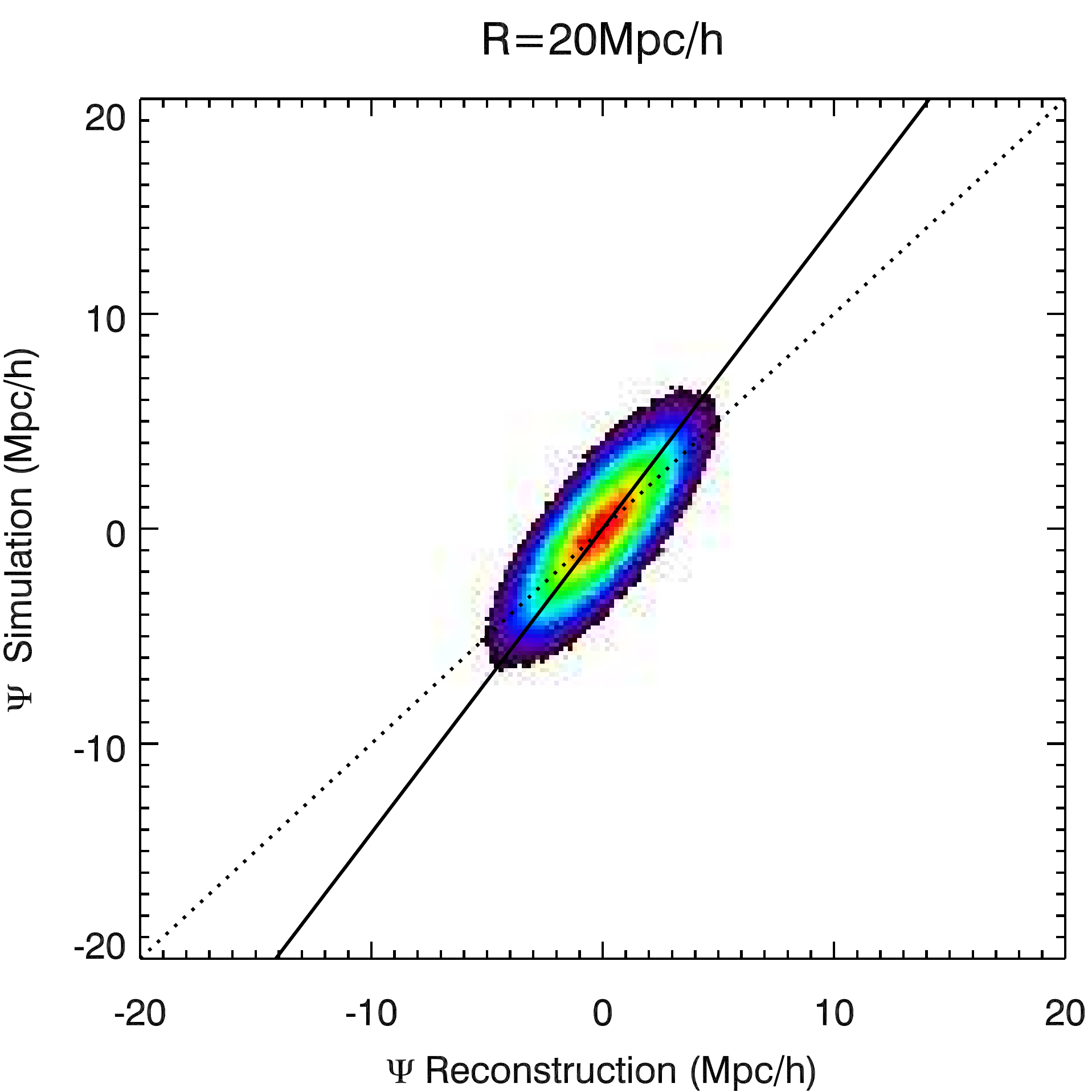}\includegraphics[width=1.1in]{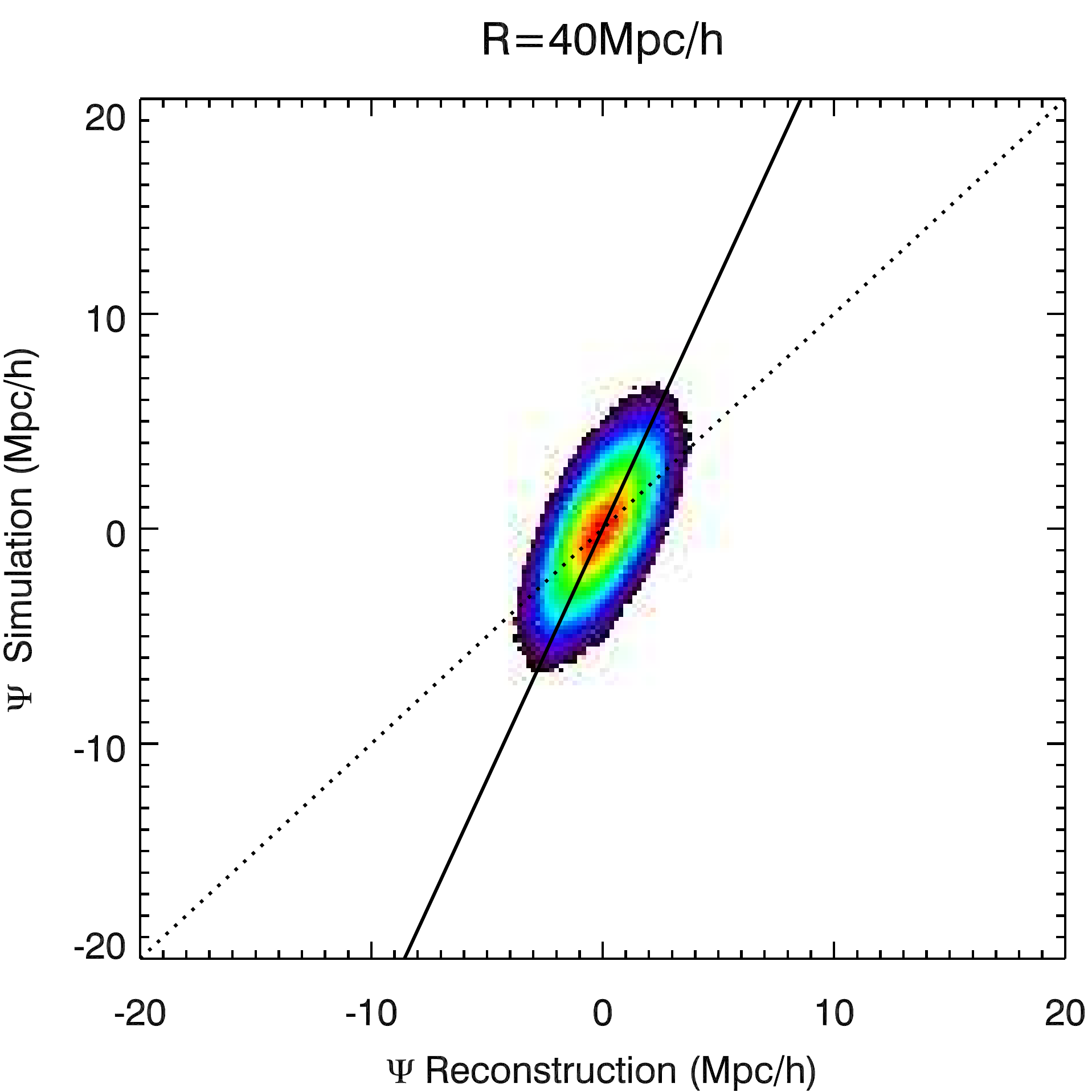}
\includegraphics[width=1.1in]{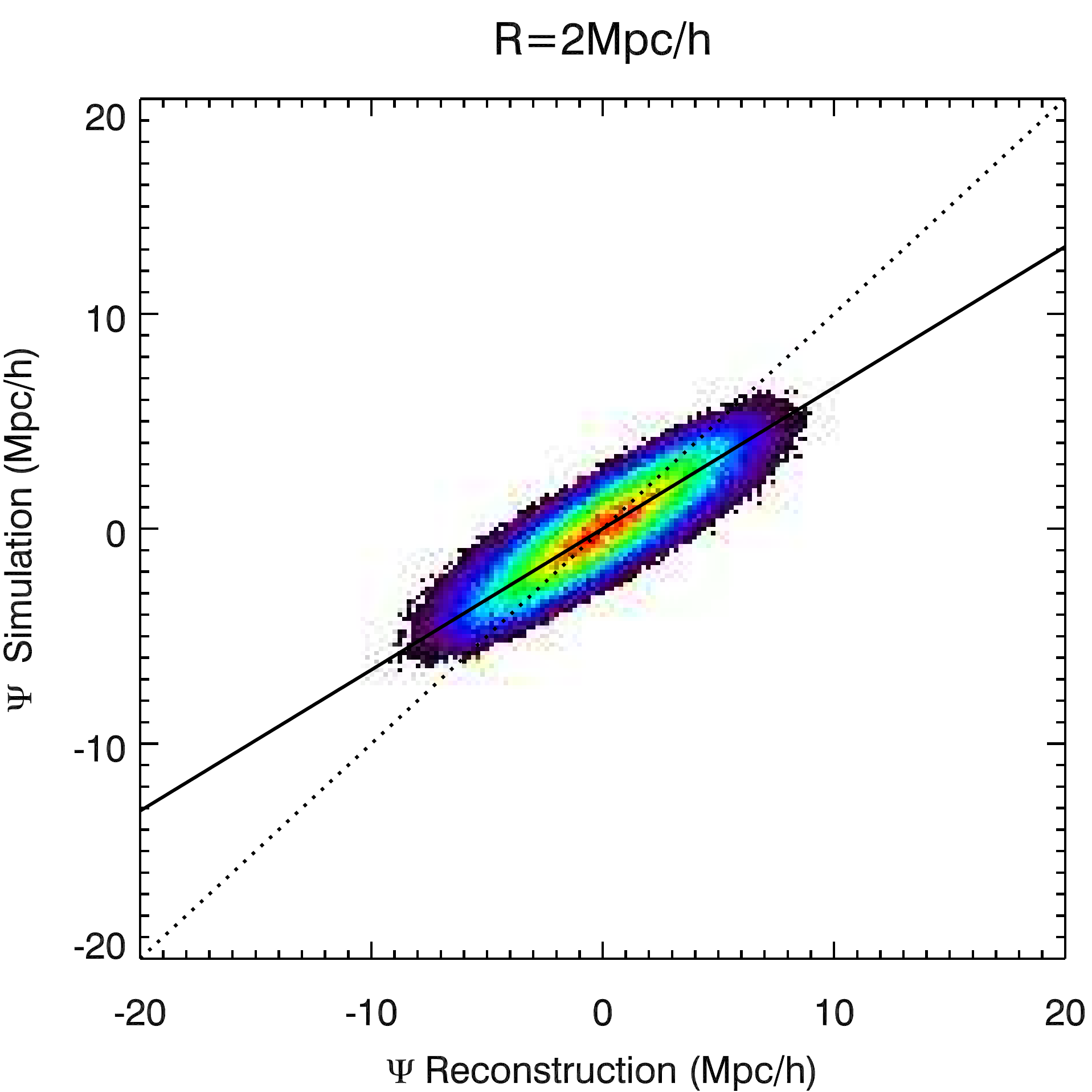}\includegraphics[width=1.1in]{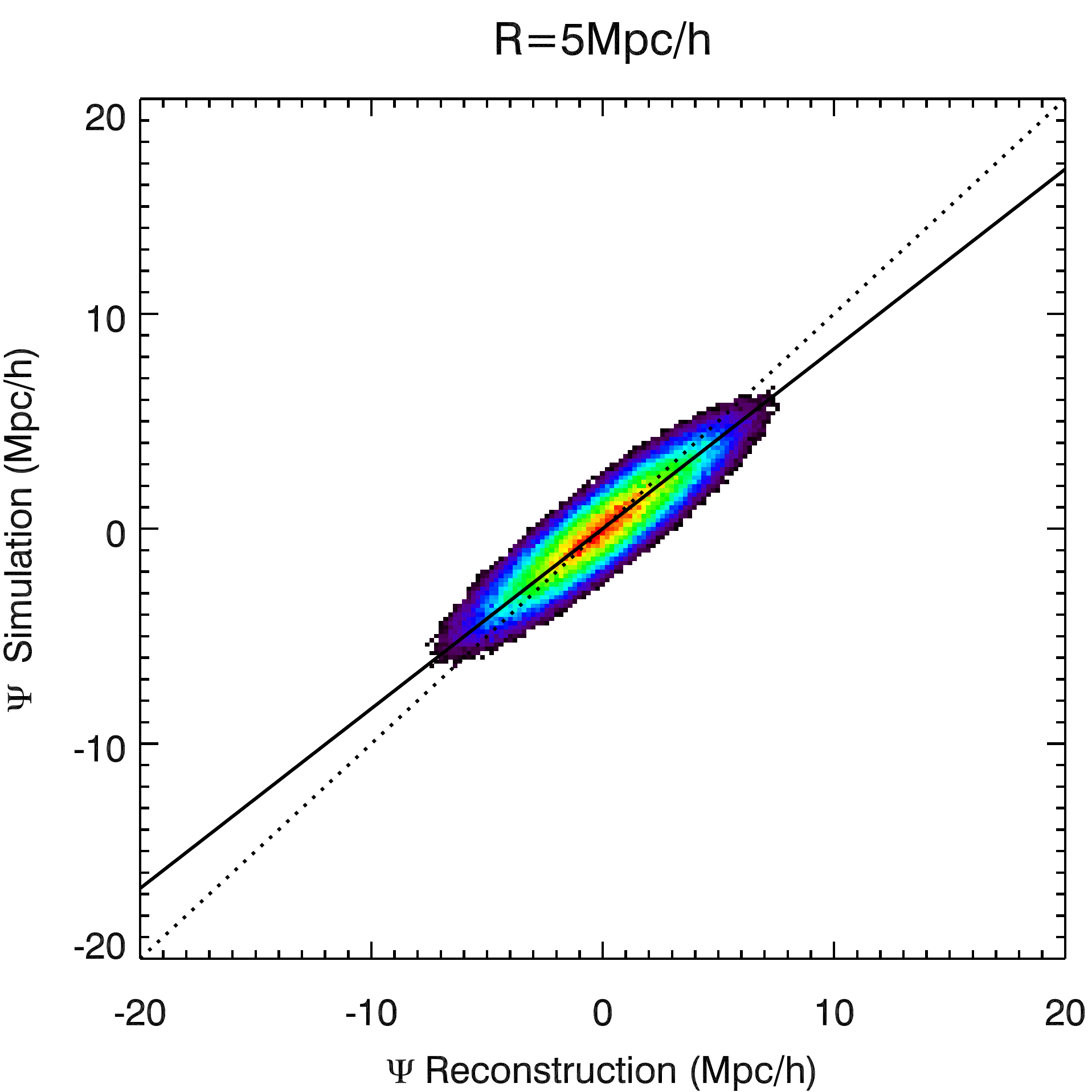}\includegraphics[width=1.1in]{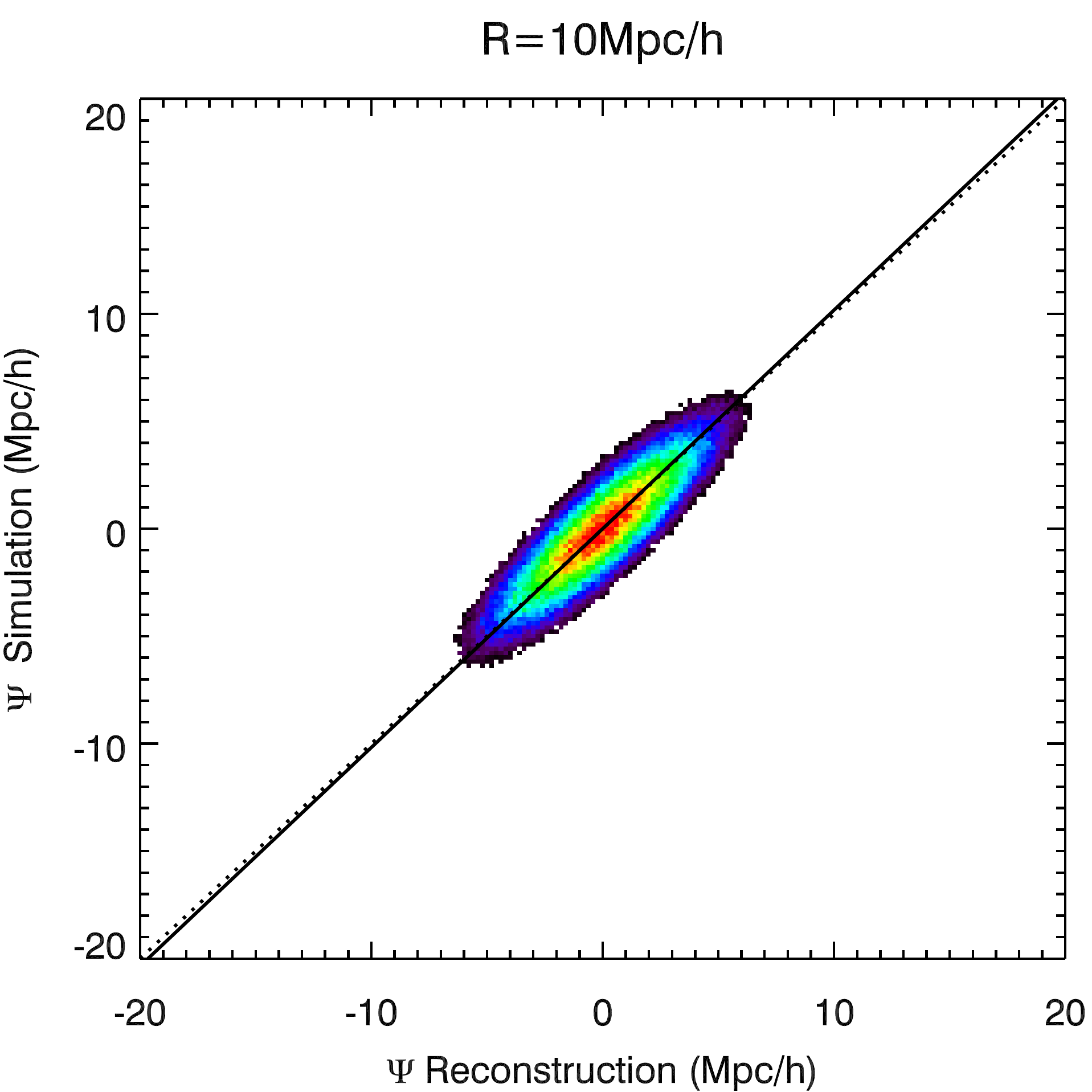}\includegraphics[width=1.1in]{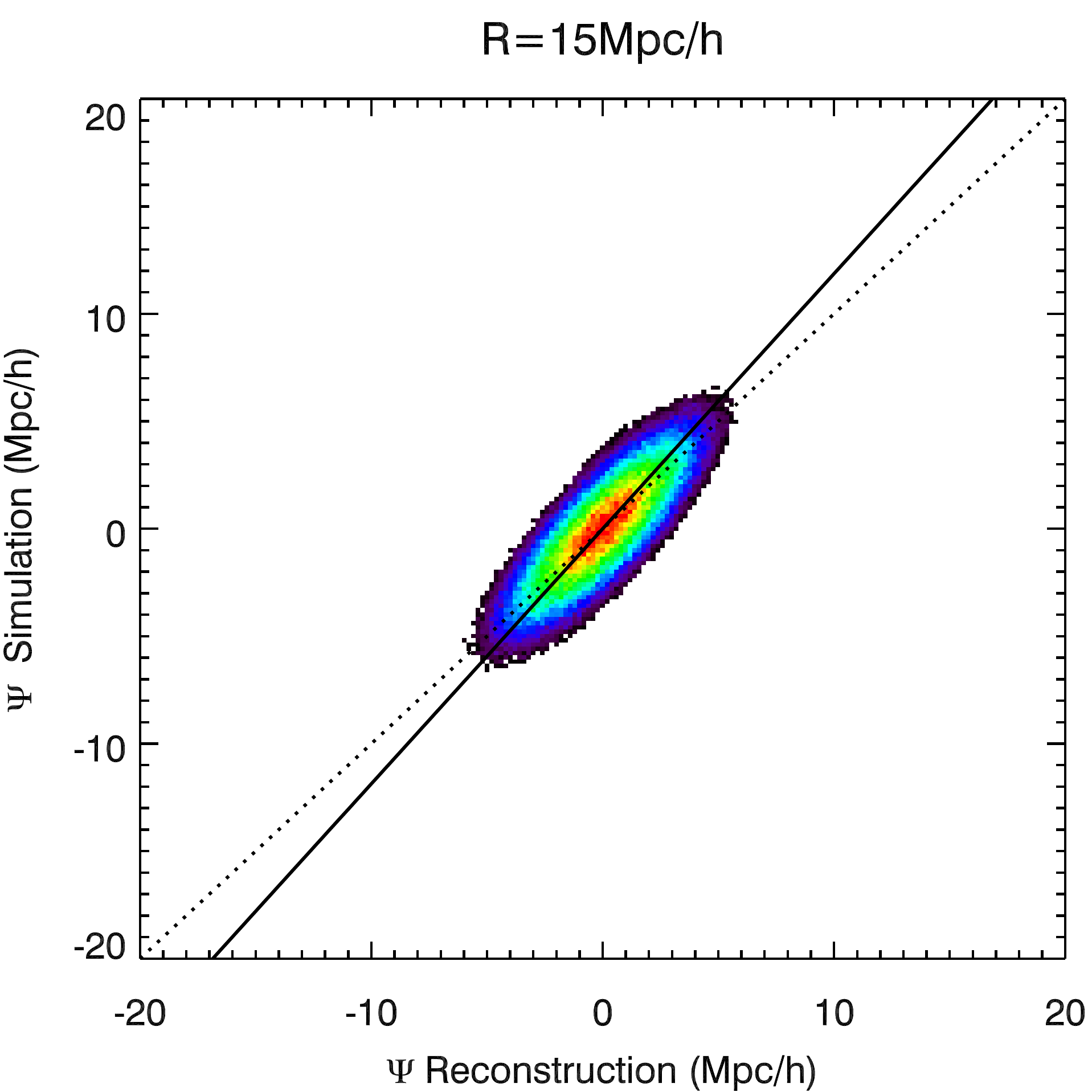}\includegraphics[width=1.1in]{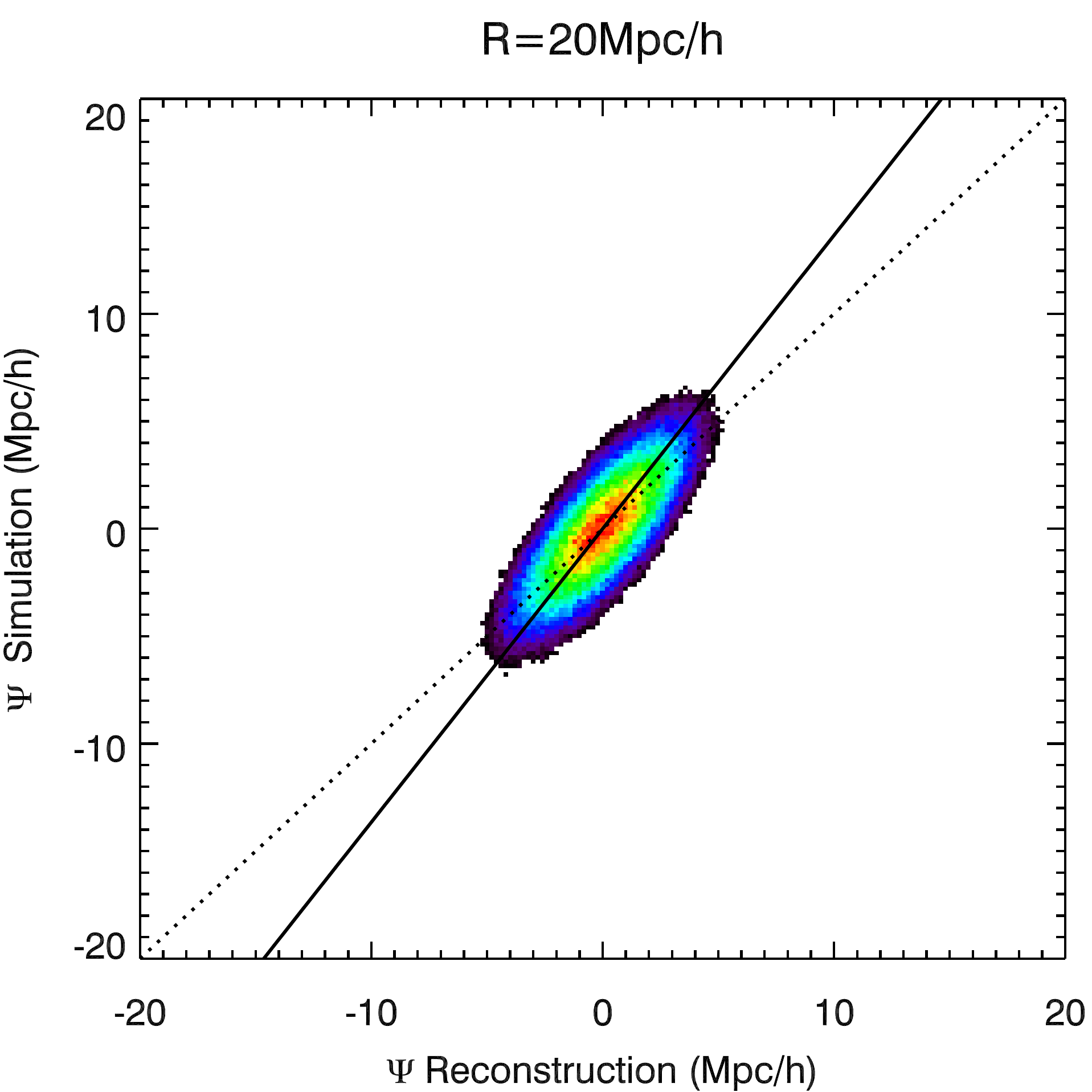}\includegraphics[width=1.1in]{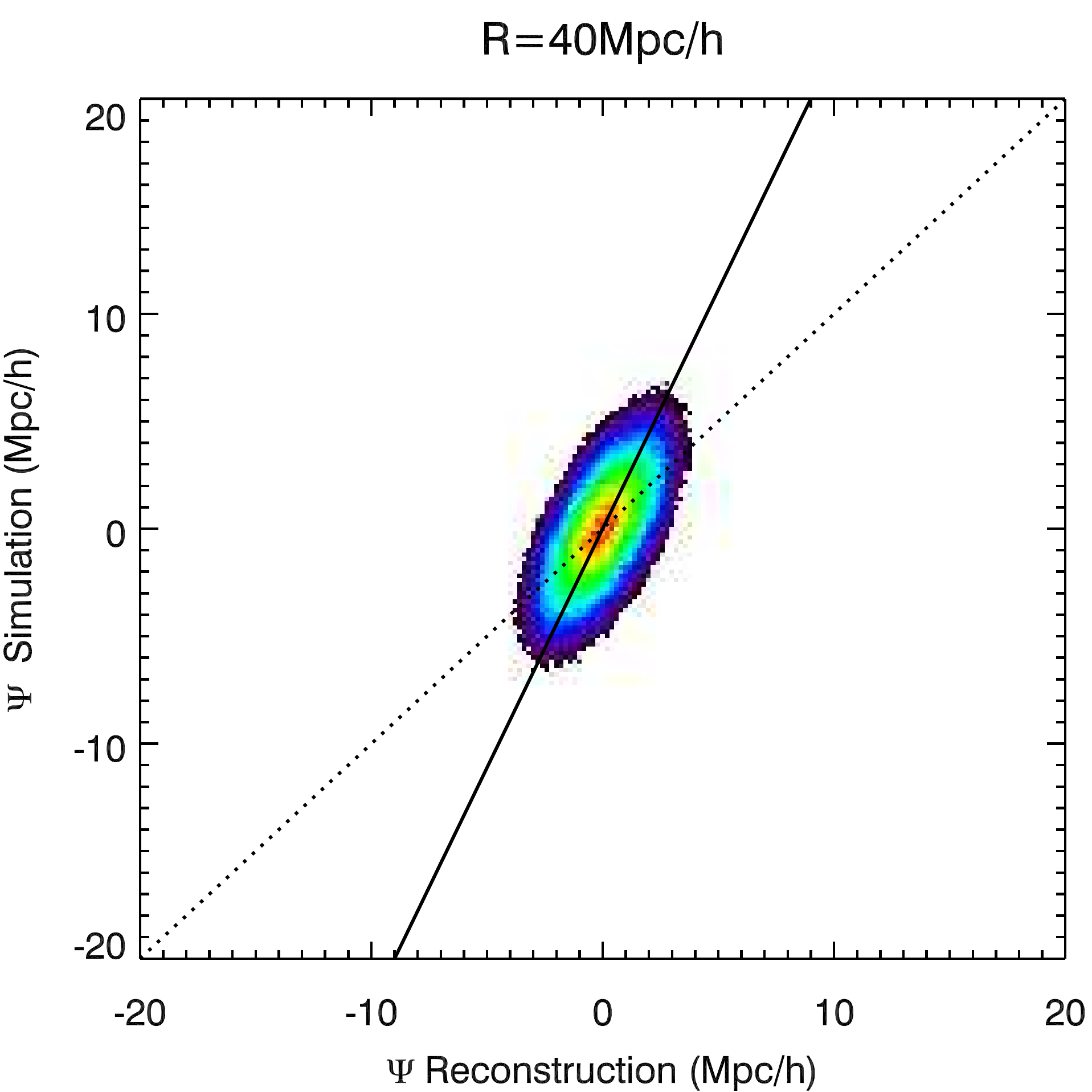}
\includegraphics[width=1.1in]{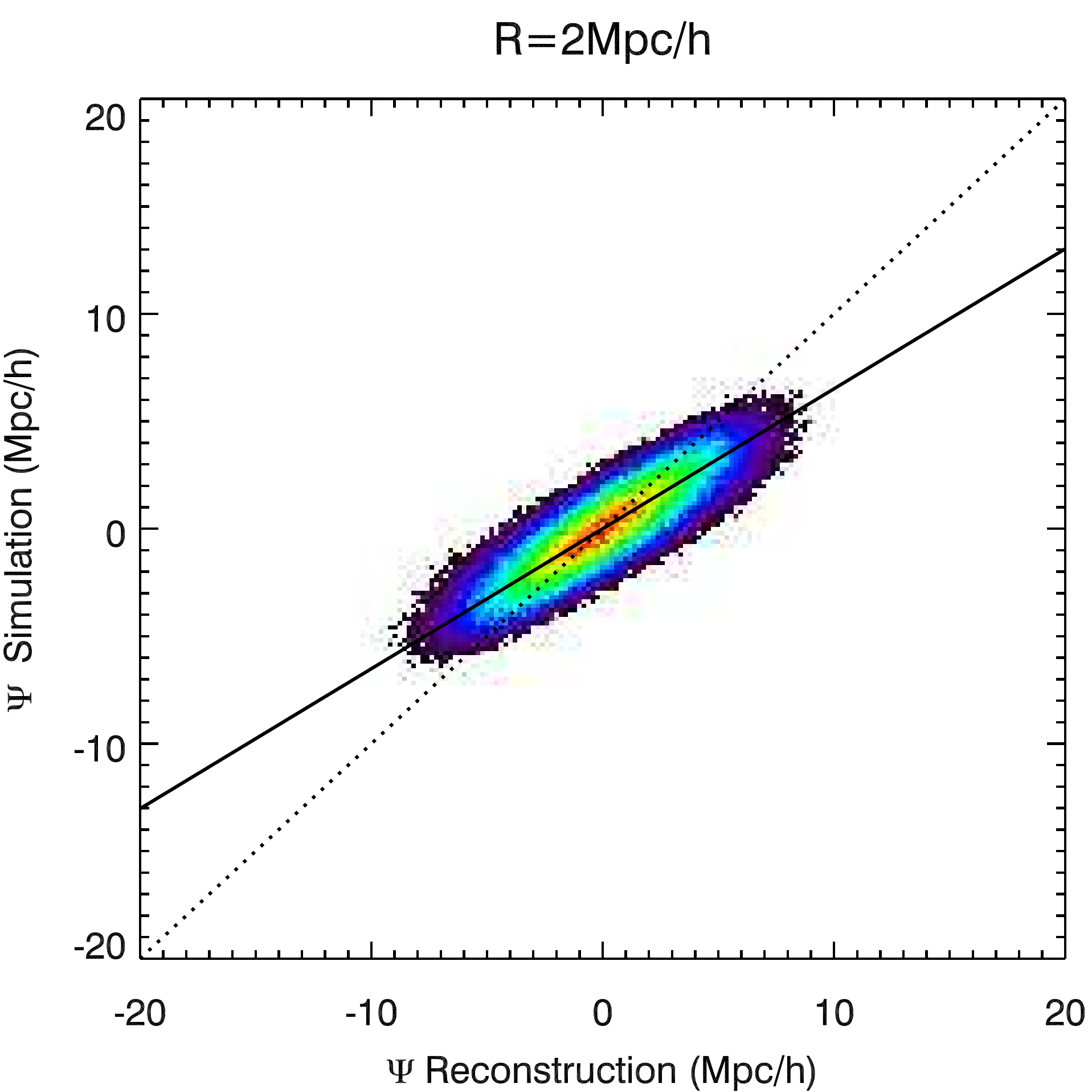}\includegraphics[width=1.1in]{Paper_Gauss2Dfit_R=5_mcut_1_0e13.pdf}\includegraphics[width=1.1in]{Paper_Gauss2Dfit_R=10_mcut_1_0e13.pdf}\includegraphics[width=1.1in]{Paper_Gauss2Dfit_R=15_mcut_1_0e13.pdf}\includegraphics[width=1.1in]{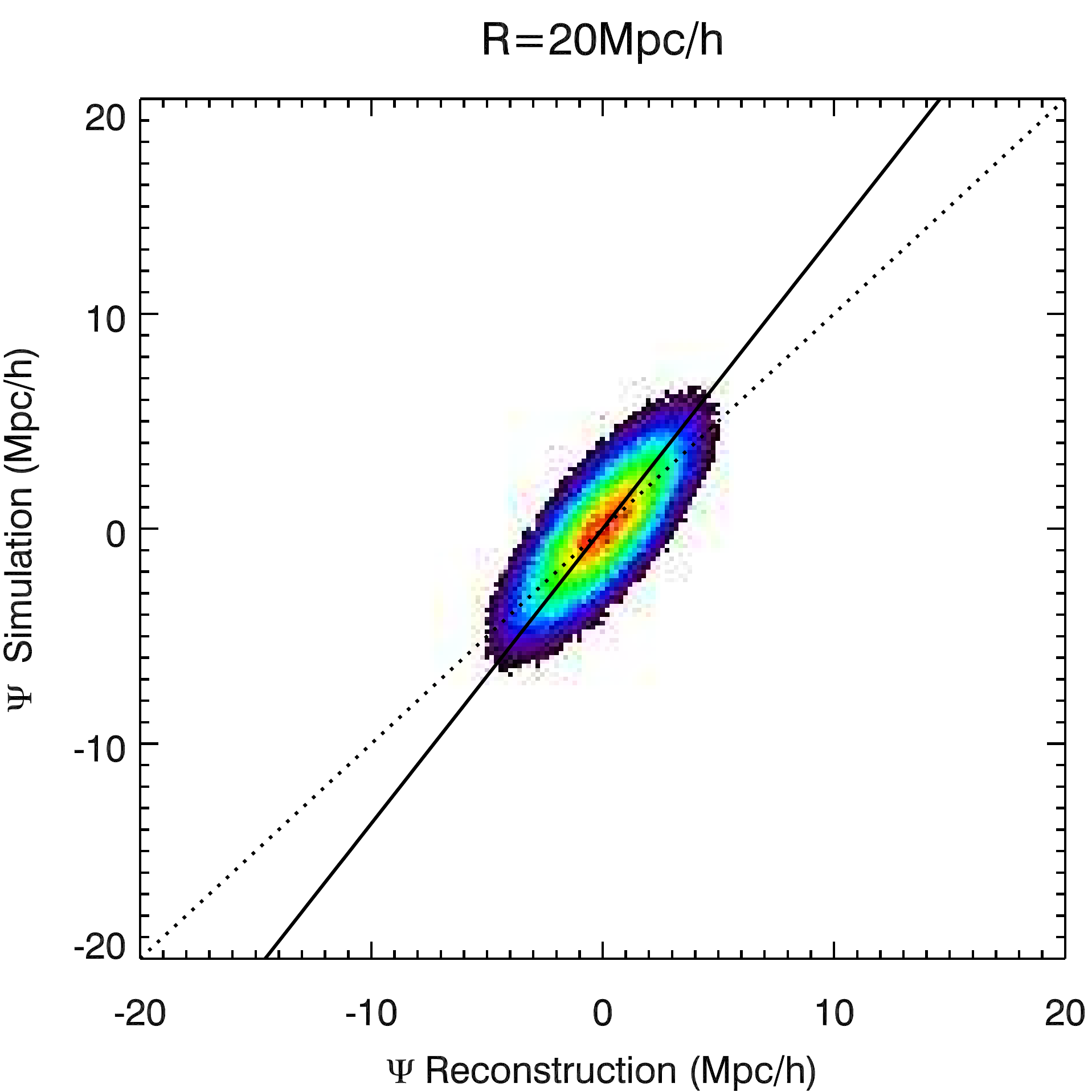}\includegraphics[width=1.1in]{Paper_Gauss2Dfit_R=40_mcut_1_0e13.pdf}
\includegraphics[width=1.1in]{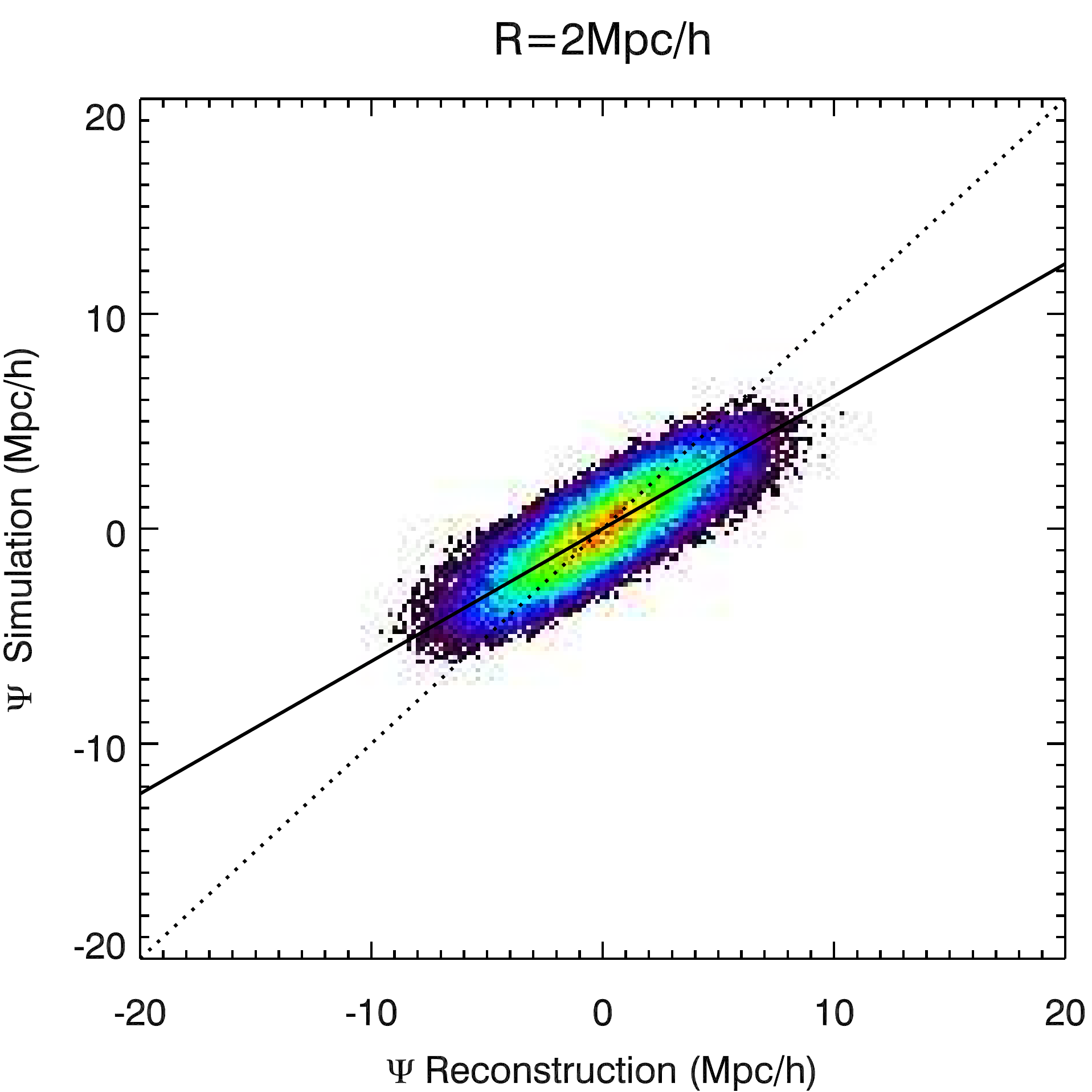}\includegraphics[width=1.1in]{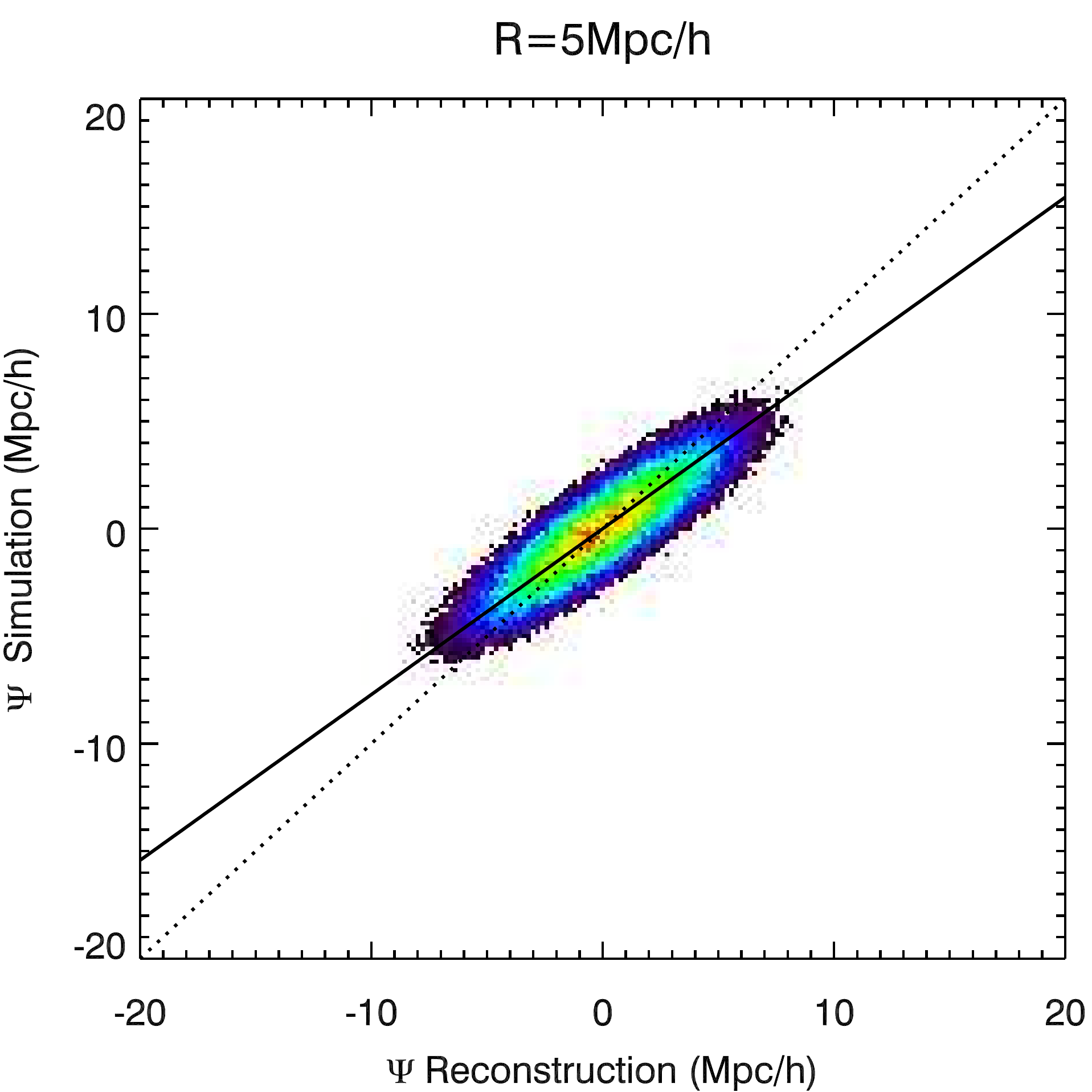}\includegraphics[width=1.1in]{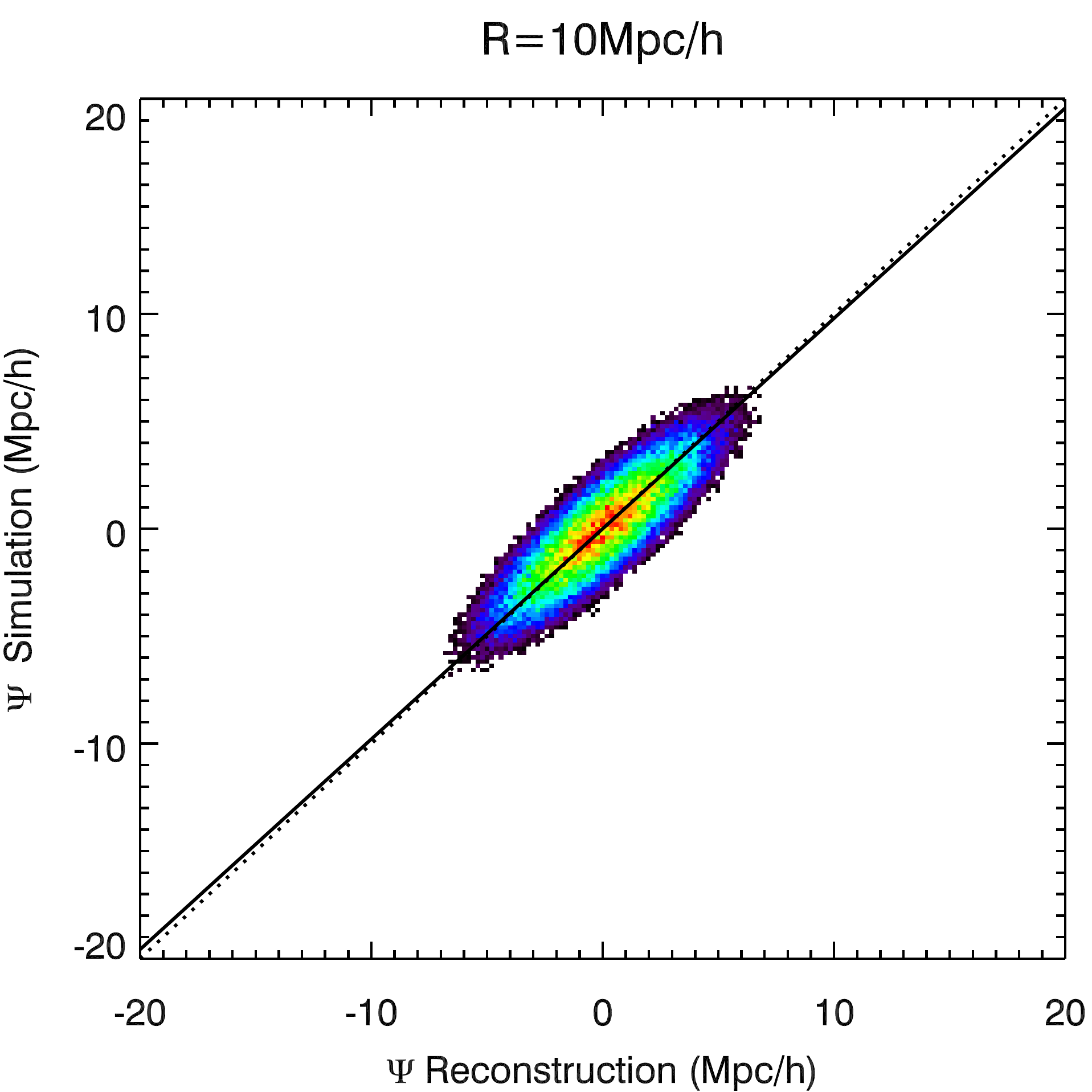}\includegraphics[width=1.1in]{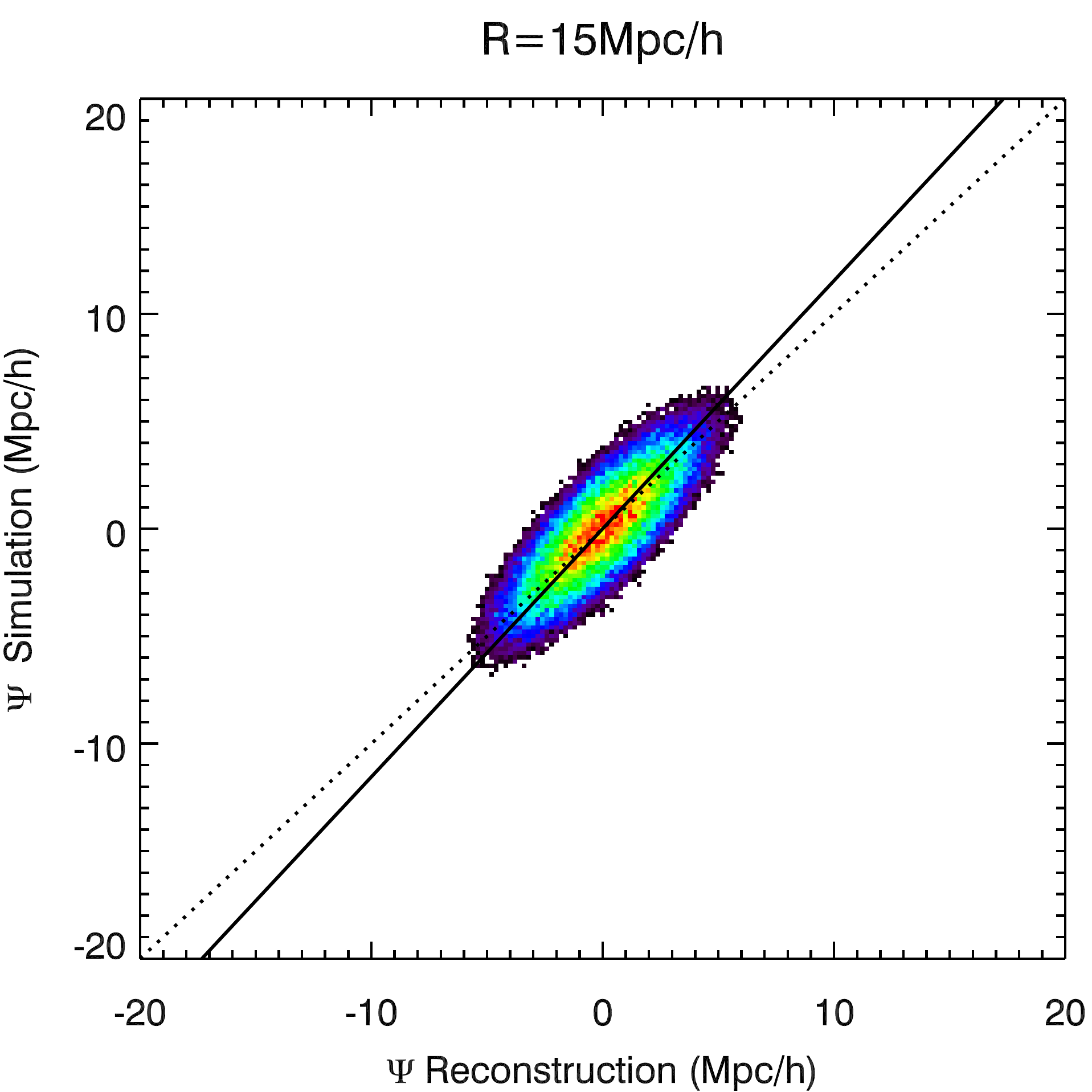}\includegraphics[width=1.1in]{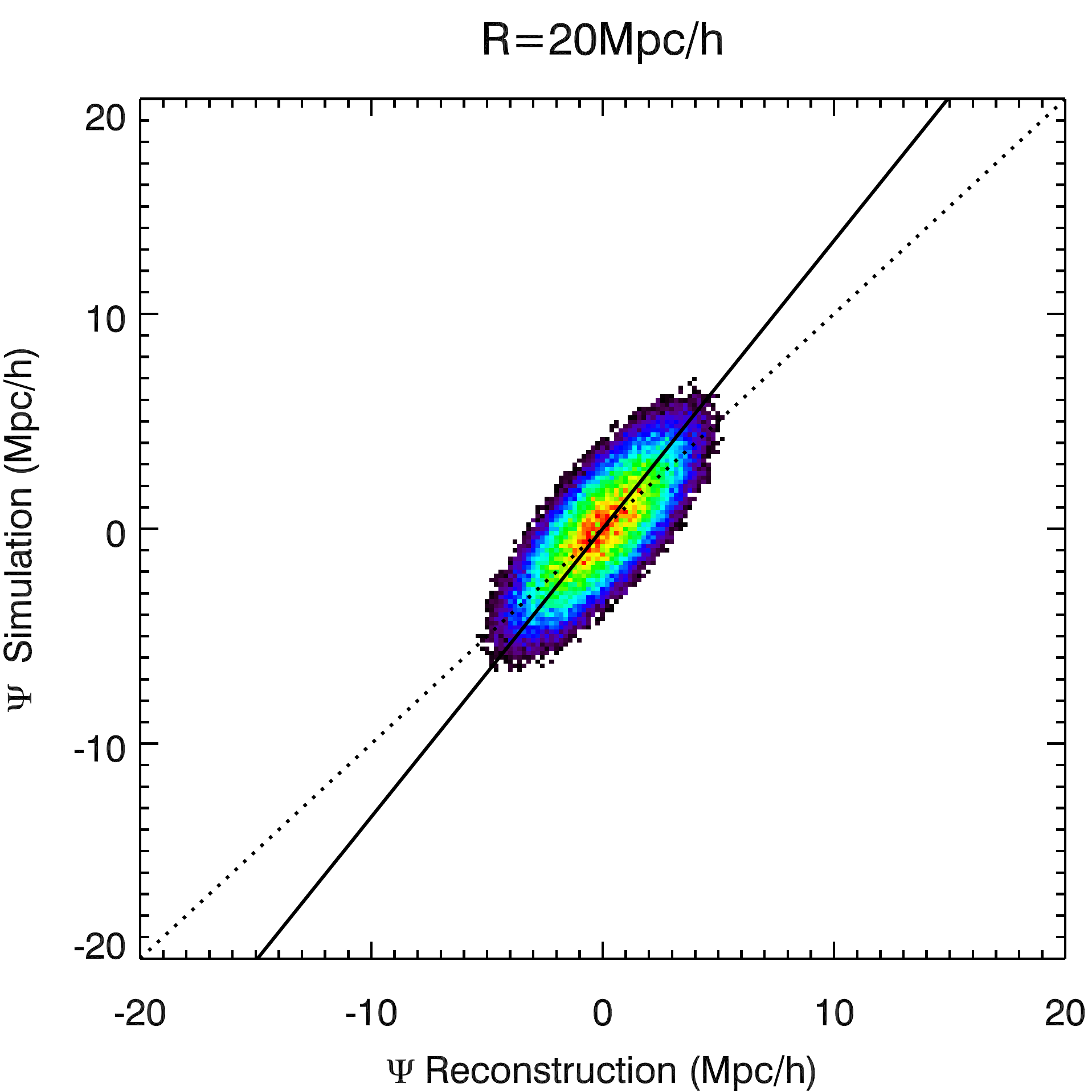}\includegraphics[width=1.1in]{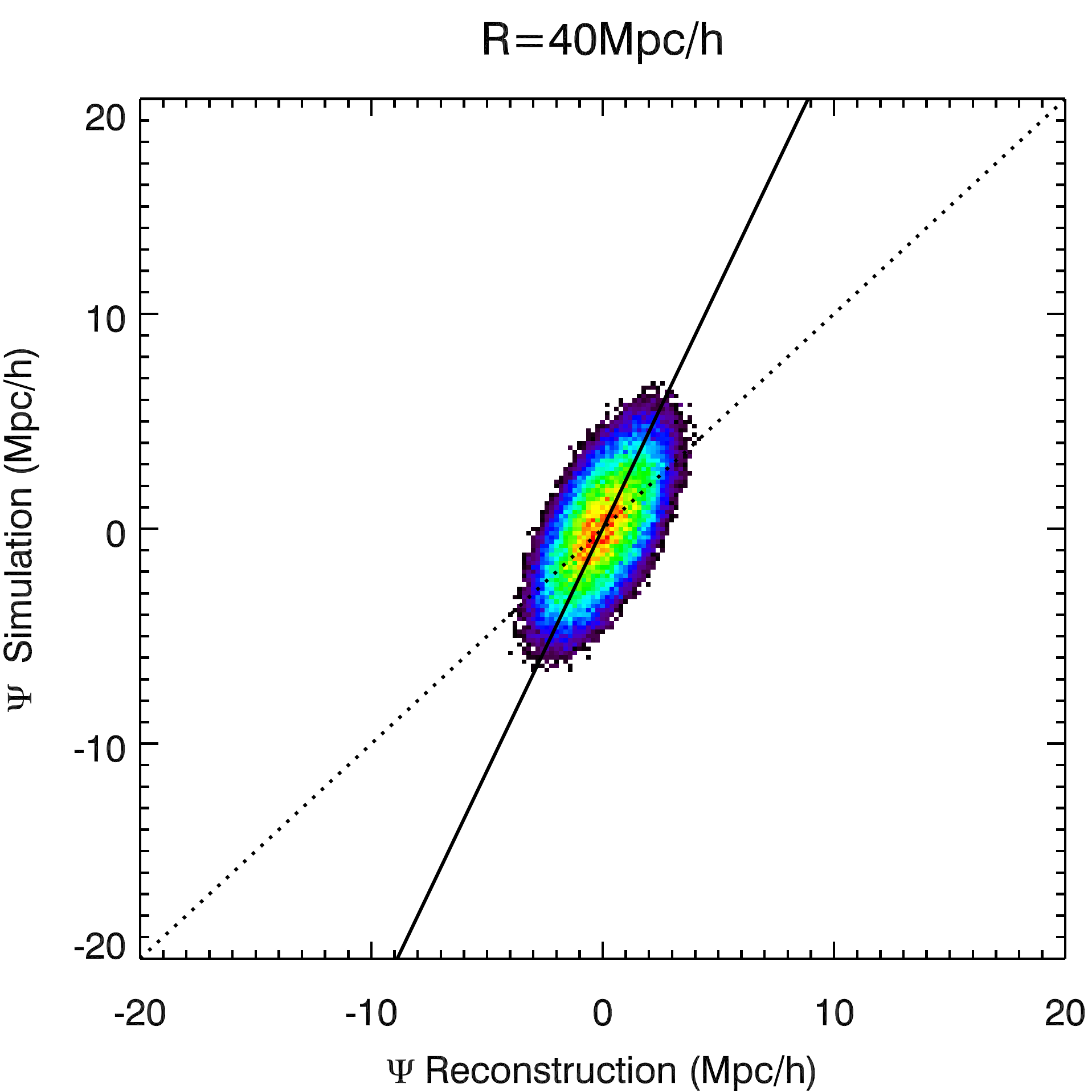}
\includegraphics[width=1.1in]{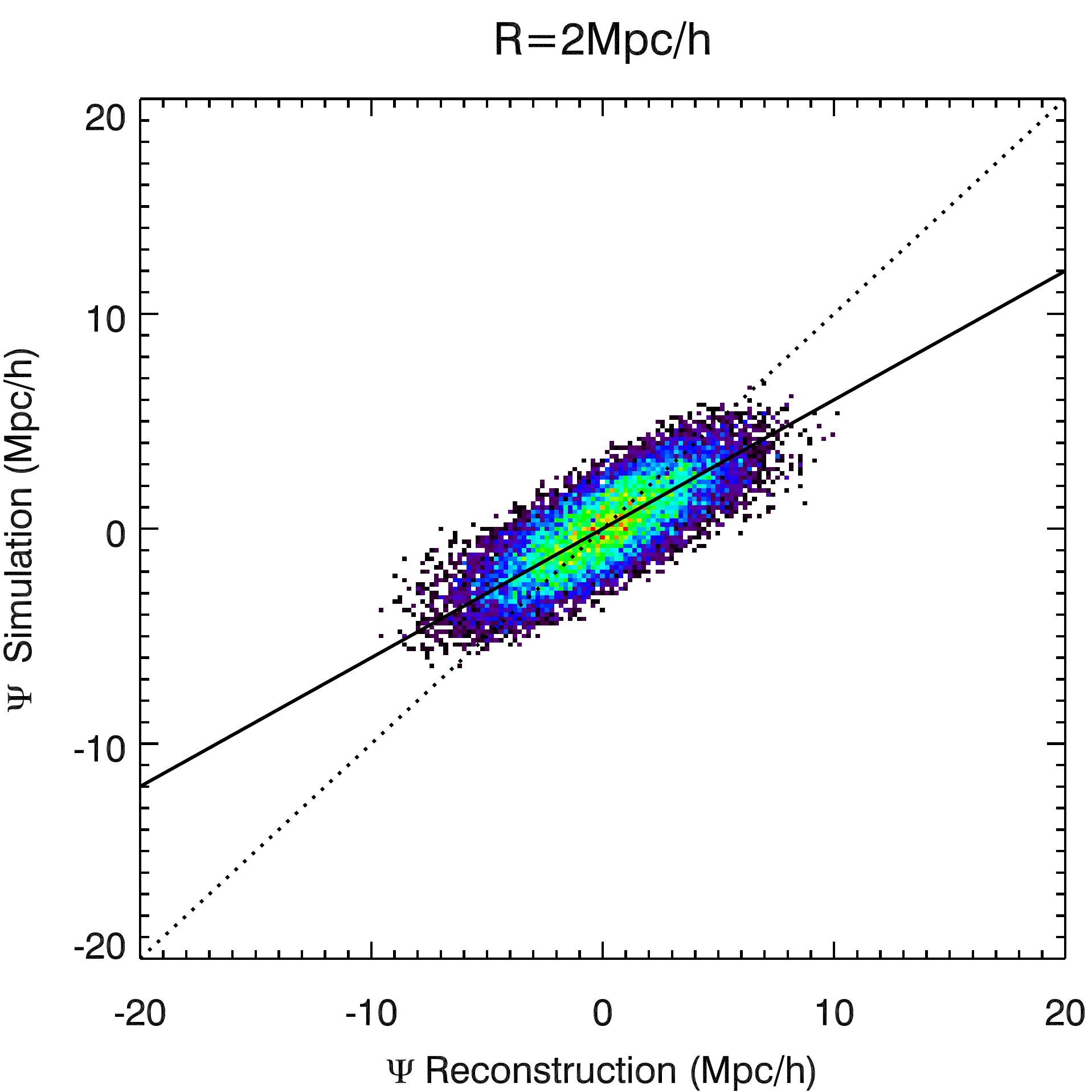}\includegraphics[width=1.1in]{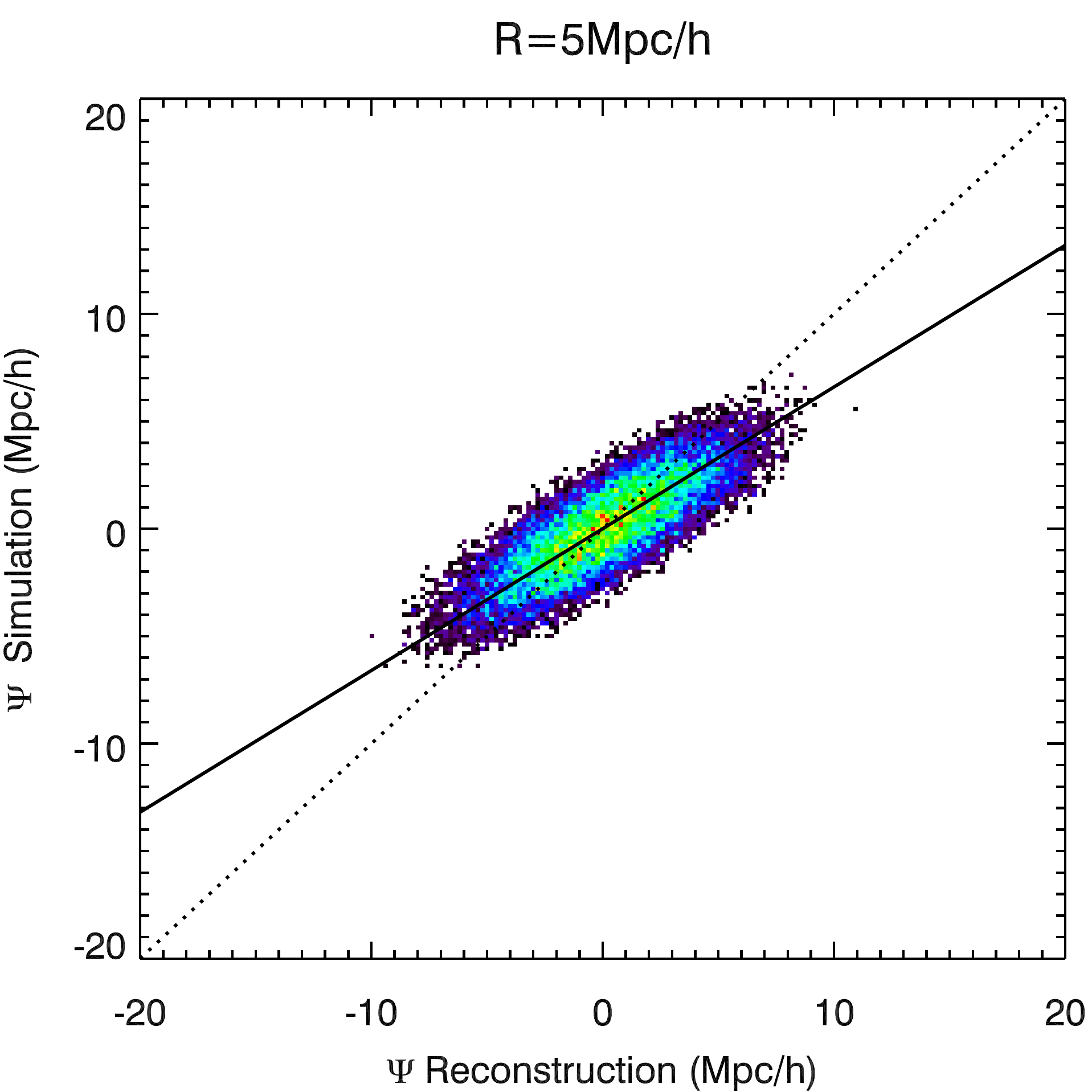}\includegraphics[width=1.1in]{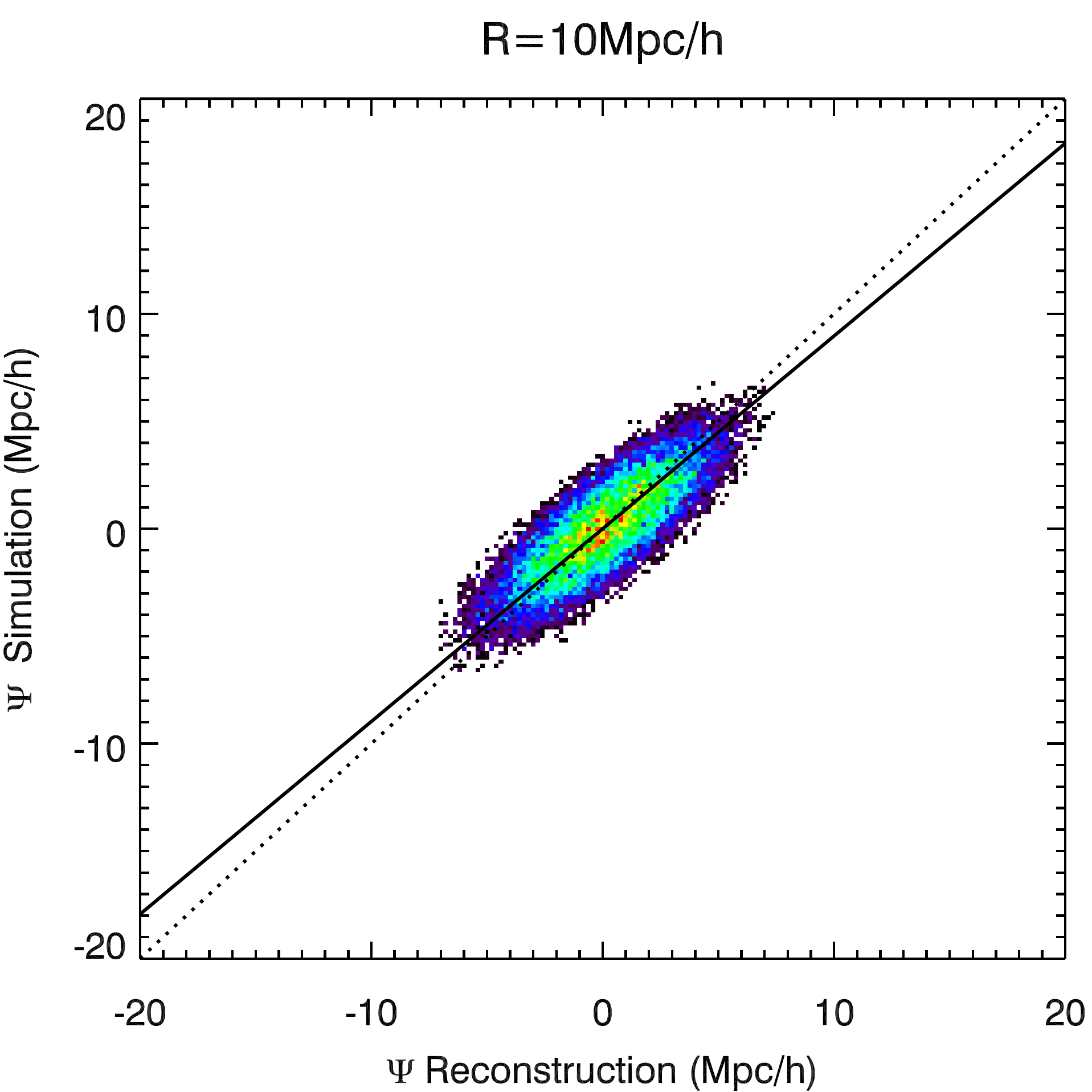}\includegraphics[width=1.1in]{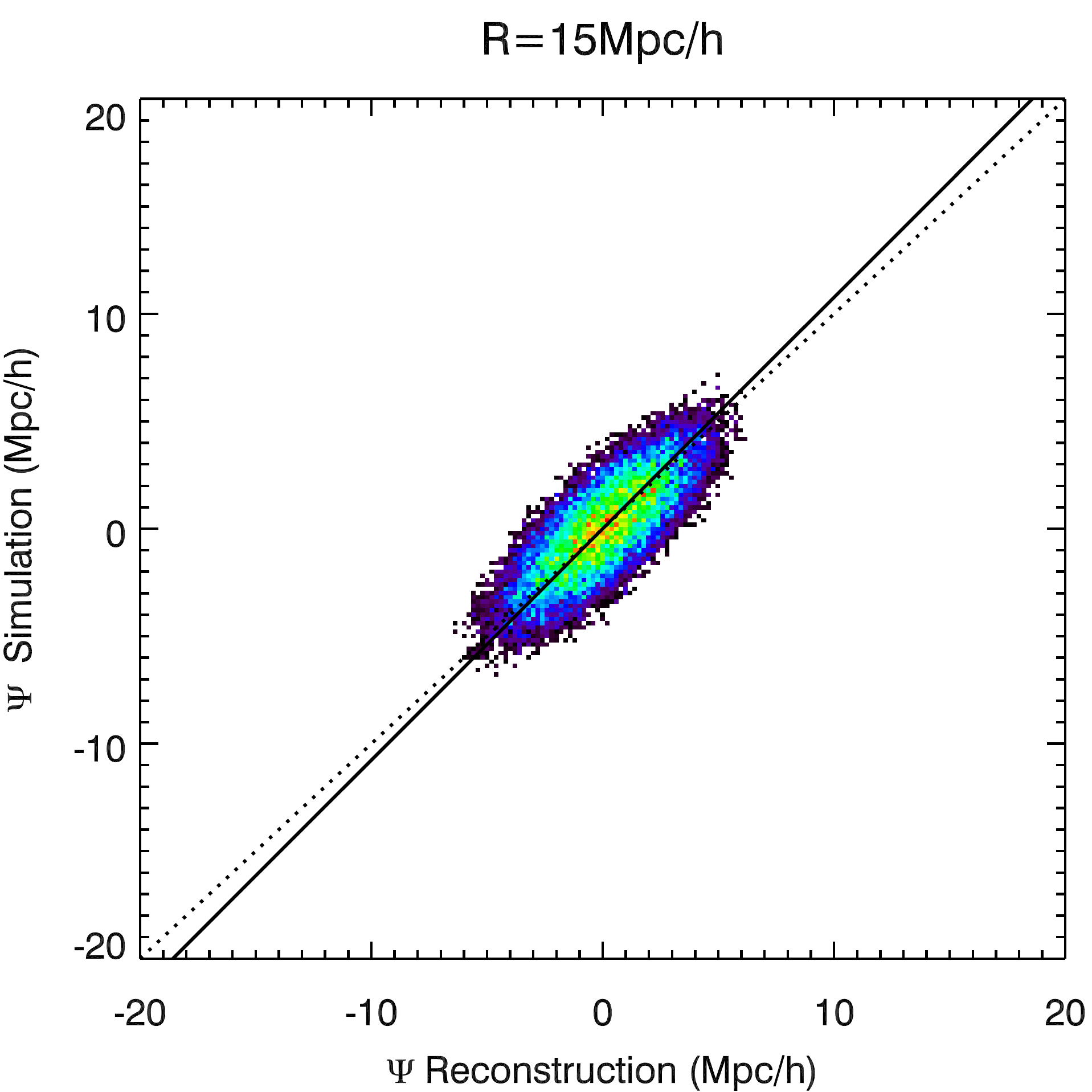}\includegraphics[width=1.1in]{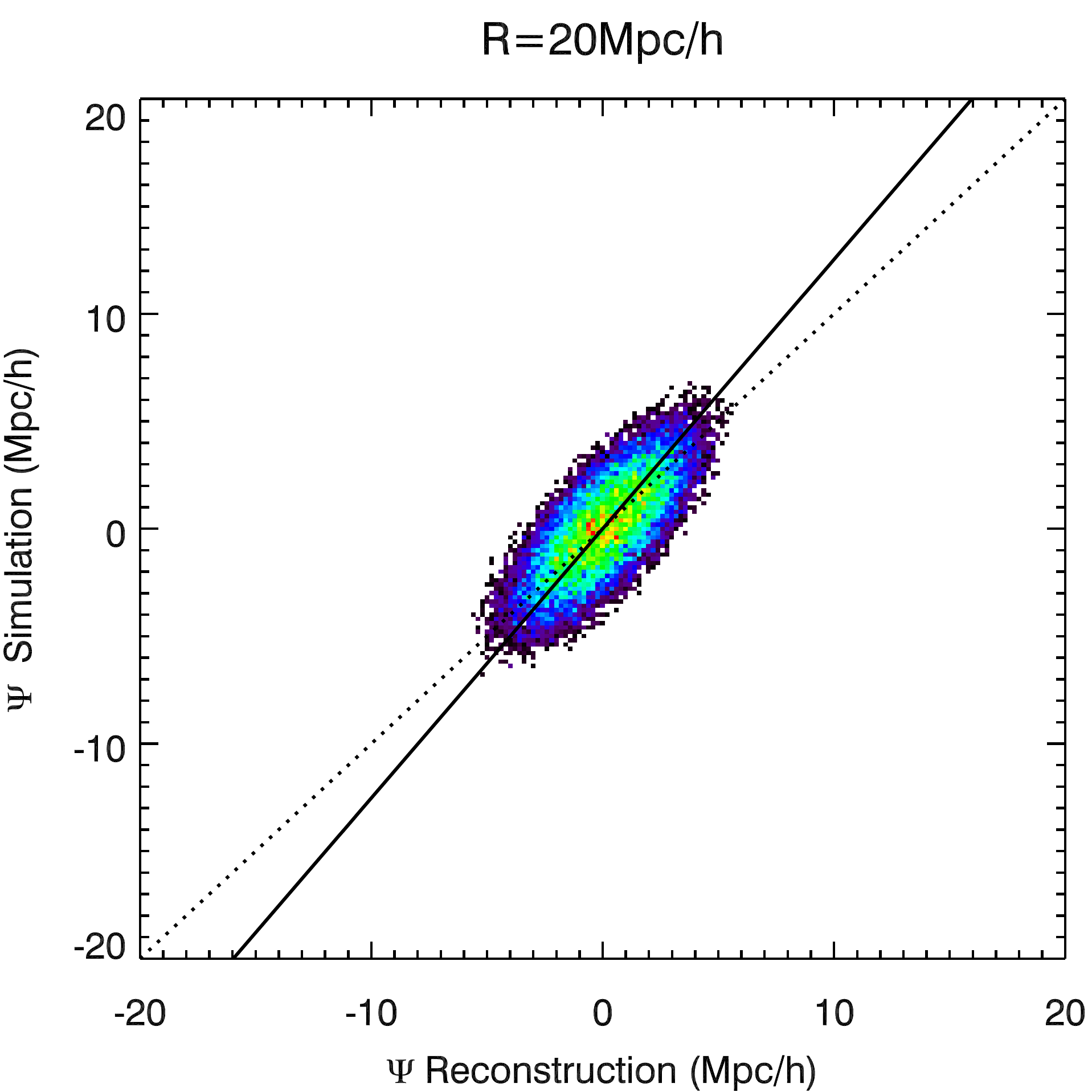}\includegraphics[width=1.1in]{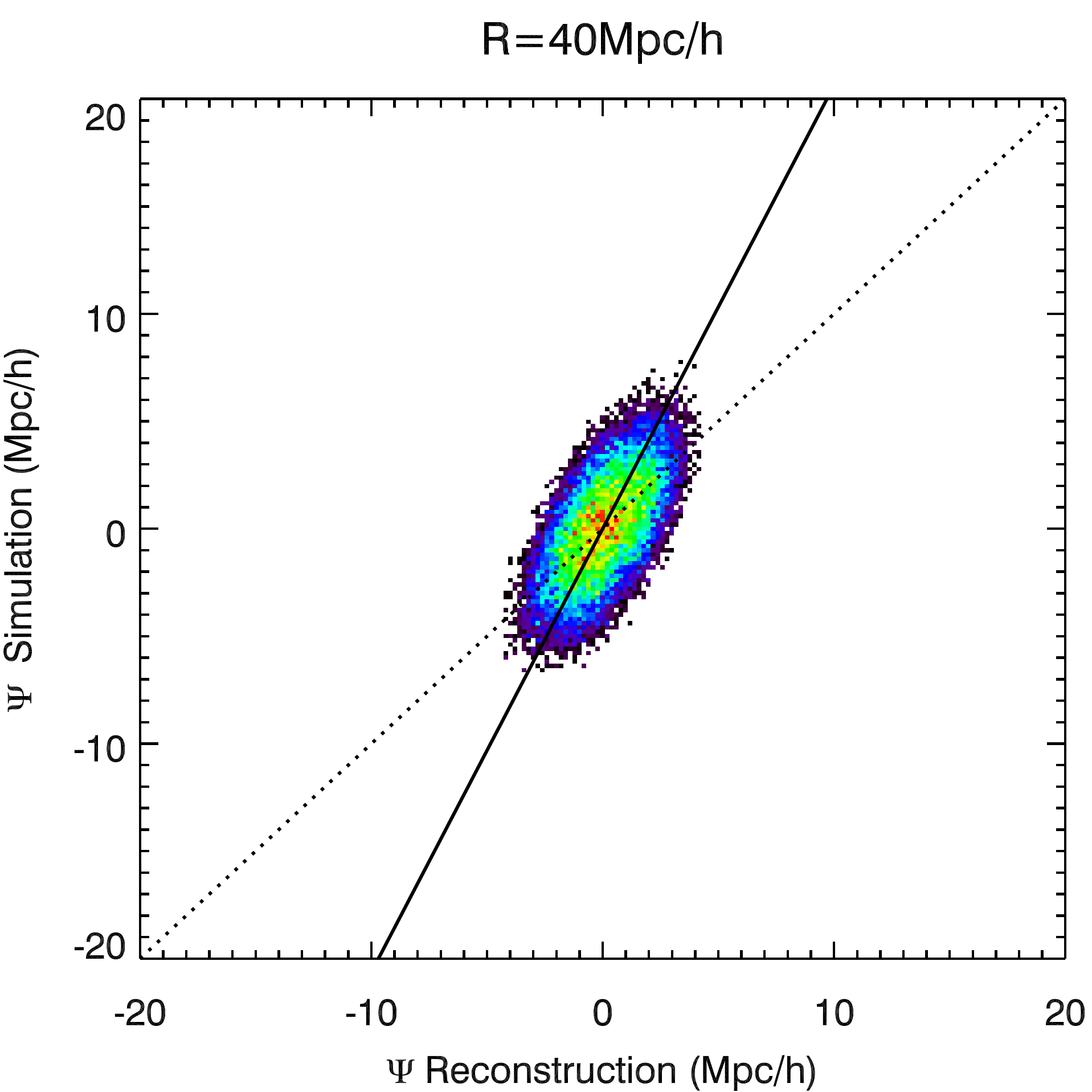}
\includegraphics[width=1.1in]{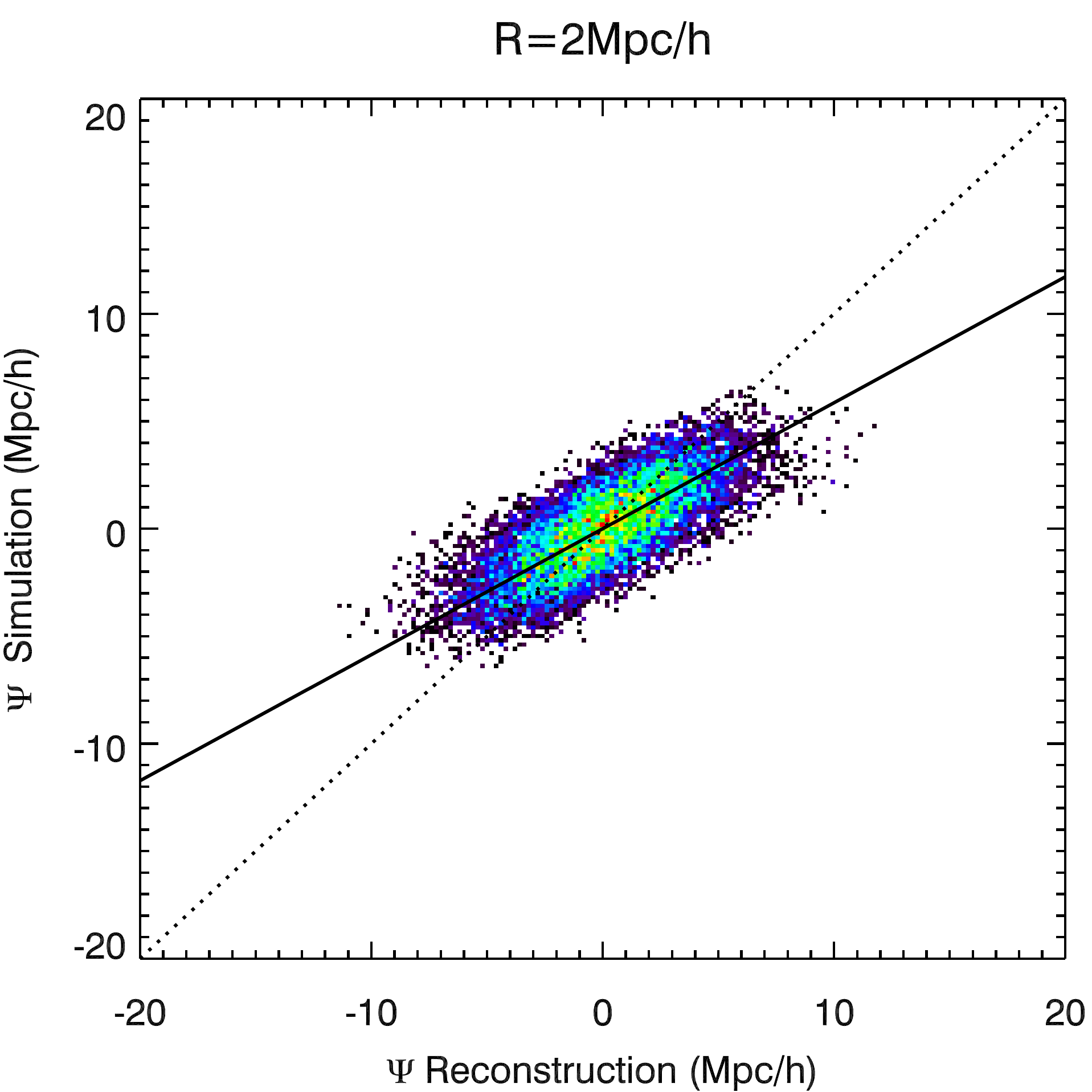}\includegraphics[width=1.1in]{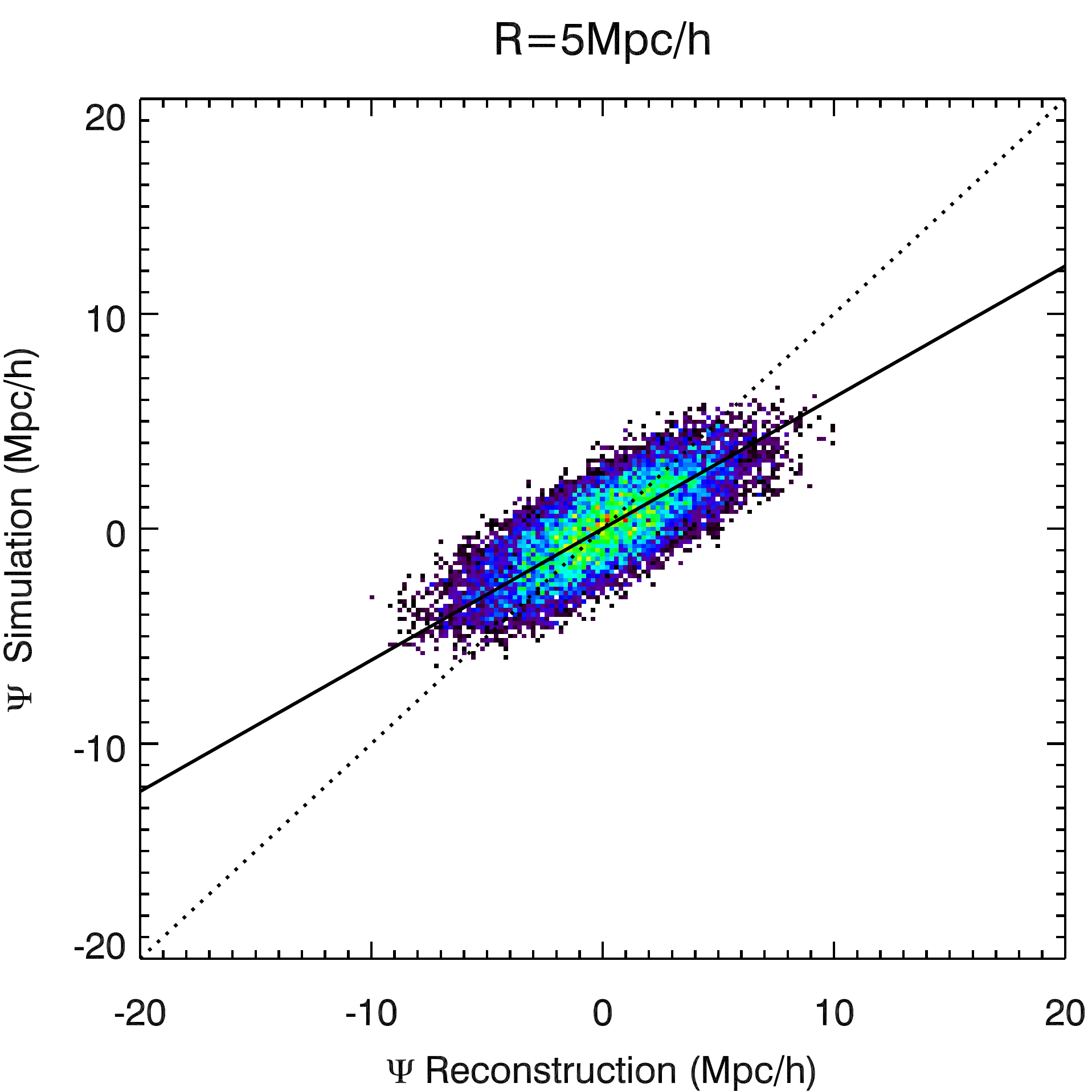}\includegraphics[width=1.1in]{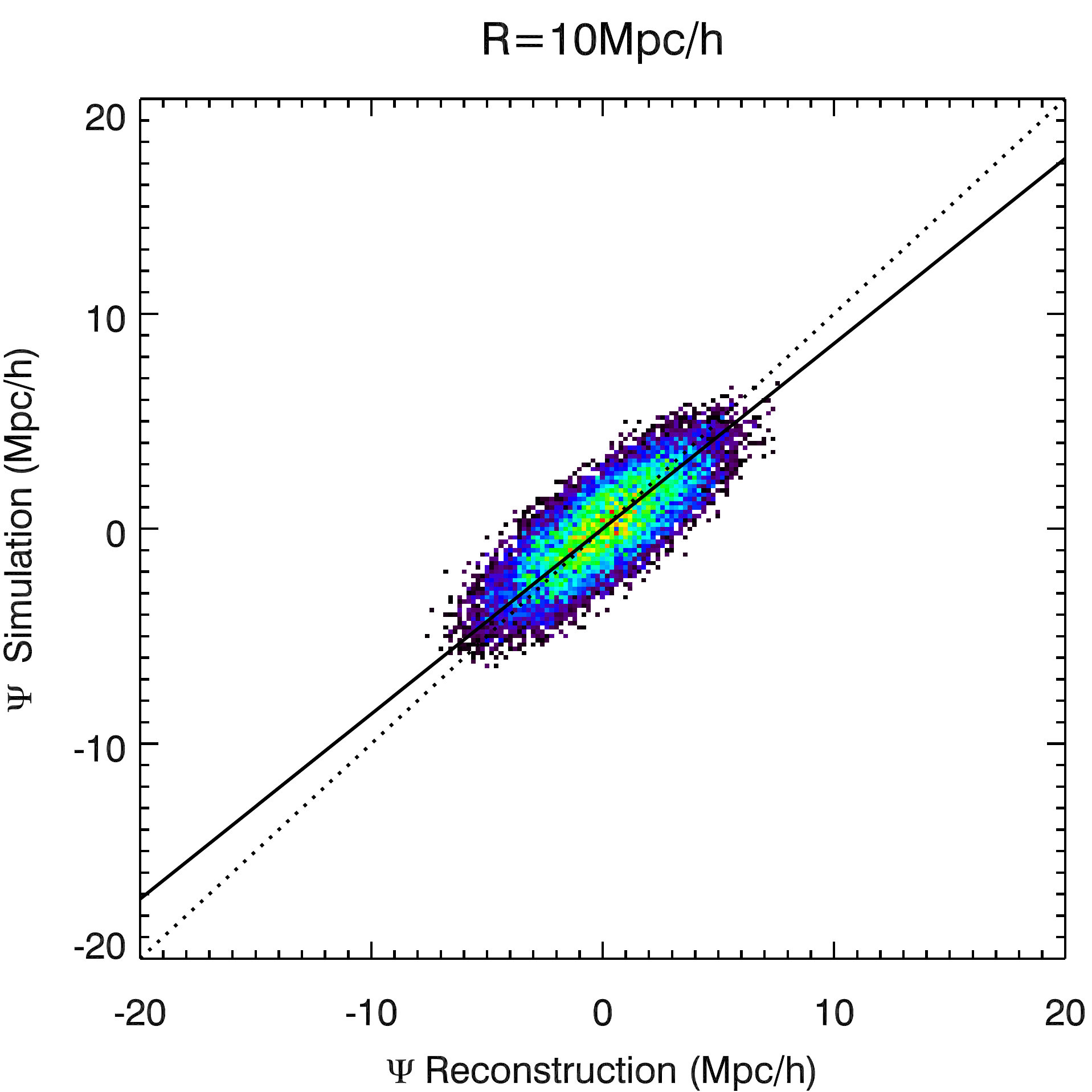}\includegraphics[width=1.1in]{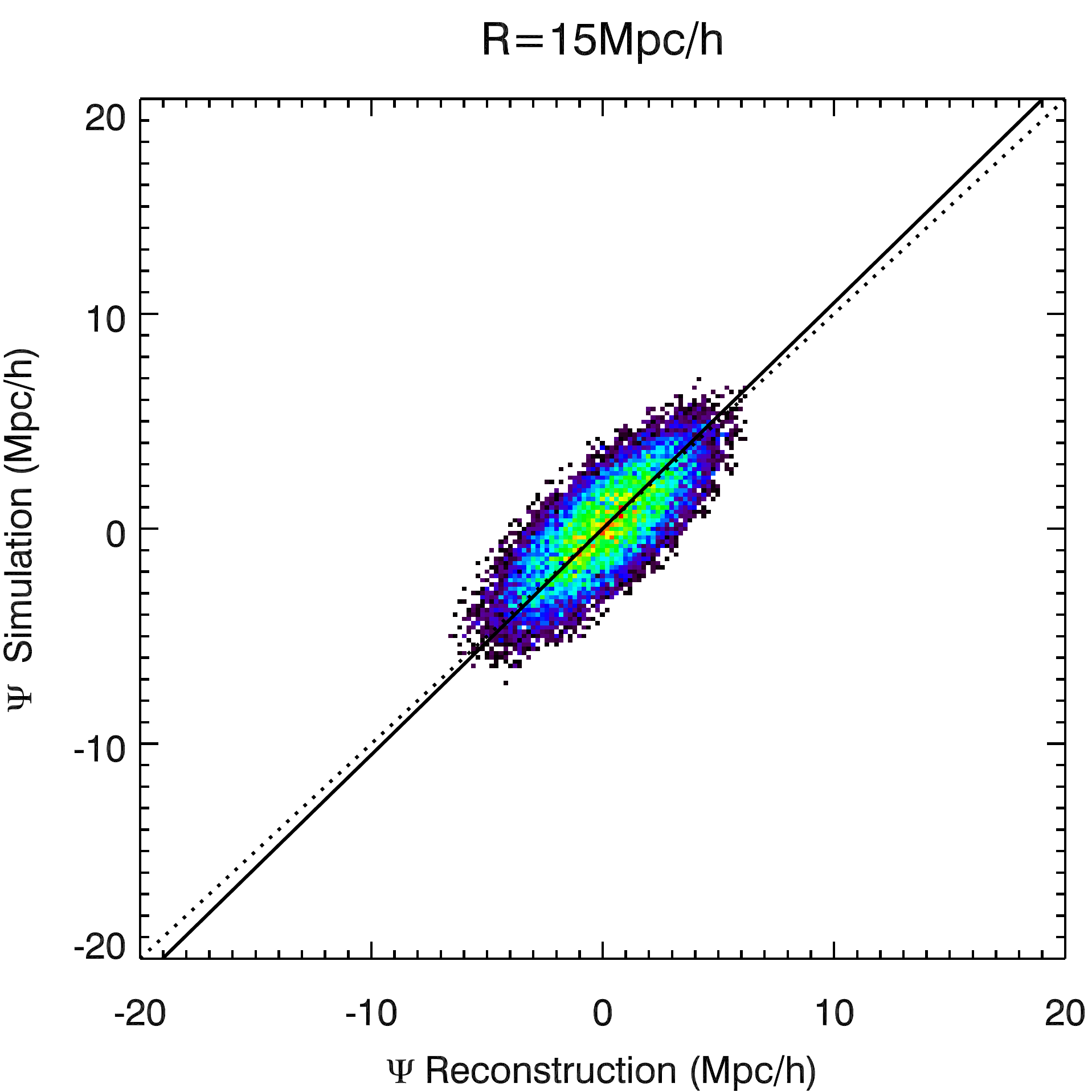}\includegraphics[width=1.1in]{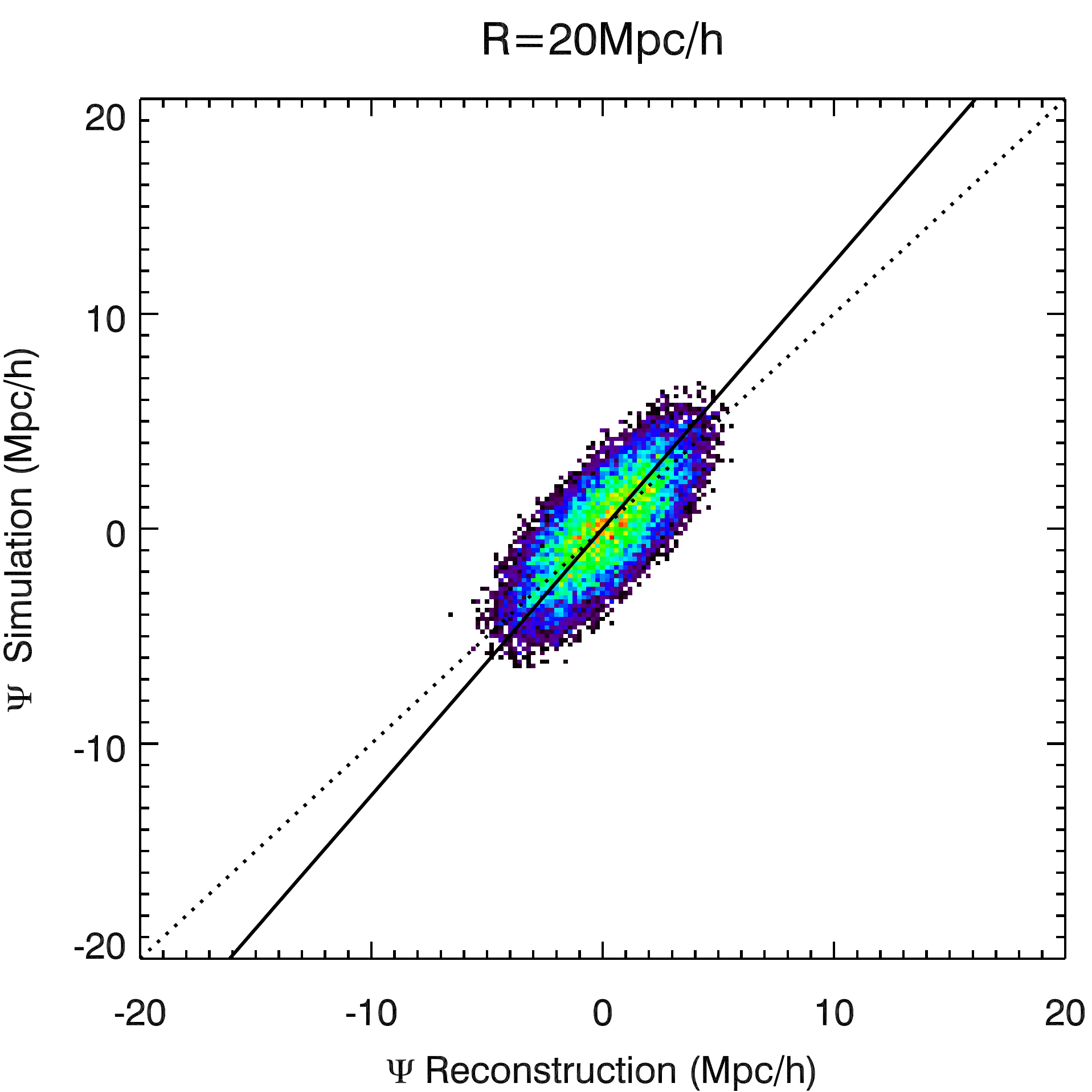}\includegraphics[width=1.1in]{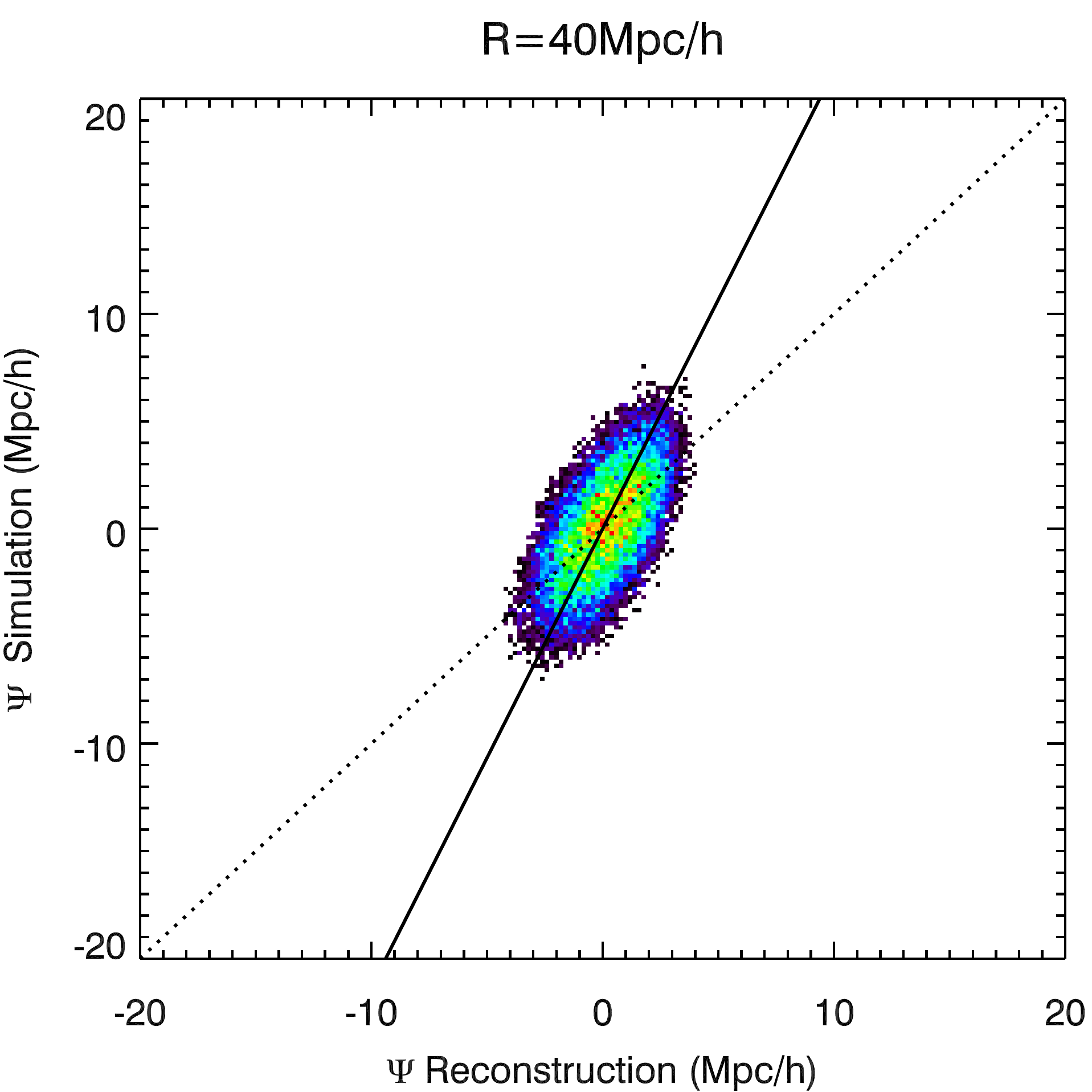}

	 \caption{We show the scatter plots between true and reconstructed displacements for RunPB mocks. Each line of plots represents different halo cuts (i.e  represents a sample with a different halo bias). The different columns represent the different smoothing scales, from left to right R=2, 5, 10, 15, 20, 40$h^{-1}$ Mpc. The mocks used in this test were in real space. The dashed black line indicates the 1-1 relation.
	 The solid line comes from fitting a 2D Gaussian \modiff{distribution} to the 2D histogram, \modiff{representing} the angle between the major axis and the ordinate axis. 
	  This angle does not  indicate the bias of the reconstructed displacements compared with the true displacements but provides an illustration of the effect.	  
	  The results of the bias and noise are presented in the Table \ref{tab:halocut}. The  higher correlation coefficient and the lower quality factor indicates the best result for each case.
	  } 

   	\label{fig:hist2dpbruncuts}
\end{figure*}

 We show the scatter plots between true and reconstructed displacements for RunPB Mocks in Figure \ref{fig:hist2dpbruncuts}. Each line of plots \modiff{represents a different halo cut}, i.e represents a sample with a different halo bias. From top to bottom, $M_{halo}=[1e12, 5e12, 9e12, 1e13, 2e13, 4e13, 5e13 ]h^{-1} M_{\odot}$. The different columns show the different smoothing scales, from left to right R=[2,5,10,15,20,40]$h^{-1}$ Mpc. The dashed black line indicates the 1-1 relation. The solid line comes from fitting a 2D Gaussian distribution to the 2D histogram; it represents the angle between the major axis and the ordinate axis. This angle does not  indicate the bias of the reconstructed displacements compared with the true displacements but provides an illustration of the effect. 

\begin{figure*}
   \centering    
\includegraphics[width=3.3in]{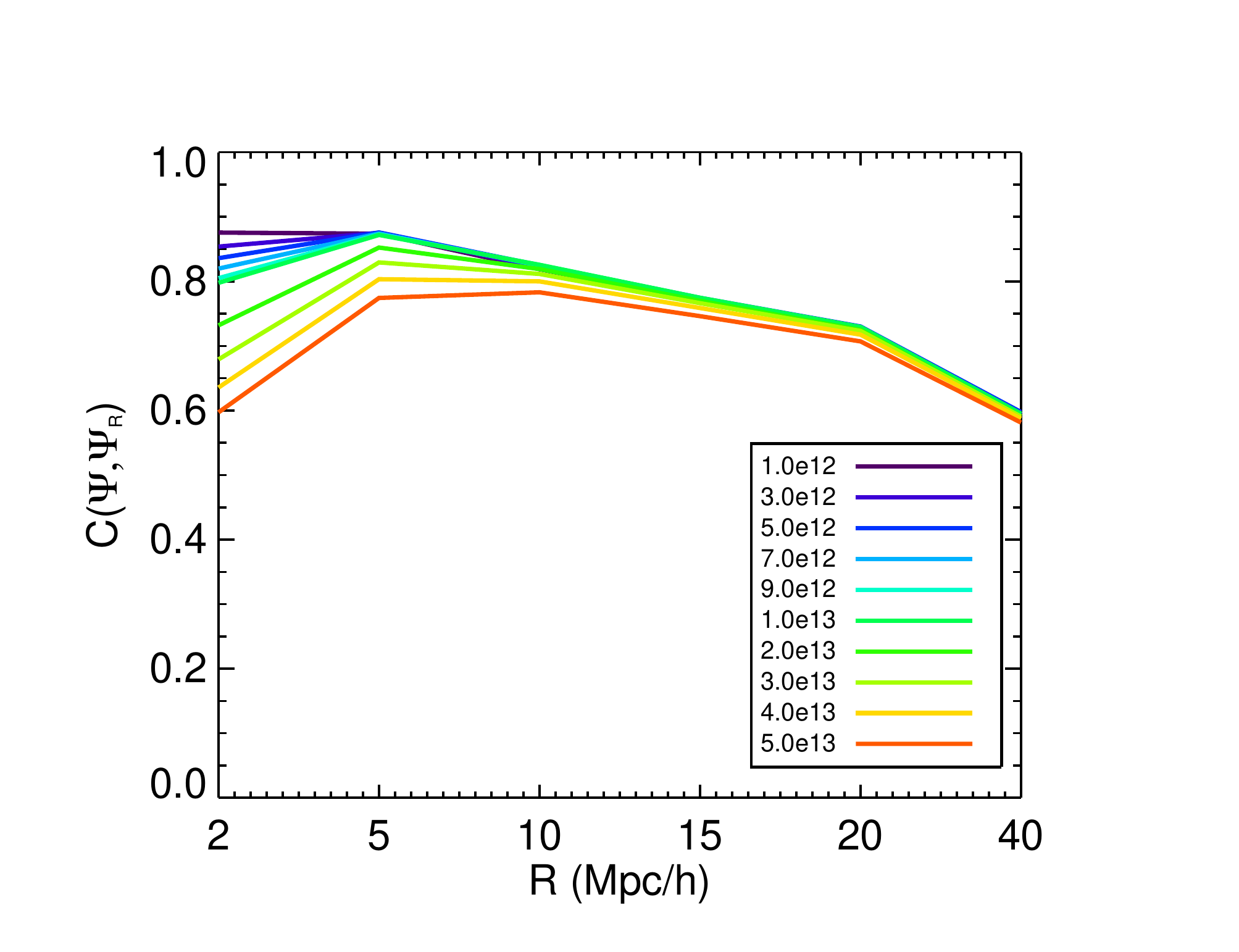}\includegraphics[width=3.3in]{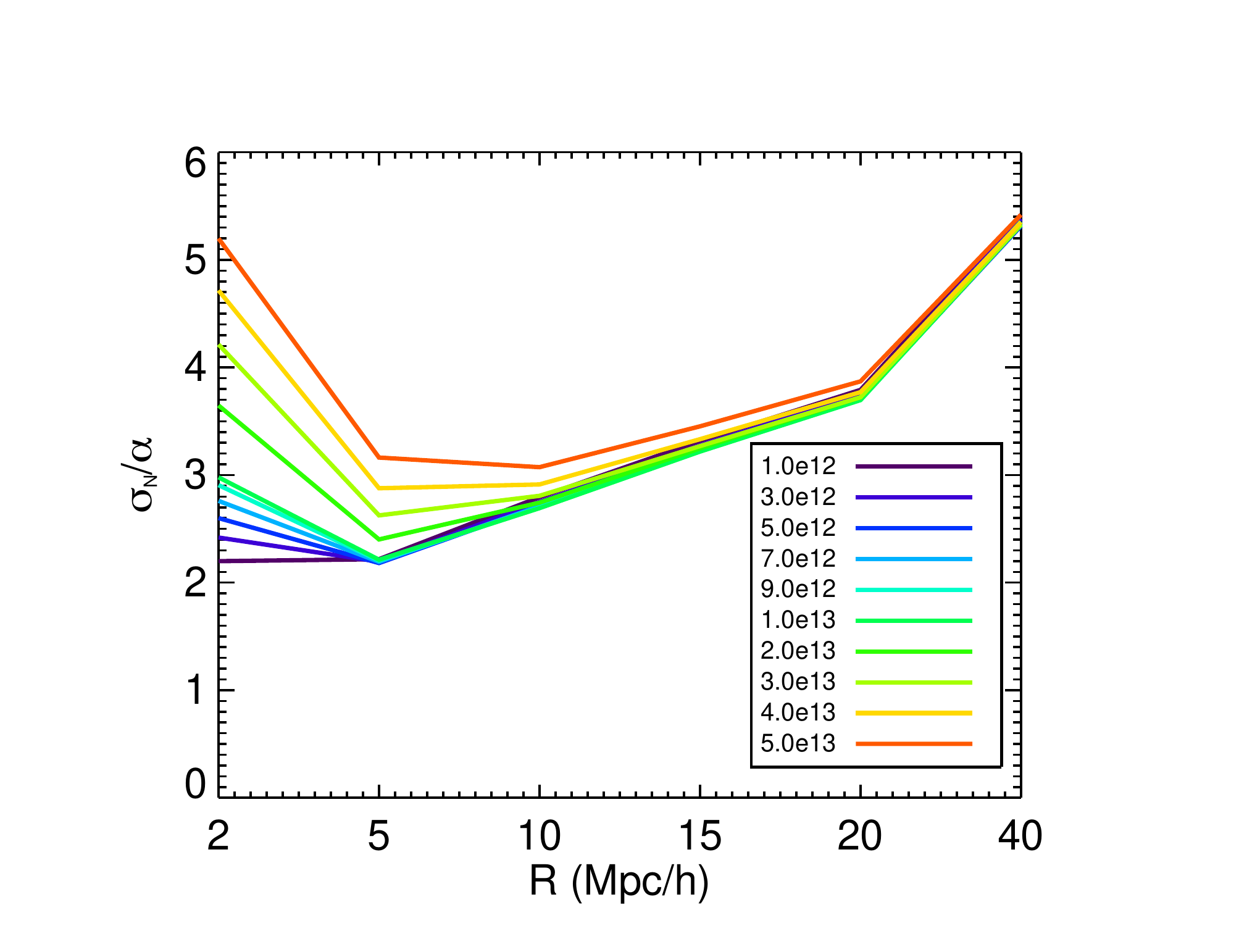}
\caption{ The left panel shows the correlation coefficient $<\mathcal{C}(\tilde{\psi}_R,\psi)>$ and the right panel the quality factor $\sigma_N / \alpha$ as a function of the smoothing scale. 
The different samples follow similar trends: the best results are obtained for the higher correlation coefficient and the lower quality factor; using the  5 $h^{-1}$ Mpc smoothing scale gives the best results for the most of the cases. The more biased samples are more affected by the shot noise, and for those cases, a smoothing scale from 5-10$h^{-1}$ Mpc is preferable, while for low-biased tracers, a smaller smoothing scale, from 2-5$h^{-1}$ Mpc, also gives a good result. On the other hand, we can see that the result is similar for all the biases for the large smoothing scale, indicating that the smoothing is compensating for the shot noise effect on the reconstructed displacements. }
	\label{fig:sumbiasnoisecut}
\end{figure*}

 The results of the bias and noise (equations \ref{eq:alpha} and \ref{eq:sigma}) are presented in Table \ref{tab:halocut}. The quality factor and coefficient of correlation indicate what is the best smoothing scale for each case. 
The best results are for the higher correlation coefficient and the lower quality factor.
Figure \ref{fig:sumbiasnoisecut} summarises bias and noise results for the samples generated with different halo mass cuts. \modiff{The left} panel shows the correlation coefficient $<\mathcal{C}(\tilde{\psi}_R,\psi)>$ and \modiff{the} right panel the quality factor $\sigma_N / \alpha$ as a function of the smoothing scale.
 
 The results for the correlation coefficient and the quality factor show two regimes: 1) for a small smoothing scale the result  depends strongly on the bias; 2) for a large smoothing scale the results converge in the correlation coefficient and quality factor for all the biases.
These two behaviours could be understood considering number density and shot noise. 
The greater the mass cut is, the greater the bias is and also the greater the shot noise level of the samples. The samples that are more biased are more affected by the shot noise; for those cases, a smoothing scale from 5-10$h^{-1}$ Mpc is preferable, while for low-biased tracers, a smaller smoothing scale, from 2-5$h^{-1}$ Mpc, also gives a good result. 
On the other hand we can see that the result is similar for all the biases for the large smoothing scale, indicating that the smoothing is compensating for the shot noise effect on the reconstructed displacements. 

\begin{table}
\begin{center}
\caption{ RunPB mocks samples for several halo mass cuts.  Error and bias from reconstructed RunPB cubic mocks with different smoothing lengths. } 
\label{tab:halocut}
\begin{tabular}{@{}lccccc}
\hline
R & $\mathcal{C}(\tilde{\psi}_R,\psi)$ & $\alpha$ & 
$\sigma_N$ & 
$\frac{\sigma_N}{\alpha}$   \\
\hline
Mass cut = 1.0e12\\
\hline
5 & 0.87 & 0.87 & 1.94 & 2.21 \\
10 & 0.81 & 0.70 & 1.95 & 2.79 \\
15 & 0.76 & 0.58 & 1.93 & 3.31 \\
20 & 0.72 & 0.49 & 1.88 & 3.79 \\
40 & 0.59 & 0.29 & 1.59 & 5.41 \\
\hline
Mass cut = 3.0e12\\
\hline
5 & 0.87 & 0.91 & 2.00 & 2.19 \\
10 & 0.82 & 0.72 & 1.97 & 2.74 \\
15 & 0.77 & 0.59 & 1.95 & 3.27 \\
20 & 0.72 & 0.50 & 1.90 & 3.75 \\
40 & 0.59 & 0.29 & 1.61 & 5.38 \\
\hline
Mass cut = 5.0e12\\
\hline
5 & 0.87 & 0.94 & 2.06 & 2.18 \\
10 & 0.82 & 0.73 & 2.00 & 2.71 \\
15 & 0.77 & 0.60 & 1.96 & 3.23 \\
20 & 0.73 & 0.51 & 1.91 & 3.70 \\
40 & 0.59 & 0.30 & 1.61 & 5.31 \\
\hline
Mass cut = 7.0e12\\
\hline
5 & 0.87 & 0.97 & 2.13 & 2.19 \\
10 & 0.82 & 0.75 & 2.03 & 2.71 \\
15 & 0.77 & 0.61 & 2.00 & 3.23 \\
20 & 0.72 & 0.52 & 1.94 & 3.71 \\
40 & 0.59 & 0.30 & 1.65 & 5.35 \\
\hline
Mass cut = 9.0e12\\
\hline
5 & 0.87 & 0.99 & 2.19 & 2.20 \\
10 & 0.82 & 0.76 & 2.06 & 2.69 \\
15 & 0.77 & 0.62 & 2.02 & 3.22 \\
20 & 0.72 & 0.53 & 1.96 & 3.70 \\
40 & 0.59 & 0.31 & 1.67 & 5.33 \\
\hline
\end{tabular}
\end{center}
\end{table}

\begin{table}
\begin{center}
\caption{ RunPB mocks samples for several halo mass cuts.  Error and bias from reconstructed RunPB cubic mocks  with different smoothing lengths. } 
\label{tab:halocut2}
\begin{tabular}{@{}lccccc}
\hline
R & $\mathcal{C}(\tilde{\psi}_R,\psi)$ & $\alpha$ & 
$\sigma_N$ & 
$\frac{\sigma_N}{\alpha}$   \\

\hline
Mass cut = 1.0e13\\
\hline
5 & 0.87 & 1.00 & 2.22 & 2.21 \\
10 & 0.82 & 0.76 & 2.06 & 2.69 \\
15 & 0.77 & 0.62 & 2.02 & 3.21 \\
20 & 0.72 & 0.53 & 1.96 & 3.69 \\
40 & 0.59 & 0.31 & 1.67 & 5.32 \\
\hline
Mass cut = 2.0e13\\
\hline
5 & 0.85 & 1.09 & 2.62 & 2.40 \\
10 & 0.81 & 0.80 & 2.20 & 2.74 \\
15 & 0.76 & 0.65 & 2.11 & 3.24 \\
20 & 0.72 & 0.54 & 2.03 & 3.72 \\
40 & 0.59 & 0.31 & 1.70 & 5.35 \\
\hline
Mass cut = 3.0e13\\
\hline
5 & 0.82 & 1.16 & 3.05 & 2.62 \\
10 & 0.81 & 0.83 & 2.34 & 2.80 \\
15 & 0.76 & 0.66 & 2.18 & 3.26 \\
20 & 0.72 & 0.55 & 2.08 & 3.72 \\
40 & 0.59 & 0.32 & 1.73 & 5.32 \\
\hline
Mass cut = 4.0e13\\
\hline
5 & 0.80 & 1.22 & 3.53 & 2.87 \\
10 & 0.80 & 0.86 & 2.50 & 2.91 \\
15 & 0.75 & 0.68 & 2.27 & 3.33 \\
20 & 0.71 & 0.56 & 2.14 & 3.77 \\
40 & 0.58 & 0.32 & 1.75 & 5.35 \\
\hline
Mass cut = 5.0e13\\
\hline
5 & 0.77 & 1.29 & 4.09 & 3.16 \\
10 & 0.78 & 0.88 & 2.71 & 3.07 \\
15 & 0.74 & 0.69 & 2.39 & 3.45 \\
20 & 0.70 & 0.57 & 2.22 & 3.87 \\
40 & 0.58 & 0.32 & 1.77 & 5.41 \\  
\hline
\end{tabular}
\end{center}
\end{table}

\section{Conclusion}\label{sec:con}

We summarize the results presented in this paper.

\begin{itemize}
\item We test \modiff{the} reconstruction algorithm using four different smoothing scales, $5$, $10,15$, $40$ Mpc/h with QPM BOSS-like mock {catalogues} and we study the effect in BAO anisotropic analysis.
 We find that the different smoothing scales affect the multipoles. We observe variations in the amplitude of the monopole \modiff{at} low scales and in the sharpening of the BAO feature. The effects on the quadrupole are larger than \modiff{on} the monopole. The smoothing scale of 15 $h^{-1}$ Mpc produces a quadrupole not completely consistent with zero. Taking a smaller smoothing scale reduces the negative correlation observed in the quadrupole to be in agreement with zero  within the fitting range. A large smoothing scale increases the negative correlation in the quadrupole at scales between $50-100 h^{-1}$Mpc.
 \item We show that the smoothing scale affects the anisotropic clustering results.\modif{ The results indicate that the best choice for the smoothing length, in terms of anisotropic clustering, is given by the smaller smoothing scale of $5 h^{-1}Mpc$.  This smoothing scale shows the smaller bias and error  on $\alpha$ an $\epsilon$ error bars simultaneously.} The variations in the mean value for best fits parameters using $15 h^{-1}$ Mpc compared with $5h^{-1}$ Mpc are
$\sim$0.5\% for $\alpha$ and 0.3 for $\epsilon$,  { when considering the noise in the covariance the variations in $\alpha$ are 0.3 and 0.2 for $\epsilon$. The  variations produce } $\Delta D_A\sim0.3$  and for $\Delta H\sim 0.4$\%.
 The smoothing scale affects the precision at which we can measure the BAO parameters. Taking $15 h^{-1}$Mpc as a reference, the median error using 5 Mpc/h is reduced by 10\% for $\alpha$ and 45\% for $\epsilon$. The  effect on the uncertainties of $D_A(z)$ and $H(z)$ are 40\% and 30\% respectively. {The variations in the uncertainties does not consider the covariance noise. }
\item We explore the effect of the smoothing scale in the reconstructed displacement field to investigate our hypothesis, \modiff{whether} the improvement observed in the post-reconstruction anisotropic clustering was coming from a better estimation of the displacement field when we reduce the smoothing scale.\modif{ We estimated the correlation coefficient between the two fields and also we provided a methodology to estimate the noise in addition to the correlation coefficient, through the quality factor (defined as the ratio $\sigma_N/\alpha$). The correlation coefficient and the quality factor are symmetric and so provide equivalent information. However, while the correlation coefficient provides a single information, with the methodology we proposed, we have access to the bias and the noise of the estimator separately. This point is important when one wants to use the reconstruction methodology to derive the velocity field directly and associate individual errors to it. We find that the quality value, $\sigma_N/\alpha$ is minimized for $R=5 h^{-1}$ Mpc and also that the maximal correlation coefficient value ($C=0.87$) corresponds to the $R=5 h^{-1}$ Mpc. }
 \item  We explore how the conclusions we get for BOSS-like samples scale for other bias and number densities so the results could be applied to other surveys. We show the effects in terms of the displacement field (and in the appendix, the bias effect in the correlation function). For this test, we generate several halo samples from RunPB simulation, applying different halo mass cuts. The results for the correlation coefficient and the quality factor show two regimes: 1) For small smoothing scales, the result  depends strongly on the bias;  the samples that are more biased are more affected by the shot noise; for those cases, a smoothing scale from 5-10$h^{-1}$ Mpc is preferable, while for low-biased tracers, a smaller smoothing scale, from 2-5$h^{-1}$ Mpc also gives a good result. 2) For large smoothing scales, the results converge in the correlation coefficient and quality factor for all the biases, indicating that the smoothing is compensating for the shot noise effect on the reconstructed displacements, and the dependence on bias is weaker.
 
 \end{itemize}
 Further work should be done to provide a theoretical framework \modiff{for} the empirical results shown in this article. 
 The optimal smoothing scale could vary for different tracers (different kinds of galaxies) or could \modiff{possibly} also depend on the redshift and environmental properties. A future work will \modiff{focus on} exploring these different dependencies to look for ways to improve the reconstruction technique for future surveys such as eBOSS, DESI, EUCLID. 
\section{Acknowledgements}
Many thanks to Nikhil Padmanabhan, Angela Burden and Ross O'Connell for useful discussions and correspondence.

MV is partially supported by Programa de Apoyo a Proyectos de Investigaci\'on e Innovaci\'on Tecnol\'ogica (PAPITT) No IA102516, Proyecto Conacyt Fronteras No 281 and Proyecto DGTIC SC16-1-S-120. 
Funding for SDSS-III has been provided by the Alfred P. Sloan Foundation, the Participating Institutions, the National Science Foundation, and the U.S. Department of Energy Office of Science. The SDSS-III web site is http://www.sdss3.org/.
SDSS-III is managed by the Astrophysical Research Consortium for the Participating Institutions of the SDSS-III Collaboration including the University of Arizona, the Brazilian Participation Group, Brookhaven National Laboratory, Carnegie Mellon University, University of Florida, the French Participation Group, the German Participation Group, Harvard University, the Instituto de Astrofisica de Canarias, the Michigan State/Notre Dame/JINA Participation Group, Johns Hopkins University, Lawrence Berkeley National Laboratory, Max Planck Institute for Astrophysics, Max Planck Institute for Extraterrestrial Physics, New Mexico State University, New York University, Ohio State University, Pennsylvania State University, University of Portsmouth, Princeton University, the Spanish Participation Group, University of Tokyo, University of Utah, Vanderbilt University, University of Virginia, University of Washington, and Yale University.

This research used resources of the National Energy Research Scientific Computing Center, a DOE Office of Science User Facility supported by the Office of Science of the U.S. Department of Energy under Contract No. DE-AC02-05CH11231.

\appendix
\section{Comparison of Different Reconstruction Implementations.}\label{ap:imp}
\begin{figure}
   \centering    
              \includegraphics[width=1.7in]{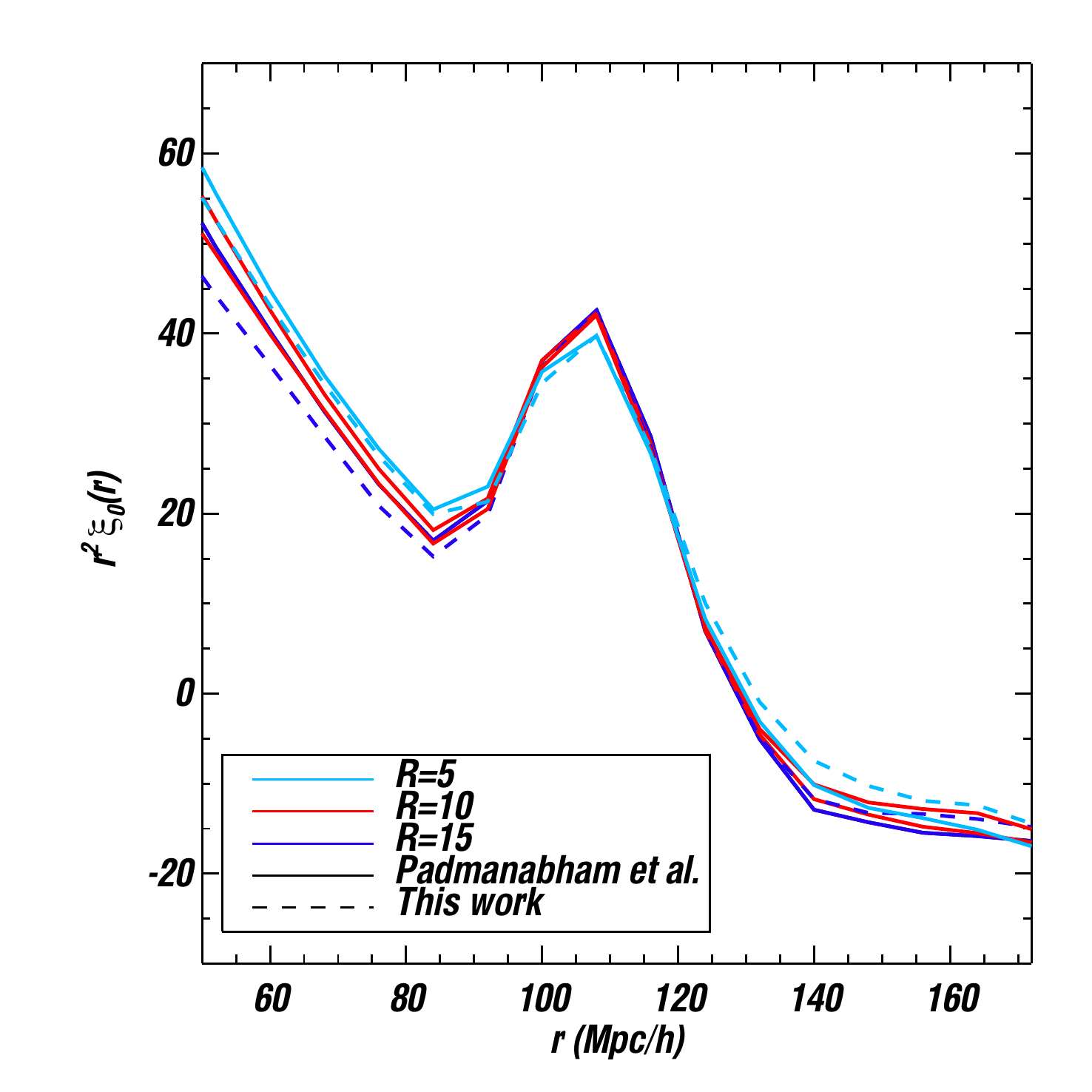}\includegraphics[width=1.7in]{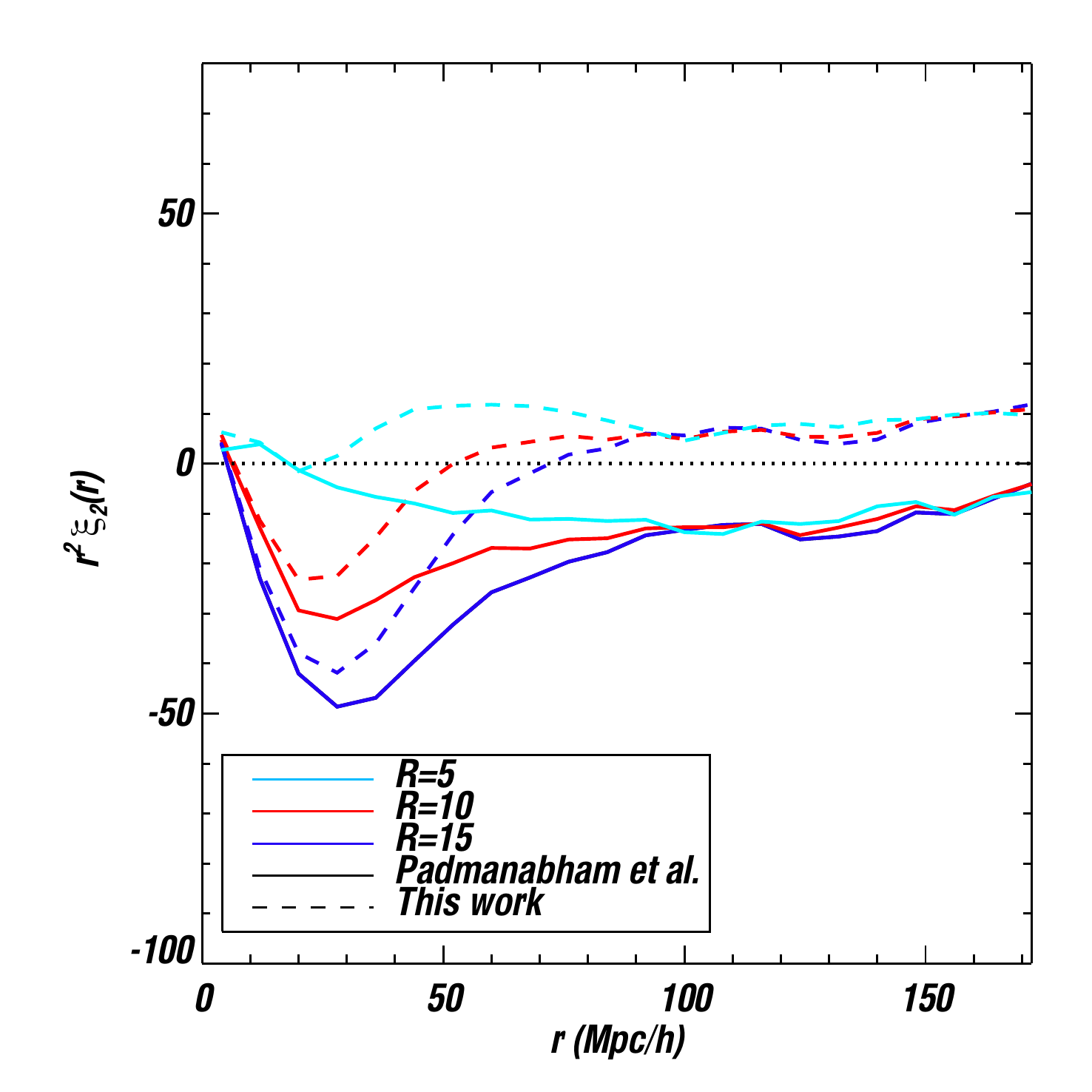} 
  \caption{Multipoles for different smoothing scales: $5,10$ and  15 $h^{-1}$ Mpc for different implementations. \modiff{In our implementation: 5 $h^{-1}$ Mpc in cyan, 10 $h^{-1}$ Mpc in blue and 15 $h^{-1}$ Mpc in black. For Padmanabhan et al.'s implementation: 5 $h^{-1}$ Mpc in yellow, 10 $h^{-1}$ Mpc  in green and 15 $h^{-1}$ Mpc in red}. In the monopole, we observe that, in the fitting region, both implementations perform similarly; however, there are some differences at very small scales ($<20 $$h^{-1}$ Mpc). For the quadrupole, we observe  important differences in the amplitude. The quadrupole for Padmanabhan et al. seems to be kind of over-correcting, while in our implementation seems to be under-correcting; however, both quadrupoles are very similar in shape. Differences in the quadrupole are generated by two effects: redshift distortions correction and \modiff{the} effects of the angular and radial selection functions that are implemented in slightly different ways. Further exploration and quantification of these two contributions, which generate differences in the quadrupole amplitude, are treated in Vargas-Maga\~na et al. (in prep).}
   \label{fig:mean_smooth_bias21}
\end{figure}

In this section we show that the results obtained in this work are independent of the particular implementation of \modiff{the} reconstruction algorithm. We compare our implementation of the reconstruction algorithm (hereafter ``RV") to the method implemented by \cite{Pad12}, hereafter denoted ``RP." The differences between the two methods could be summarised as follows:
\begin{itemize}
\item Solution of Displacement Field. RV implementation solves the 
Poisson equation in Fourier Space assuming \modiff{a} real space density field, RP in configuration space using a finite differences technique.
\item Redshift Space Corrections. RV neglects the redshift space corrections, while RP takes them into account. This choice is tested to give the accuracy required for BAO anisotropic analysis at sub-percent precision in Vargas-Maga\~na (in prep). The corrections are not important because the effect on the quadrupole is a change of the amplitude, and the peak position and contrast are not affected.
\item Survey Mask and Density Estimate. Both methods follow different \modiff{approaches} for dealing with the effect of geometry. 
\end{itemize}
We used the same 100 mocks described before and we run reconstruction with RP and RV implementation using the same parameters. We use for this test a bias=2.1 instead of bias=1.87 because the reconstructed {\bf catalogues} were available only for this value of the bias for the RP implementations. We compare results in terms of the multipoles and anisotropic fits. We show in Figure \ref{fig:mean_smooth_bias21} the mean multipoles for the different implementations using three different smoothing lengths, 5, 10, and 15 $h^{-1}$ Mpc. We observe that the monopole behaviour is pretty similar between different implementations in the range of the fitting; the differences observed are only important at scales smaller than 20 $h^{-1}$ Mpc. In the case of the quadrupole, the amplitude of the quadrupole between different implementations is significantly different. The quadrupole for RP implementation seems to over-correct the anisotropy, while it seems to be under-correcting in our implementation; however, both quadrupoles are very similar in shape. Differences in the quadrupole are generated by two effects: redshift distortions correction and the effects of the angular and radial selection functions that are implemented in slightly different ways. Further exploration and quantification of these two contributions, which generate differences in the quadrupole amplitude are addresses in Vargas-Maga\~na et al. (in prep).

\begin{figure}
   \centering   
\includegraphics[width=1.6in]{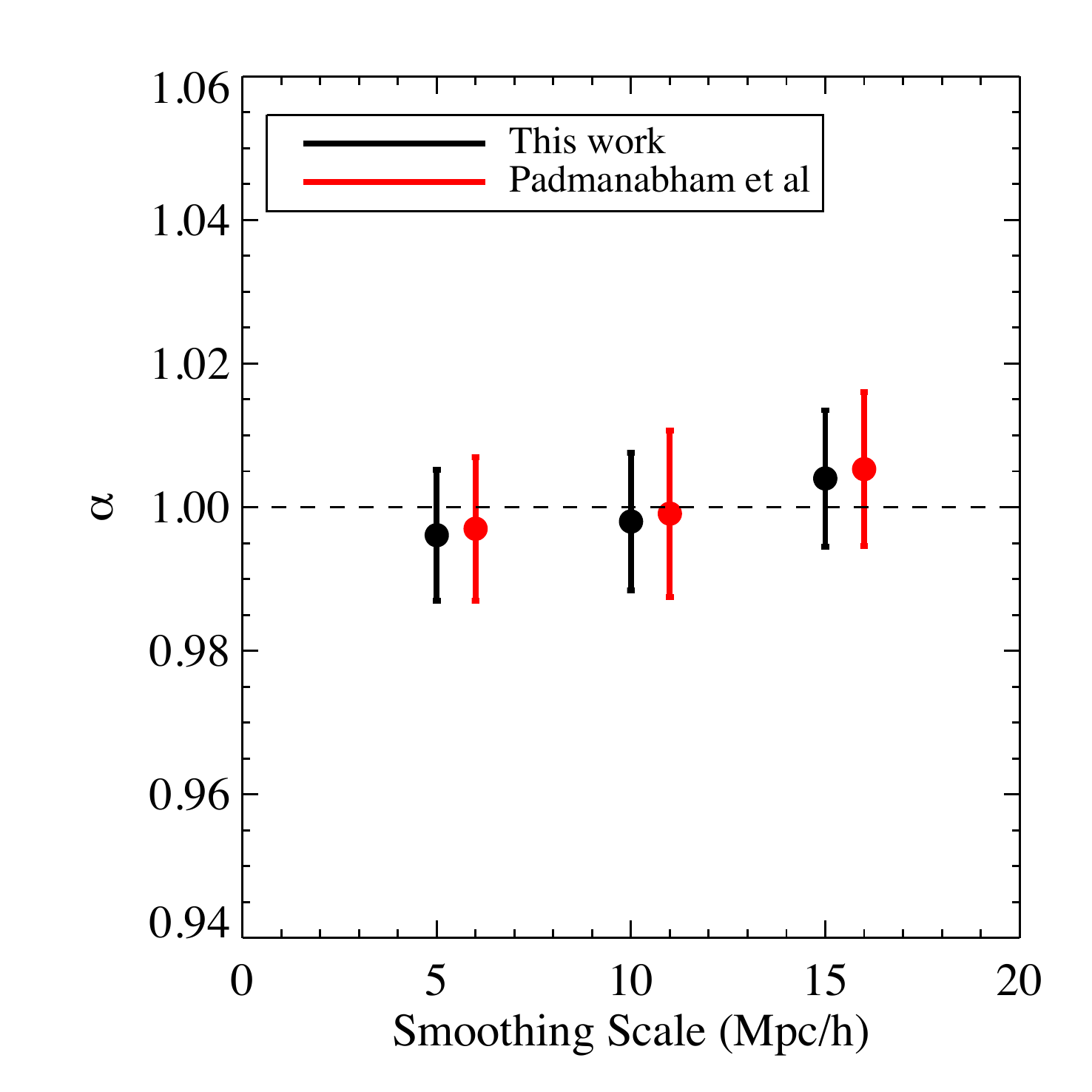}
\includegraphics[width=1.6in]{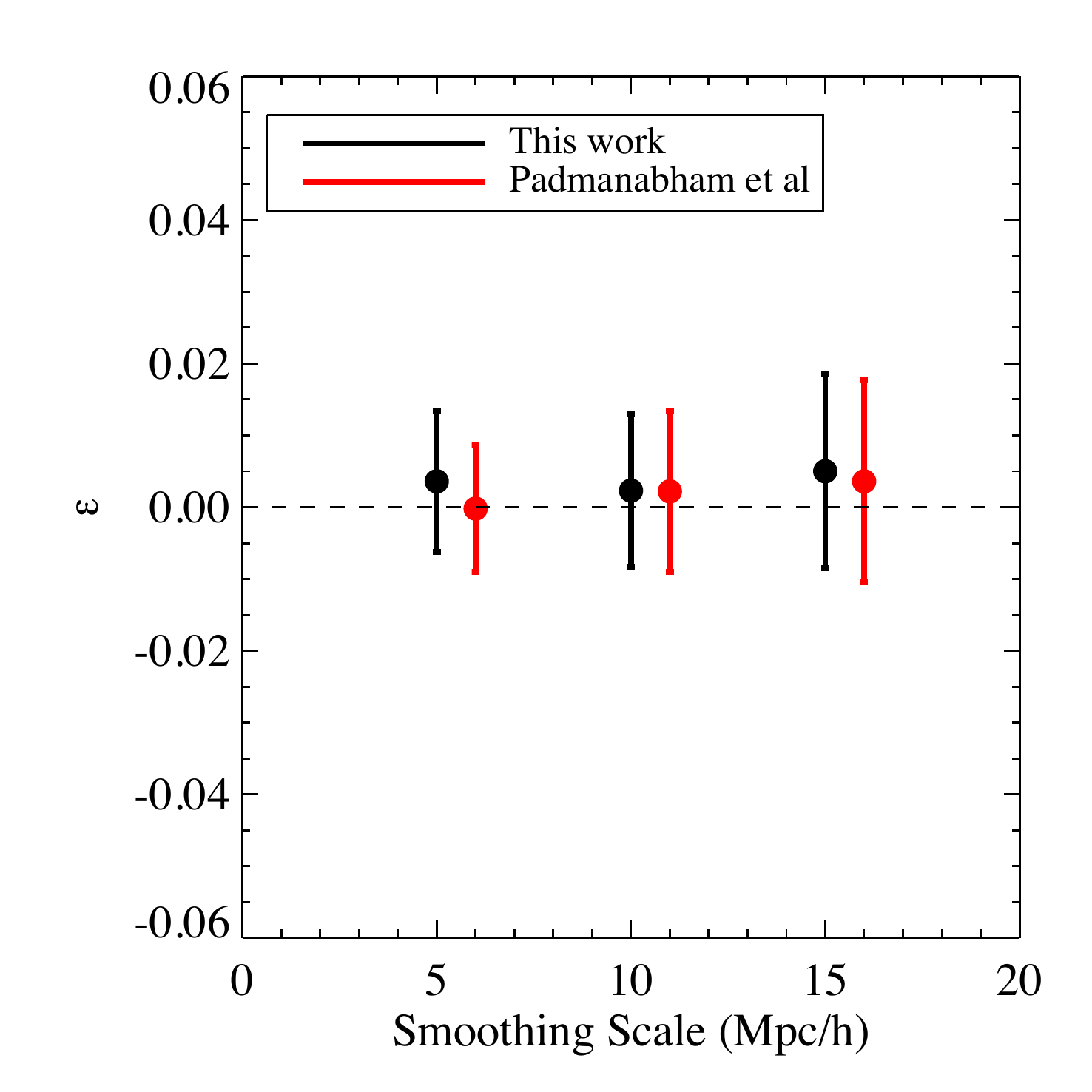} 
\caption{Smoothing Fitting Results. Right panel, mean $\alpha$ and RMS of $\alpha$ distribution [right]. Left panel, mean $\epsilon$ and RMS for both implementations. We use a value of the bias=2.1 different than the value used in the article, i.e bias=1.87}   
\label{fig:smoothing_fits}
\end{figure}

\begin{figure}
   \centering   

\includegraphics[width=1.6in]{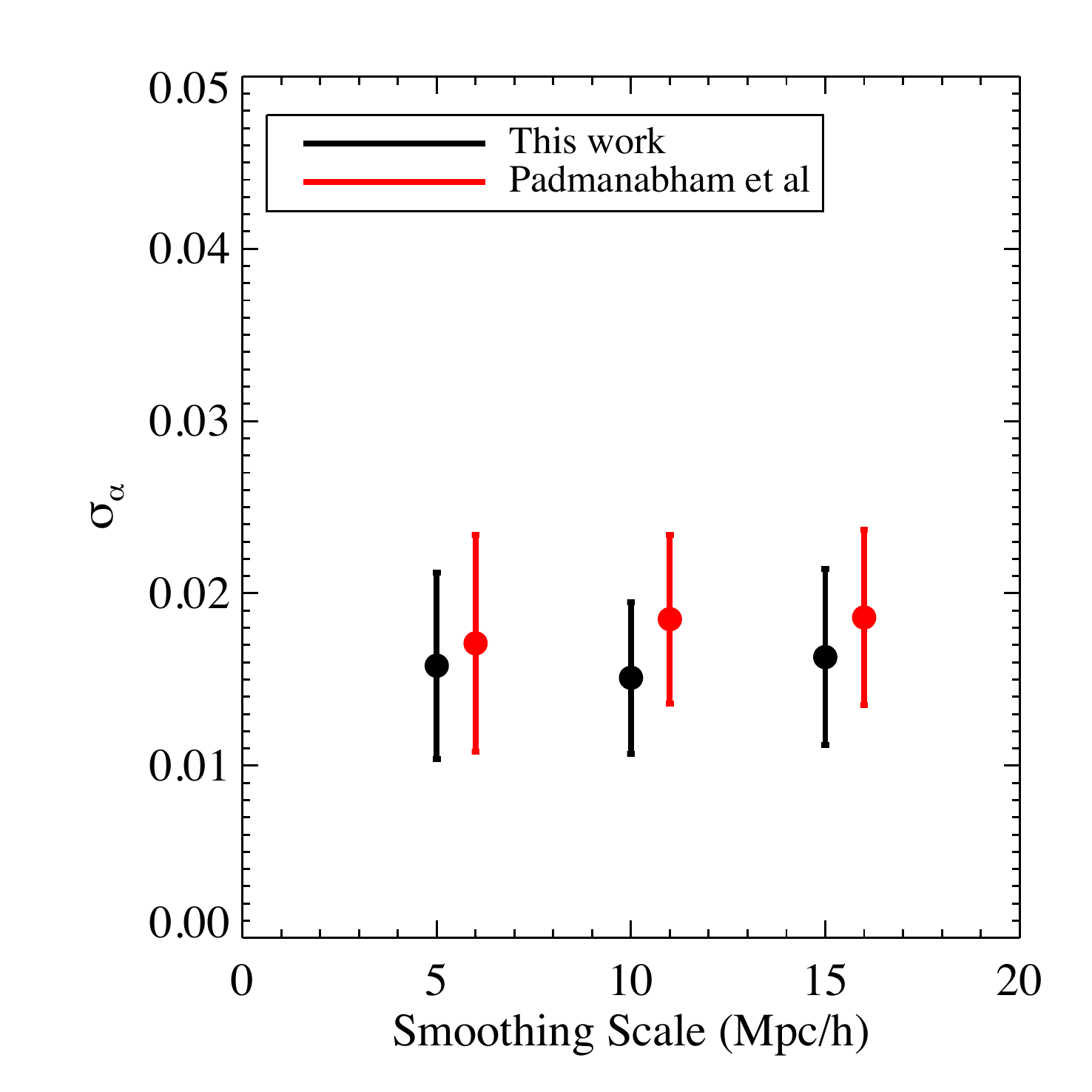}\includegraphics[width=1.6in]{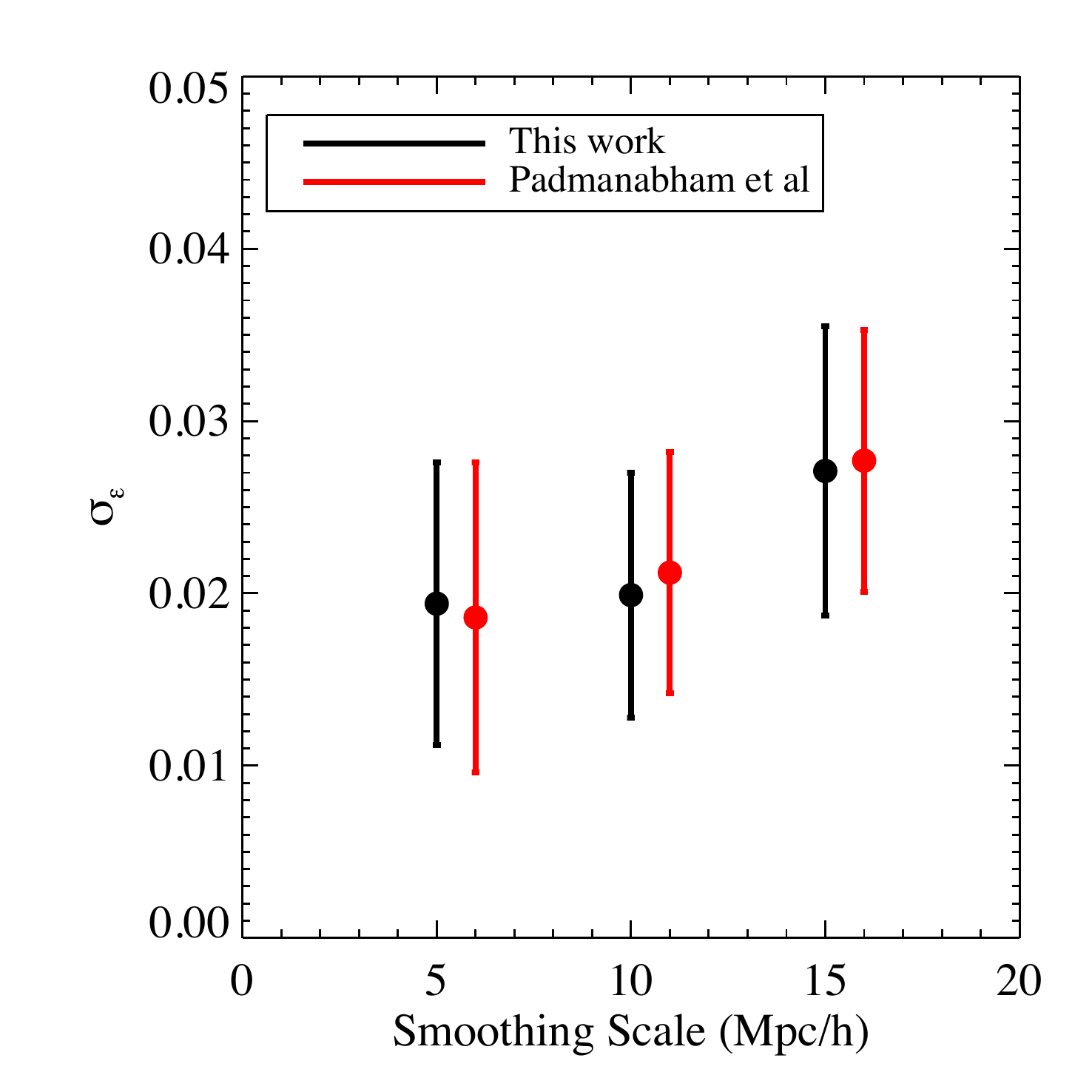}

\caption{Smoothing Fitting Results.Right, mean $\sigma_\alpha$ and RMS of $\sigma_\alpha$ distribution [right]. Left mean $\sigma_\epsilon$ and RMS for both implementations. We use a value of the bias=2.1 different than the value used in the article, i.e bias=1.87}.

   \label{fig:smoothing_fitserrors}
\end{figure}

Tables \ref{tab:fittingcomp} and \ref{tab:fittingerrors} summarise the best fitting results. The first table refers to the best fits of $\alpha$ and $\epsilon$ as well as their variance. In the second, table we show the mean values of the errors and the RMS of the error distributions.  Additionally, Figures \ref{fig:smoothing_fits} and \ref{fig:smoothing_fitserrors} show the comparison between both implementations' best fit parameters and errors (mean and RMS). The differences between the mean of two different implementations is  $\sim0.1\%$ for $\alpha$ and $\epsilon$, indicating the results are completely consistent between both implementations. The RMS of the best fits are \modiff{also} similar at 0.1\% in both quantities. 

Concerning the errors, the dispersion of the distributions are about the same; however, it seems  to be a systematic larger error in RP implementation, \modiff{as} compared to RV in $\alpha$ and $\epsilon$. Although the trends are observed in both implementations, the error in $\epsilon$ is monotonically decreasing as we apply a smaller smoothing scale.  In the case of $\alpha$ the errors we get with \modiff{the} 5 and 10 $h^{-1}$ Mpc smoothing scales are very similar \modiff{to} but smaller than using the 15 $h^{-1}$ Mpc that is regularly applied in BAO analysis.

\begin{table}
\begin{center}
\caption{Best fits. Mean and RMS from 100 reconstructed QPM NGC mocks for different Reconstruction Implementations. } 
\label{tab:fittingcomp}
\begin{tabular}{@{}lccccccccccccc}
\hline
$\Sigma$&
$\widetilde{\alpha}$&RMS&
$\widetilde{\epsilon}$&RMS\\
\hline
This work \\
\hline
5&
$0.9961$&$0.0091$&
$0.0036$&$0.0098$\\
\\[-1.5ex]

10&
$0.9980$&$0.0096$&
$0.0023$&$0.0107$\\
\\[-1.5ex]

15&
$1.0040$&$0.0095$&
$0.0050$&$0.0135$\\

\hline
Padmanabhan et al \\
\hline
5&
$0.9970$&$0.0100$&
$-0.0002$&$0.0088$\\
\\[-1.5ex]

10&
$0.9991$&$0.0116$&
$0.0022$&$0.0112$\\
\\[-1.5ex]

15&
$1.0053$&$0.0107$&
$0.0036$&$0.0141$\\

\hline

 \\[-1.5ex]
\end{tabular}
\end{center}
\end{table}

\begin{table}
\begin{center}
\caption{  Uncertainties on the best fits. Mean, RMS from 100 reconstructed QPM NGC mocks different Reconstruction Implementations. } 
\label{tab:fittingerrors}
\begin{tabular}{@{}lccccccccccccc}
\hline
$\Sigma$&
$\widetilde{\sigma_\alpha}$&RMS&
$\widetilde{\sigma_\epsilon}$&RMS\\
\hline
This work. \\
\hline
5&
$0.0158$&$0.0054$&
$0.0194$&$0.0082$& \\
\\[-1.5ex]
10&
$0.0151$&$0.0044$&
$0.0199$&$0.0071$\\
\\[-1.5ex]
15&
$0.0163$&$0.0051$&
$0.0271$&$0.0084$\\

\hline
Padmanabhan et al \\
\hline
5&
$0.0171$&$0.0063$&
$0.0186$&$0.0090$\\
\\[-1.5ex]
10&
$0.0185$&$0.0049$&
$0.0212$&$0.0070$\\
\\[-1.5ex]
15&
$0.0186$&$0.0051$&
$0.0277$&$0.0076$&\\

\hline
\end{tabular}
\end{center}
\end{table}

\begin{figure}
   \centering   
\includegraphics[width=2.5in]{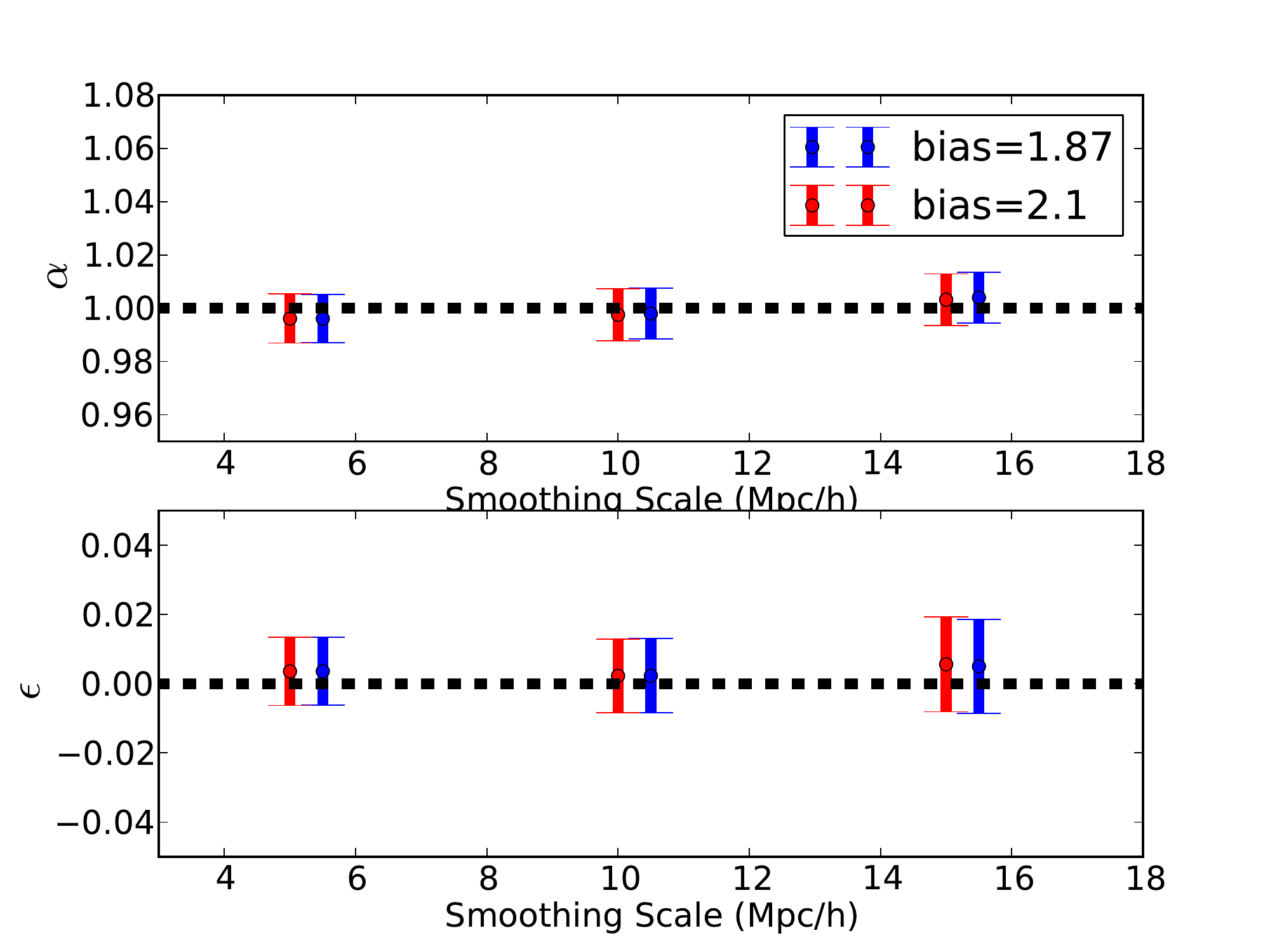}
\caption{Smoothing Fitting Results. Top panel, mean $\alpha$ and RMS of $\alpha$ distribution [right]. Bottom panel, mean $\epsilon$ and RMS for two values of the bias=1.87 and bias=2.1. Best fits and dispersion are similar for both values of the bias for all the smoothing scales.} 
   \label{fig:bias_fits_dep}
\end{figure}

\begin{figure}
   \centering   
\includegraphics[width=2.5in]{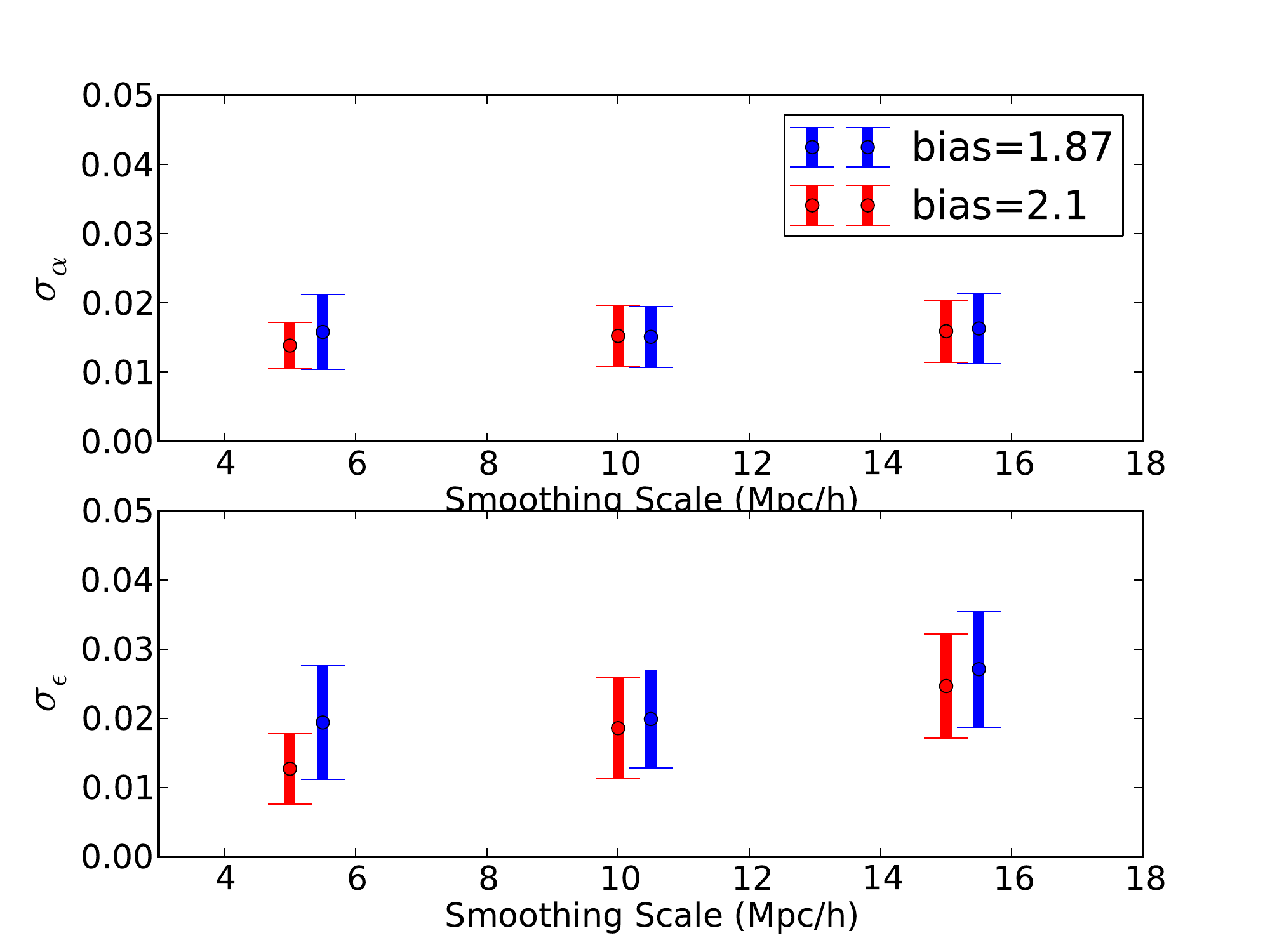}
\caption{Smoothing Fitting Results. Top panel, mean $\sigma_\alpha$ and RMS of $\sigma_\alpha$ distribution [right]. Bottom panel, mean $\sigma_\epsilon$ and RMS $\sigma_\epsilon$ distribution  for for two values of the bias=1.87 and bias=2.1. There is a large difference in the dispersion of the $\alpha$ and $\epsilon$ distributions for the lower smoothing scale; however, for 10-15 $h^{-1}$ Mpc, the results are consistent between both bias values.} 
   \label{fig:bias_fits_dep2}
\end{figure}

Finally, in Figure \ref{fig:bias_fits_dep}, we put together the results from Section \ref{sec:anisofits} and the results from this section for the anisotropic fits using the same methodology but varying the bias from b=1.87 to b=2.1. The results show that the bias is not affecting the best fits, i.e, the conclusions we get for the smoothing scales in terms of anisotropic results are valid even considering other bias values. 
It is interesting to mention that even when comparing Figures \ref{fig:meanSmooth} and \ref{fig:mean_smooth_bias21}, we observe that the bias affects the quadrupole amplitude post-reconstruction, this does not affect the best fitting values. 
The uncertainties are shown in Figure \ref{fig:bias_fits_dep2} for both bias values. We observe that the distributions of the uncertainties are quite similar for 10-15 $h^{-1}$ Mpc smoothing scale for both $\alpha$ and $\epsilon$ distributions; however, the lower smoothing scale shows a large dispersion using the bias=1.87 compared to the bias=2.1. This result seems to show that the bias can play a role  in the uncertainties' determination when using small smoothing scales. 

\section{Bias and Noise Determination}\label{ap:biasnoise}
While the equation (\ref{eq:postulat}) provides us a relation between the variances of reconstructed and true displacement, we need to break the degeneracy between the bias $\alpha$ and the noise effect. We propose to use the correlation plot between $\tilde{\Psi}_R$ and $\Psi$.
We will express the theoretical probability distribution $\mathcal{P}(\tilde{\psi}_R,\psi)$ knowing\footnote{$\sigma_{\psi}$ is a measured quantity and so is known too.} the bias $\alpha$, the noise standard deviation $\sigma_N$.

\begin{equation}
\mathcal{P}(\tilde{\psi}_R,\psi) = \mathcal{P}(\psi)\times\mathcal{P}(\tilde{\psi}_R|\psi).
\end{equation}
Assuming a Gaussian distribution, we can express the probability distribution of the ``true" displacements  by: 
\begin{equation}
\mathcal{P}(\psi) = \frac{1}{\sqrt{2\pi}\sigma_{\psi}}\exp\left\{-\frac{\psi^2}{2\sigma_{\psi}^2}\right\}.
\end{equation}

Using the relation expressed in equation (\ref{eq:postulat}) we can get the expression for the conditional probability distribution $\mathcal{P}(\tilde{\psi}_R,\psi)$:

\begin{equation}
\mathcal{P}(\tilde{\psi}_R,\psi) = \frac{1}{2\pi\sigma_{\psi}\sigma_{N}}\exp\left\{-\frac{1}{2}\left[\frac{\psi^2}{\sigma_{\psi}^2}+\frac{(\tilde{\psi}_R-\alpha\psi)^2}{\sigma_{N}^2}\right]\right\}.
\end{equation}

The relation we expect to measure between $\tilde{\psi}_R$ and $\psi$ is where $\mathcal{P}(\tilde{\psi}_R,\psi)$ is maximum depending on $\tilde{\psi}_R$ value, i.e. we are interested in the value $\psi_{max}$, which maximises the exponential term for a given $\tilde{\psi}_R$. Then we use the partial derivation:
\begin{equation}
\left.\frac{\partial }{\partial_{\psi}} \left(\frac{\psi^2}{\sigma_{\psi}^2}+\frac{(\tilde{\psi}_R-\alpha\psi)^2}{\sigma_{N}^2}\right)\right|_{\tilde{\psi}_R} = 0.
\end{equation}

Developing this calculus, we obtain the following expression for $\psi_{max}$:
\begin{equation}
\psi_{max} = \frac{\alpha\tilde{\psi}_R}{\frac{\sigma_N^2}{\sigma_{\psi}^2} + \alpha^2 },
\label{eq:rel_psi_max}
\end{equation}
and so $\tilde{\alpha}$ is given by:
\begin{equation}
\tilde{\psi}_R = \tilde{\alpha} \psi_{max} \Rightarrow\tilde{\alpha} = \frac{ \frac{\sigma_N^2}{\sigma_{\psi}^2} + \alpha^2  }{\alpha}.
\label{eq:second_rel}
\end{equation}

We find the relation between $\tilde{\psi}_R$ and $\psi_{max}$, which can be thought of as representing the apparently most correlated value.  

The 2D correlation provides us an estimation of the bias $\tilde{\alpha}$ which is different than $\alpha$ except for a null noise term $\sigma_N = 0$. 

We have two independent equations (Eq. \ref{eq:first_rel} and Eq. \ref{eq:second_rel} ) which connect ($\alpha$, $\sigma_N$) to the measurable quantities ($\tilde{\alpha}$, $\sigma_{\psi}$, $\sigma_{\tilde{\psi}_R}$).

\begin{figure} 
\centering
 \includegraphics[width=1.6in]{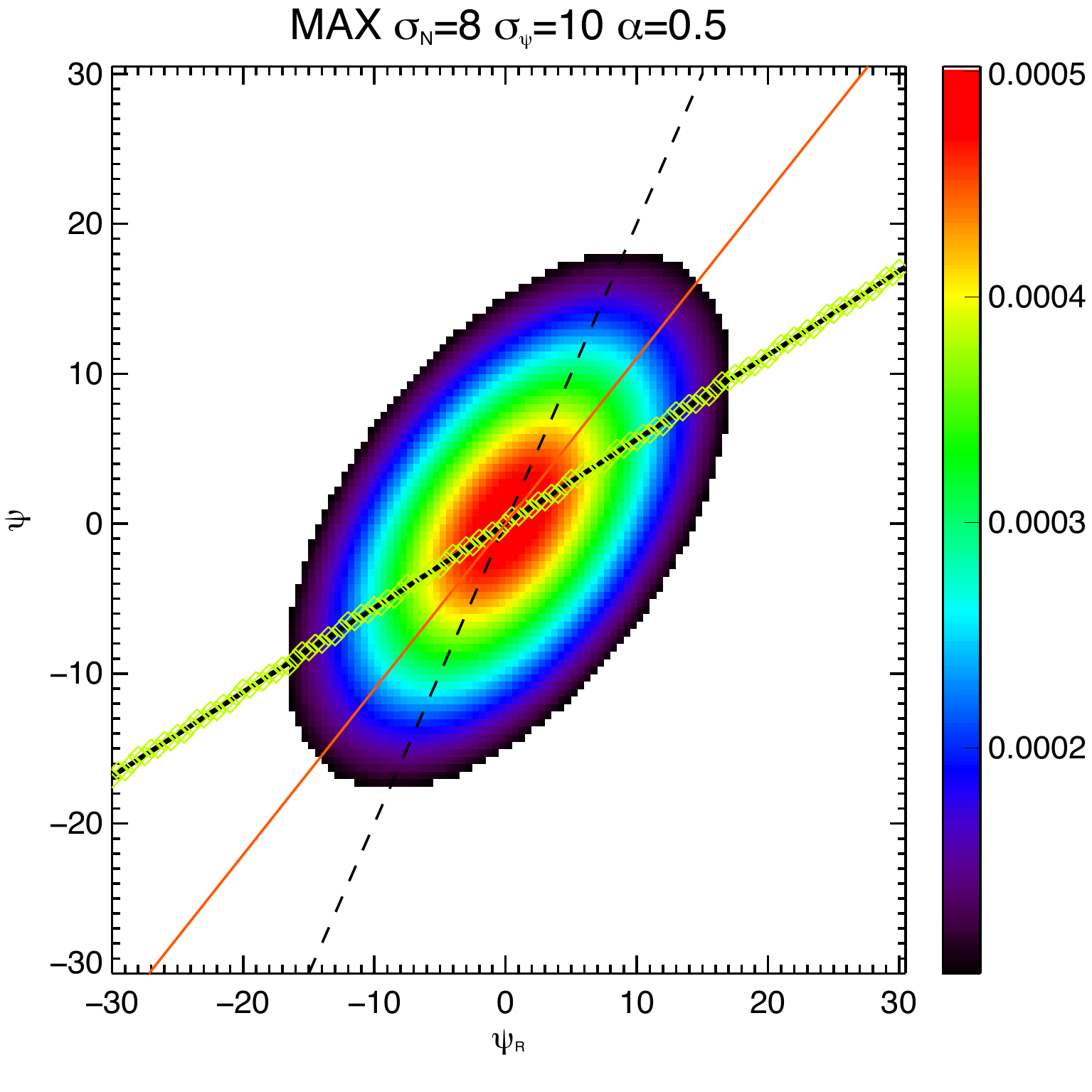}\includegraphics[width=1.6in]{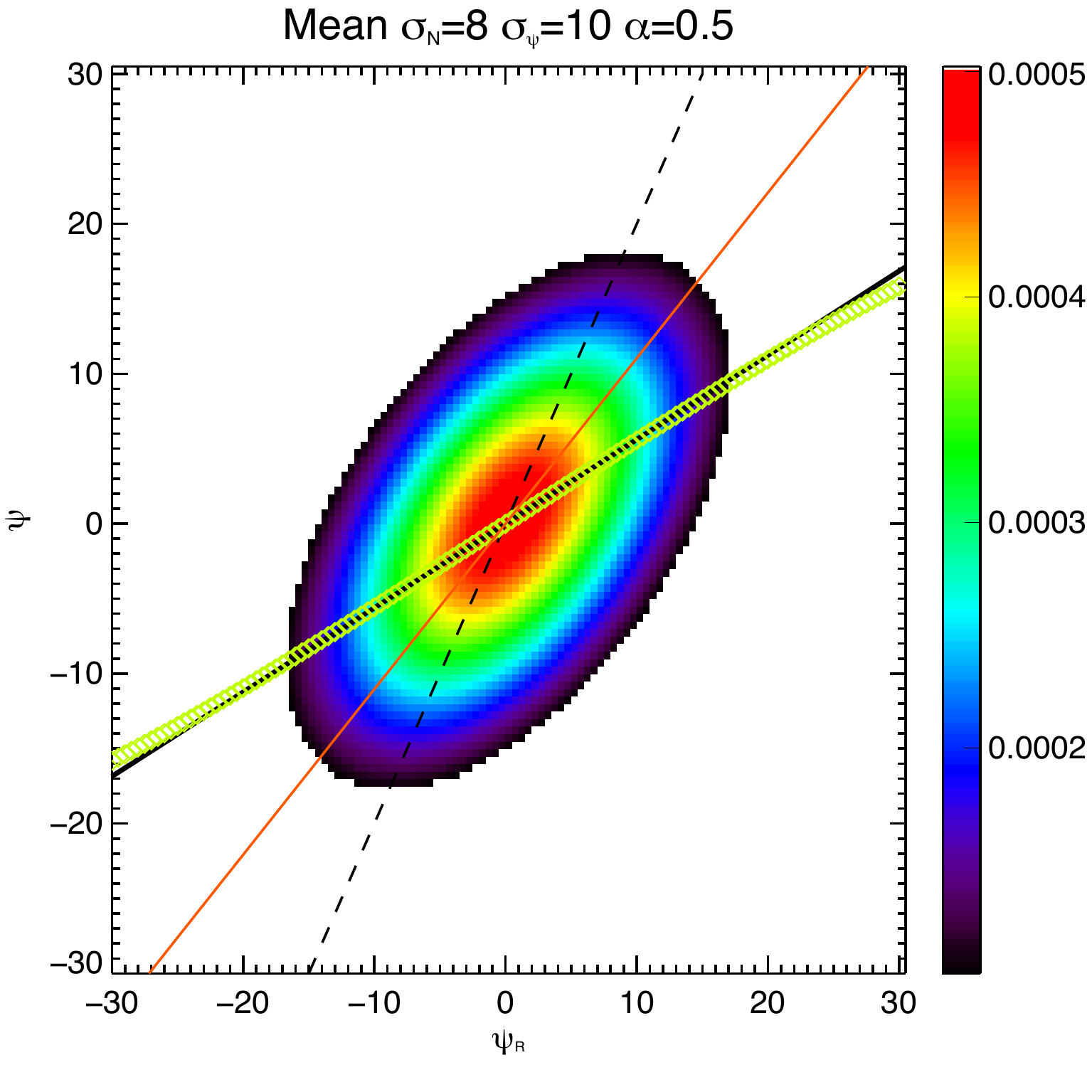}
        
  \caption{$\mathcal{P}(\tilde{\psi}_R,\psi)$ for $\alpha=0.5$, $\sigma_N=8$ and $\sigma_{\psi}=10$. The dashed black line represents the relation between $\tilde{\psi}_R$ and $\psi$ without noise term (so $\alpha$), the solid black line represents the theoretical relation obtained using Eq. \ref{eq:second_rel}  (so $\tilde{\alpha}$); the solid red line represents the major axis orientation using a 2D-Gaussian fit. The yellow diamonds correspond respectively to the measured $\psi_{max}(\tilde{\psi}_R)$ on the left and the mean value $\left<\psi(\tilde{\psi}_R)\right>$ on the right panel.}
   \label{fig:plot2D}
\end{figure}

Combining  Eq. \ref{eq:second_rel}  and Eq. \ref{eq:first_rel}, we obtain:
\begin{equation}
\label{eqa:alpha}
\alpha = \frac{\sigma_{\tilde{\psi}_R}^2}{\tilde{\alpha}\sigma_{\psi}^2}, 
\end{equation}
and

\begin{equation}
\sigma_N^2 = \tilde{\sigma}_R^2 - \frac{\tilde{\sigma}_R^4}{\tilde{\alpha}^2\sigma_{\psi}^2},
\label{eqa:sigma}
\end{equation}
for which, we present the comparison with measurements in Figure \ref{fig:plot2D}. The plot shows the $\mathcal{P}(\tilde{\psi}_R,\psi)$ for $\alpha=0.5$, $\sigma_N=8$ and $\sigma_{\psi}=10$. The dashed black line represents the relation between $\tilde{\psi}_R$ and $\psi$ without noise term (so $\alpha$), the solid black line represents the theoretical relation obtained using Eq. \ref{eq:second_rel}  (so $\tilde{\alpha}$); the solid red line represent the major axis orientation using a 2D-Gaussian fit. The yellow diamonds correspond respectively to the measured $\psi_{max}(\tilde{\psi}_R)$ on the left and the mean value $\left<\psi(\tilde{\psi}_R)\right>$ on the right panel. From these tests, we conclude that the bias and noise estimation described below describes the bias and noise levels observed in our Monte-Carlo realisations.

We can write the de-biased estimator $\overset{D}{\psi_R}$ inverting Eq. (\ref{eq:postulat}): 
\begin{equation}
\overset{D}{\psi_R} = \psi+\frac{\mathcal{N}(0,\sigma_N)}{\alpha}.
\end{equation}
Also we can write the dispersion on the de-biased reconstruction as:
\begin{equation}
\overset{D}{\psi_R} - \psi =  \frac{\mathcal{N}(0,\sigma_N)}{\alpha}.
\end{equation}

\begin{figure} 
\centering
 \includegraphics[width=1.6in]{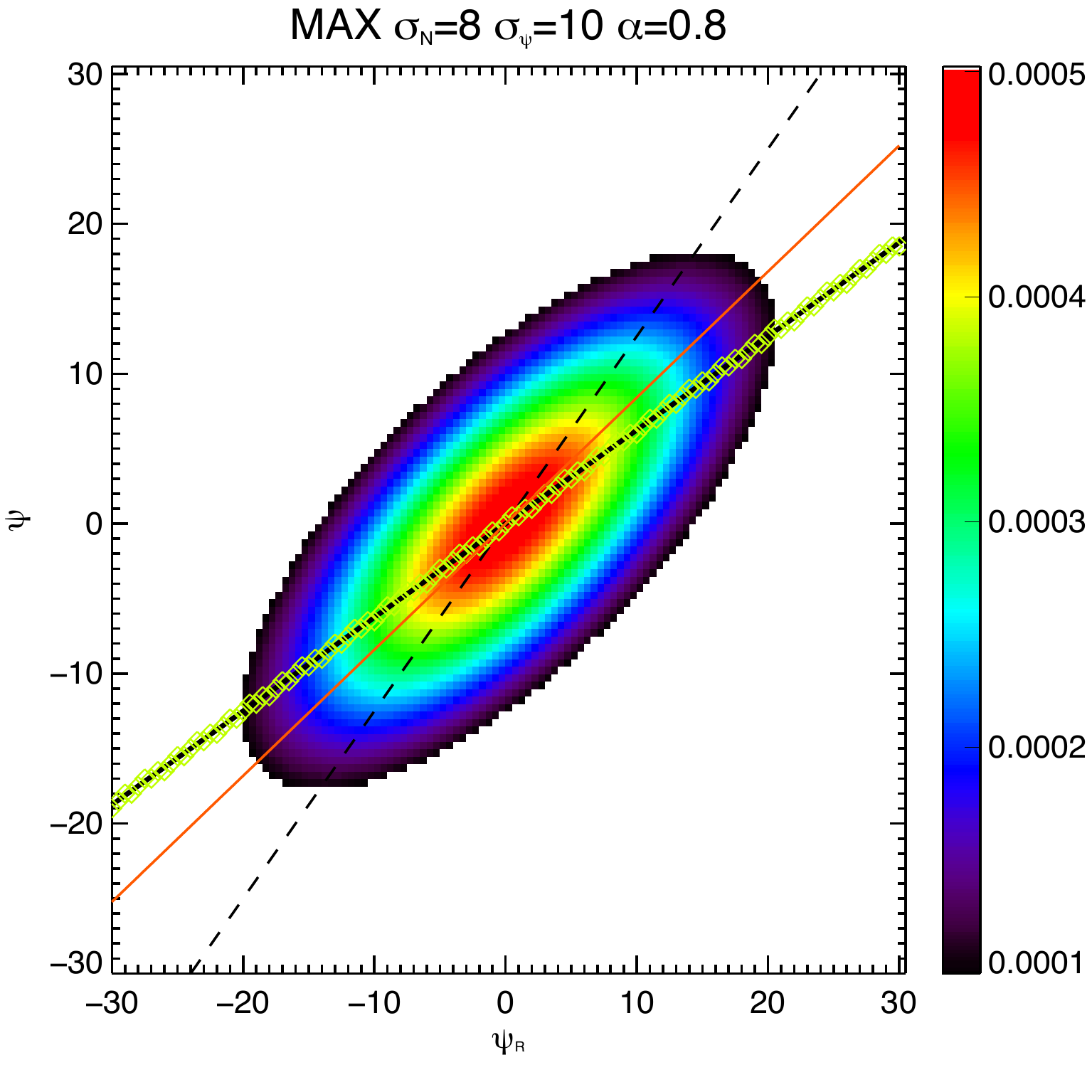}\includegraphics[width=1.6in]{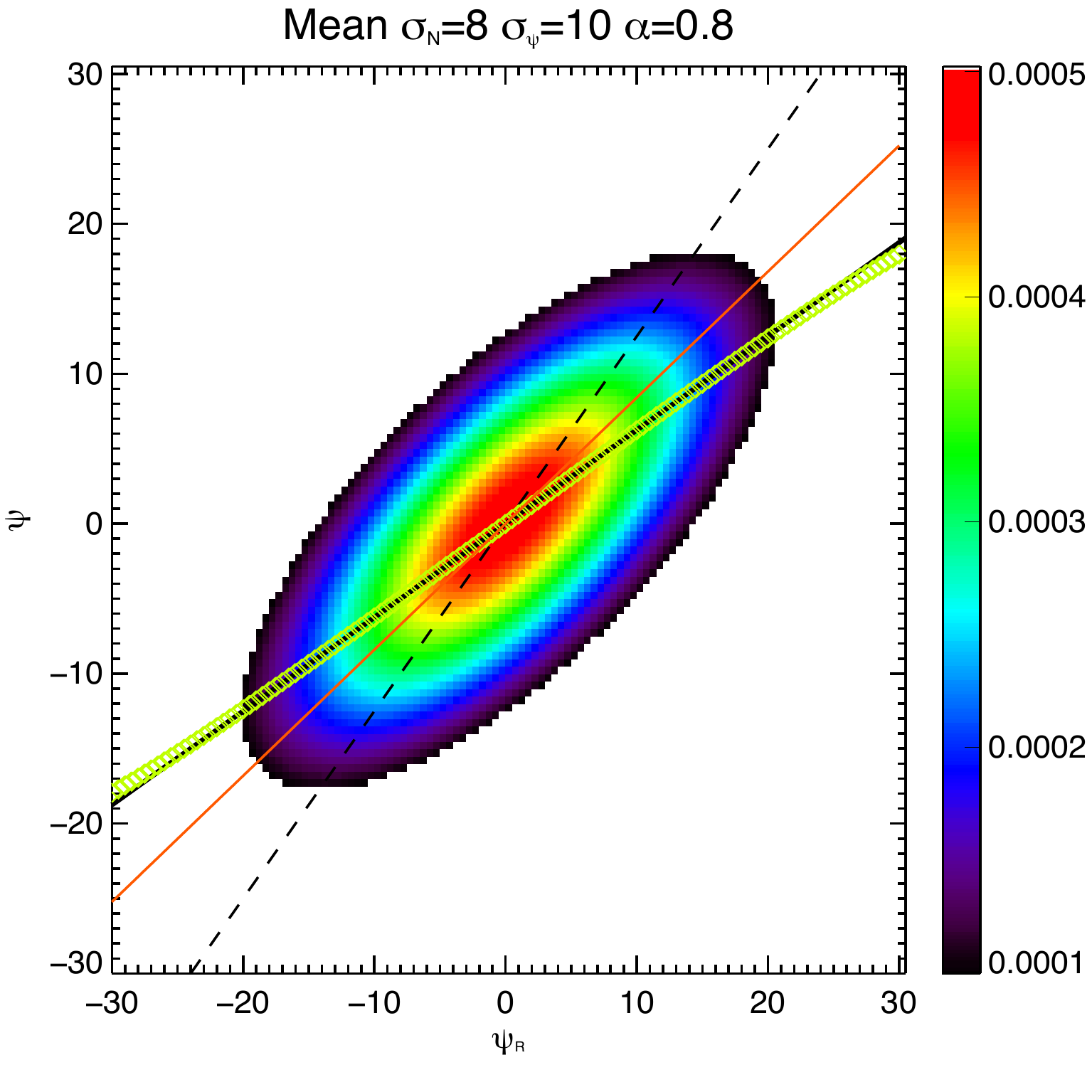}
        
  \caption{$\mathcal{P}(\tilde{\psi}_R,\psi)$ for $alpha=0.8$, $\sigma_N=8$ and $\sigma_{\psi}=10$. The dashed black line represents the relation between $\tilde{\psi}_R$ and $\psi$ without noise term (so $\alpha$), the solid black line represents the theoretical relation obtained using Eq. \ref{eq:second_rel}  (so $\tilde{\alpha}$); the solid red line represents the major axis orientation using a 2D-Gaussian fit. The yellow diamonds correspond respectively to the measured $\psi_{max}(\tilde{\psi}_R)$ on the left and the mean value $\left<\psi(\tilde{\psi}_R)\right>$ on the right panel.}
   \label{fig:plot2D_08}
\end{figure}

\end{document}